\documentclass[english,jcp,preprint,endfloats,groundaddress]{revtex4-1} 
\usepackage{epsfig} 
\usepackage{amsmath} 
\usepackage{graphicx} 
\usepackage{longtable} 
\usepackage{hyperref} 
\usepackage{amssymb} 
\usepackage{dcolumn} 
\usepackage{mhchem} 
\usepackage{subfigure} 
\usepackage{array} 
\usepackage{multirow} 
\usepackage{tabularx} 
\usepackage{cleveref} 
\begin{document} 

\title{Role of exact exchange in thermally-assisted-occupation density functional theory: A proposal of new hybrid schemes} 

\author{Jeng-Da Chai} 
\email[Electronic mail: ]{jdchai@phys.ntu.edu.tw} 
\affiliation{Department of Physics, Center for Theoretical Sciences, and Center for Quantum Science and Engineering, National Taiwan University, Taipei 10617, Taiwan} 

\date{\today} 

\begin{abstract} 

We propose hybrid schemes incorporating exact exchange into thermally-assisted-occupation density functional theory (TAO-DFT) [J.-D. Chai, J. Chem. Phys. {\bf 136}, 154104 (2012)] 
for an improved description of nonlocal exchange effects. With a few simple modifications, global and range-separated hybrid functionals in Kohn-Sham density functional theory (KS-DFT) 
can be combined seamlessly with TAO-DFT. In comparison with global hybrid functionals in KS-DFT, the resulting global hybrid functionals in TAO-DFT yield promising performance for 
systems with strong static correlation effects (e.g., the dissociation of H$_2$ and N$_2$, twisted ethylene, and electronic properties of linear acenes), while maintaining similar performance 
for systems without strong static correlation effects. Besides, a reasonably accurate description of noncovalent interactions can be efficiently achieved through the inclusion of dispersion 
corrections in hybrid TAO-DFT. Relative to semilocal density functionals in TAO-DFT, global hybrid functionals in TAO-DFT are generally superior in performance for a wide range of 
applications, such as thermochemistry, kinetics, reaction energies, and optimized geometries. 

\end{abstract} 

\maketitle 

\section{Introduction} 

Over the past two decades, Kohn-Sham density functional theory (KS-DFT) \cite{HK,KS} has emerged as one of the most popular electronic structure methods for the study of large 
ground-state systems, due to its low computational cost and reasonable accuracy \cite{Parr,DFTreview1,DFTreview2,DFTreview3}. Nevertheless, the essential ingredient of KS-DFT, the 
exact exchange-correlation (XC) energy functional $E_{xc}[\rho]$ remains unknown, and needs to be approximated. Consequently, density functional approximations (DFAs) for 
$E_{xc}[\rho]$ have been continuously developed to improve the accuracy of KS-DFT for a broad range of applications. 

Functionals based on the conventional semilocal DFAs, such as the local density approximation (LDA) \cite{LDAX,LDAC} and generalized gradient approximations 
(GGAs) \cite{B88,LYP,PBE}, can yield reasonably accurate predictions of the properties governed by short-range XC effects, and possess high computational efficiency for very large 
systems (for brevity, hereafter we use ``DFAs" for ``the conventional semilocal DFAs"). 
Nonetheless, owing to the inappropriate treatment of nonlocal XC effects \cite{Perdew09,SciYang}, KS-DFAs can perform very poorly in situations where the self-interaction error 
(SIE) \cite{PZ-SIC,SIE,SciYang,Perdew09}, noncovalent interaction error (NCIE) \cite{Dobson,D3review,SLR-vdW}, or static correlation error (SCE) \cite{SciYang,SCE,MR1,TAO1,TAO2} 
is pronounced. Over the years, considerable efforts have been made to resolve the qualitative failures of KS-DFAs at a reasonable computational cost. 

To date, global hybrid functionals \cite{hybrid1,hybrid2,B3LYP,DFA0,DFA0a,PBE0,PBE0a,SCAN0} and range-separated hybrid functionals \cite{RSH,wB97X,wB97X-2}, which incorporate 
the Hartree-Fock (HF) exchange energy into KS-DFAs, are perhaps the most successful schemes that provide an improved description of nonlocal exchange effects. Relative to KS-DFAs, 
the hybrid schemes, which greatly reduce the SIE problems, are reliably accurate for a wide variety of applications, such as thermochemistry and kinetics \cite{LCAC,EB}. 

To properly describe noncovalent interactions, a reasonably accurate treatment of middle- and long-range dynamical correlation effects is critical. Accordingly, KS-DFAs and hybrid 
functionals may be combined with the DFT-D (KS-DFT with empirical dispersion corrections) schemes \cite{D3review,DFT-D1,DFT-D2,DFT-D3,Sherrill,DFT-D3appl} and the double-hybrid 
(mixing both the HF exchange energy and the second-order M\o ller-Plesset (MP2) correlation energy \cite{MP2} into KS-DFAs) schemes \cite{DH,SCAN0,wB97X-2}, showing an overall 
satisfactory accuracy for the NCIE problems. 

In spite of their computational efficiency, KS-DFAs, hybrid functionals, and double-hybrid functionals can perform very poorly for systems with strong static correlation effects (i.e., 
multi-reference systems) \cite{SciYang,SCE,MR1,TAO1,TAO2}. Within KS-DFT, fully nonlocal XC functionals, such as those based on the random phase approximation (RPA), may be 
adopted for a reliably accurate description of strong static correlation effects. However, RPA-type functionals remain computationally very demanding for large 
systems \cite{DFTreview1,Perdew09,RPA_Yang,RPA_Yang2}. 

To reduce the SCE problems with low computational complexity, we have recently developed thermally-assisted-occupation density functional theory (TAO-DFT) \cite{TAO1,TAO2}, 
an efficient electronic structure method for studying the ground-state properties of very large systems (e.g., containing up to a few thousand electrons) with strong static correlation 
effects \cite{TAO3,TAO4,TAO5,TAO6}. Unlike finite-temperature DFT \cite{Mermin}, TAO-DFT is developed for ground-state systems at zero temperature. In contrast to KS-DFT, TAO-DFT 
is a DFT with fractional orbital occupations given by the Fermi-Dirac distribution (controlled by a fictitious temperature $\theta$), wherein strong static correlation is explicitly described by 
the entropy contribution (e.g., see Eq.\ (26) of Ref.\ \cite{TAO1}). Interestingly, TAO-DFT is as efficient as KS-DFT for single-point energy and analytical nuclear gradient calculations, and 
is reduced to KS-DFT in the absence of strong static correlation effects. Besides, existing DFA XC functionals in KS-DFT may also be adopted in TAO-DFT. The resulting TAO-DFAs have 
been shown to consistently improve upon KS-DFAs for multi-reference systems. Nevertheless, TAO-DFAs perform similarly to KS-DFAs for single-reference systems (i.e., systems without 
strong static correlation). In addition, the SIEs and NCIEs of TAO-DFAs may remain enormous in situations where these failures occur. 

In this work, we aim to improve the accuracy of TAO-DFAs for a wide variety of single-reference systems. 
Specifically, we develop hybrid schemes that incorporate exact exchange into TAO-DFAs for an improved description of nonlocal exchange effects. 
Hybrid functionals (e.g., global and range-separated hybrids) in KS-DFT can be easily modified, and seamlessly combined with TAO-DFT. 
The rest of the paper is organized as follows. A brief review of the essentials of TAO-DFT is provided in Section II. 
In Section III, the exact exchange in TAO-DFT is defined, and the corresponding global and range-separated hybrid schemes are proposed. In Section IV, the optimal $\theta$ values 
for global hybrid functionals in TAO-DFT are defined, and the performance of global hybrid functionals in TAO-DFT (with the optimal $\theta$ values) is examined for various single- 
and multi-reference systems. Our conclusions are given in Section V.

\section{TAO-DFT} 

\subsection{Rationale for fractional orbital occupations} 

Consider an interacting $N$-electron Hamiltonian for an external potential $v({\bf r})$ at zero temperature, the exact ground-state density $\rho({\bf r})$ is interacting $v$-representable, 
as it can be obtained from the ground-state wavefunction calculated using the full configuration interaction (FCI) method at the complete basis set limit \cite{H2_NOON}: 
\begin{equation}\label{eq:rhoci} 
\rho({\bf r}) = \sum_{i=1}^{\infty} n_{i} |\chi_{i}({\bf r})|^{2}, 
\end{equation} 
which can be expressed in terms of the natural orbitals (NOs) $\{\chi_{i}({\bf r})\}$ and natural orbital occupation numbers (NOONs) $\{n_{i}\}$ (i.e., the eigenfunctions and eigenvalues, 
respectively, of one-electron reduced density matrix (1-RDM)) \cite{NO}. Here, the NOONs $\{n_{i}\}$, obeying the following two conditions: 
\begin{equation}\label{eq:rhoci2} 
\sum_{i=1}^{\infty} n_{i} = N,\  \ 0\le n_{i} \le 1, 
\end{equation} 
are related to the variationally determined coefficients of the FCI expansion. As shown in Eq.\ (\ref{eq:rhoci}), the exact ground-state density $\rho({\bf r})$ can be represented by orbitals 
and their occupation numbers, showing the significance of an ensemble representation (via fractional orbital occupations) of the ground-state density. 

By contrast, in KS-DFT, the ground-state density $\rho({\bf r})$ is assumed to be noninteracting pure-state $v_s$-representable, as it belongs to a one-determinant ground-state 
wavefunction of a noninteracting $N$-electron Hamiltonian for some local potential $v_s({\bf r})$ at zero temperature \cite{v-rep,Levy,Lieb}. Accordingly, the Kohn-Sham (KS) orbital 
occupation numbers should be either 0 or 1. Due to the search over a restricted domain of densities, some ground-state densities cannot be obtained within the framework of KS-DFT 
(i.e., even with the exact $E_{xc}[\rho]$) \cite{Levy,Lieb,Baerends,Morrison,Katriel}. Baerends and co-workers \cite{Baerends} argued that the ground-state density $\rho({\bf r})$ of a 
system with strong static correlation effects may not be noninteracting pure-state $v_s$-representable, wherein an ensemble representation of the ground-state density is essential. 
Arguments supporting this are also available from other studies \cite{Morrison,SDFTYang}. 

To rectify the above situation, KS-DFT has been extended to ensemble DFT \cite{EDFT,EDFT2}, wherein $\rho({\bf r})$ is assumed to be noninteracting ensemble $v_s$-representable, 
as it is associated with an ensemble of pure determinantal states of the noninteracting KS system at zero temperature. Accordingly, the orbital occupation numbers in ensemble DFT are 
0, 1, and fractional (between 0 and 1) for the orbitals above, below, and at the Fermi level, respectively. Within the framework of ensemble DFT, the development of DFT 
fractional-occupation-number (DFT-FON) method \cite{R1,DFT-FON2,DFT-FON3,R9,SDFTYang}, spin-restricted ensemble-referenced KS (REKS) method \cite{REKS,REKS2}, and 
fractional-spin DFT (FS-DFT) method \cite{SciYang,SCE} has yielded great success for some systems with strong static correlation effects. Nevertheless, the practical implementation 
of DFT-FON and related methods has been hindered by several factors, such as a possible double-counting of correlation effects and the sharp increase of computational cost for large 
systems. 

On the other hand, the inclusion of fractional occupation numbers (FONs) in electronic structure calculations has a long history \cite{R1,R2,R3,R4,R5,R6,FTHF5,R8,R9,R10,R11,
Mermin,FTHF7,FTHF8,FTHF1,FTHF2,FTHF3,FTHF4,FTHF6,FTHF9,EDFT,EDFT2,DFT-FON2,DFT-FON3,SDFTYang,REKS,REKS2,SciYang,SCE,TAO1,TAO2}. 
In particular, the Fermi-Dirac distribution, which appears in finite-temperature DFT \cite{Mermin} and finite-temperature HF 
schemes \cite{FTHF1,FTHF2,FTHF3,FTHF4,FTHF5,FTHF6,FTHF7,FTHF8,FTHF9}, has been a popular distribution function for the FON-related schemes. For example, 
finite-temperature techniques have been developed for improving self-consistent field (SCF) convergence \cite{R5}. The grand canonical orbitals have been used for subsequent 
complete active space configuration interaction (CASCI) calculations \cite{R6,FTHF5,R8}. Recently, a fractional occupation number weighted electron density has been adopted for 
a real-space measure and visualization of static correlation effects \cite{FTHF8}. 

In TAO-DFT \cite{TAO1,TAO2}, the representation of the ground-state density from the exact theory (see Eq.\ (\ref{eq:rhoci})) has been highlighted. In contrast to the orbital occupation 
numbers in KS-DFT and ensemble DFT, the NOONs can be fractional (between 0 and 1) for all the NOs. While the exact NOONs are intractable for large systems (due to the 
exponential complexity), the distribution of NOONs (the microcanonical averaging of NOONs) can, however, be approximately described by the Fermi-Dirac distribution with 
renormalized parameters (i.e., orbital energies, chemical potential, and temperature) based on the statistical arguments of Flambaum {\it et al.} \cite{Flambaum}. 
Accordingly, in TAO-DFT, the ground-state density $\rho({\bf r})$ of a system of $N$ interacting electrons moving in an external potential $v_{ext}({\bf r})$ at zero temperature is 
assumed to be noninteracting thermal ensemble $v_s$-representable, as it is expressed as the thermal equilibrium density of an auxiliary system of $N$ noninteracting electrons 
moving in some local potential $v_s({\bf r})$ at a fictitious temperature $\theta$. Consequently, $\rho({\bf r})$ can be represented by 
\begin{equation}\label{eq:taoa1} 
\rho({\bf r}) = \sum_{i=1}^{\infty} f_{i} |\psi_{i}({\bf r})|^{2}, 
\end{equation} 
where the orbital occupation number $f_{i}$ is the Fermi-Dirac distribution 
\begin{equation}\label{eq:taoa2} 
f_{i} = \{1+\text{exp}[(\epsilon_{i} - \mu)/ \theta] \}^{-1}, 
\end{equation} 
which satisfies the following two conditions: 
\begin{equation}\label{eq:taoa3} 
\sum_{i=1}^{\infty} f_{i} = N,\  \ 0\le f_{i} \le 1, 
\end{equation} 
$\epsilon_{i}$ is the orbital energy of the $i$-th orbital $\psi_{i}({\bf r})$, and $\mu$ is the chemical potential determined by the conservation of the number of electrons $N$. 

As discussed in Ref.\ \cite{TAO1}, for a given fictitious temperature $\theta$, the Hohenberg-Kohn theorems \cite{HK} and the Mermin theorems \cite{Mermin} can be employed for 
the physical and auxiliary systems, respectively, to derive a set of self-consistent equations in TAO-DFT for determining the remaining ``renormalized parameters'' (i.e., the orbital 
energies $\{\epsilon_{i}\}$ and chemical potential $\mu$) of the orbital occupation numbers $\{f_{i}\}$ and the orbitals $\{\psi_{i}({\bf r})\}$, which can then be used to represent the 
ground-state density $\rho({\bf r})$, and evaluate the ground-state energy of the physical system at zero temperature. In addition, due to the similarity of Eqs.\ (\ref{eq:rhoci}) and 
(\ref{eq:taoa1}), when the fictitious temperature $\theta$ in TAO-DFT is so chosen that the NOONs $\{n_{i}\}$ are approximately described by the orbital occupation numbers 
$\{f_{i}\}$ (in the sense of statistical average, as mentioned above), the NOs $\{\chi_{i}({\bf r})\}$ will be approximately described by the orbitals $\{\psi_{i}({\bf r})\}$. This implies 
that the exact $\rho({\bf r})$ is likely to be noninteracting thermal ensemble $v_s$-representable at this $\theta$ value (plus some range of possible other values around it). 
In addition, as discussed in Ref.\ \cite{TAO1}, strong static correlation has been shown to be properly described by the entropy contribution (e.g., see Eq.\ (26) of Ref.\ \cite{TAO1}) 
in TAO-DFT at this $\theta$ value (plus some range of possible other values around it). 

While also adopting the Fermi-Dirac distribution, TAO-DFT is developed for the ground-state density and ground-state energy of a physical system at zero temperature, which is 
different from the aforementioned finite-temperature FON-related schemes (which mostly focus on the SCF convergence, the adoption of grand canonical orbitals and density for 
different purposes, and the thermodynamic properties of a physical system at finite temperature). On the other hand, while KS-DFT, ensemble DFT, and TAO-DFT all belong to 
zero-temperature DFT, the representations of the ground-state density are, however, different in these methods (as mentioned above). While the entropy contribution in TAO-DFT 
plays an important role in simulating strong static correlation (even though at the price of adding an extra $\theta$ parameter that is related to the distribution of NOONs), 
this term is, however, absent in KS-DFT and ensemble DFT.

\subsection{Self-consistent equations} 

Consider a system of $N_{\alpha}$ up-spin and $N_{\beta}$ down-spin electrons moving in an external potential $v_{ext}({\bf r})$ at zero (physical) temperature. 
In spin-polarized (spin-unrestricted) TAO-DFT \cite{TAO1,TAO2}, two noninteracting reference systems at the same fictitious (reference) temperature $\theta$ (measured in energy units) 
are employed: one described by the spin function $\alpha$ and the other described by the spin function $\beta$, with the corresponding thermal equilibrium density distributions 
$\rho_{s,\alpha}({\bf r})$ and $\rho_{s,\beta}({\bf r})$ exactly equal to the up-spin density $\rho_{\alpha}({\bf r})$ and down-spin density $\rho_{\beta}({\bf r})$, respectively, in the original 
interacting system at zero temperature. The resulting self-consistent equations for the $\sigma$-spin electrons ($\sigma$ = $\alpha$ or $\beta$) can be 
expressed as ($i$ runs for the orbital index) 
\begin{equation}\label{eq:tao1} 
\bigg\lbrace -\frac{1}{2} {\bf \nabla}^2 \ + \ v_{s,\sigma}({\bf r}) \bigg\rbrace \psi_{i\sigma}({\bf r}) = \epsilon_{i\sigma} \psi_{i\sigma}({\bf r}), 
\end{equation} 
where 
\begin{equation}\label{eq:tao2} 
v_{s,\sigma}({\bf r}) = v_{ext}({\bf r}) + \int \frac{\rho({\bf r'})}{|{\bf r} - {\bf r'}|}d{\bf r'} + \frac{\delta E_{xc}[\rho_{\alpha},\rho_{\beta}]}{\delta \rho_{\sigma}({\bf r})} 
+ \frac{\delta E_{\theta}[\rho_{\alpha},\rho_{\beta}]}{\delta \rho_{\sigma}({\bf r})} 
\end{equation} 
is the effective potential (atomic units, i.e., $\hbar = m_e = e = 4 \pi \epsilon_0 = 1$, are adopted throughout this work). Here, 
$E_{xc}[\rho_{\alpha},\rho_{\beta}] \equiv E_{x}[\rho_{\alpha},\rho_{\beta}] + E_{c}[\rho_{\alpha},\rho_{\beta}]$ is the XC energy (i.e., the sum of the exchange energy 
$E_{x}[\rho_{\alpha},\rho_{\beta}]$ and correlation energy $E_{c}[\rho_{\alpha},\rho_{\beta}]$) defined in spin-polarized KS-DFT \cite{SDFT,SDFTYang}, and 
$E_{\theta}[\rho_{\alpha},\rho_{\beta}] \equiv A_{s}^{\theta=0}[\rho_{\alpha},\rho_{\beta}] - A_{s}^{\theta}[\rho_{\alpha},\rho_{\beta}]$ 
is the difference between the noninteracting kinetic free energy at zero temperature and that at the fictitious temperature $\theta$. The $\sigma$-spin density 
\begin{equation}\label{eq:tao3} 
\rho_{\sigma}({\bf r}) = \sum_{i=1}^{\infty} f_{i\sigma} |\psi_{i\sigma}({\bf r})|^{2} 
\end{equation} 
is expressed in terms of the thermally-assisted-occupation (TAO) orbitals $\{\psi_{i\sigma}({\bf r})\}$ and their occupation numbers $\{f_{i\sigma}\}$, 
\begin{equation}\label{eq:tao4} 
f_{i\sigma} = \{1+\text{exp}[(\epsilon_{i\sigma} - \mu_{\sigma})/ \theta] \}^{-1}, 
\end{equation} 
which are given by the Fermi-Dirac distribution. Here, the chemical potential $\mu_{\sigma}$ is determined by the conservation of the number of $\sigma$-spin electrons $N_{\sigma}$, 
\begin{equation}\label{eq:tao5} 
\sum_{i=1}^{\infty} \{1+\text{exp}[(\epsilon_{i\sigma} - \mu_{\sigma})/ \theta] \}^{-1} = N_{\sigma}. 
\end{equation} 
The two sets (one for each spin function) of self-consistent equations, \Cref{eq:tao1,eq:tao2,eq:tao3,eq:tao4,eq:tao5}, for $\rho_{\alpha}({\bf r})$ and $\rho_{\beta}({\bf r})$, respectively, 
are coupled with the ground-state density 
\begin{equation}\label{eq:tao6} 
\rho({\bf r}) = \sum_{\sigma}^{\alpha,\beta} \rho_{\sigma}({\bf r}). 
\end{equation} 

The self-consistent procedure described in Ref.\ \cite{TAO1} may be employed to obtain $\rho_{\sigma}({\bf r})$ and $\rho({\bf r})$. 
After self-consistency is achieved, the noninteracting kinetic free energy 
\begin{equation}\label{eq:tao7} 
A_{s}^{\theta}[\{f_{i\alpha}, \psi_{i\alpha} \}, \{f_{i\beta}, \psi_{i\beta} \}] = T_{s}^{\theta}[\{f_{i\alpha}, \psi_{i\alpha} \}, \{f_{i\beta}, \psi_{i\beta} \}] + E_{S}^{\theta}[\{f_{i\alpha} \}, \{f_{i\beta} \}] 
\end{equation} 
can be computed, in an exact manner, as the sum of the kinetic energy 
\begin{equation}\label{eq:tao8} 
T_{s}^{\theta}[\{f_{i\alpha}, \psi_{i\alpha} \}, \{f_{i\beta}, \psi_{i\beta} \}] 
= -\frac{1}{2} \sum_{\sigma}^{\alpha,\beta} \sum_{i=1}^{\infty} f_{i\sigma} \int \psi_{i\sigma}^{*}({\bf r}){\bf \nabla}^2\psi_{i\sigma}({\bf r}) d{\bf r} 
\end{equation} 
and entropy contribution 
\begin{equation}\label{eq:tao9} 
E_{S}^{\theta}[\{f_{i\alpha} \}, \{f_{i\beta} \}] = \theta \sum_{\sigma}^{\alpha,\beta} \sum_{i=1}^{\infty} \bigg\lbrace f_{i\sigma}\ \text{ln}(f_{i\sigma}) + (1-f_{i\sigma})\ \text{ln}(1-f_{i\sigma}) \bigg\rbrace 
\end{equation} 
of noninteracting electrons at the fictitious temperature $\theta$. 
The ground-state energy of the original interacting system at zero temperature is given by 
\begin{equation}\label{eq:tao10} 
E[\rho_{\alpha},\rho_{\beta}] = A_{s}^{\theta}[\{f_{i\alpha}, \psi_{i\alpha} \}, \{f_{i\beta}, \psi_{i\beta} \}] + \int \rho({\bf r}) v_{ext}({\bf r}) d{\bf r} + E_{H}[\rho] 
+ E_{xc}[\rho_{\alpha},\rho_{\beta}] + E_{\theta}[\rho_{\alpha},\rho_{\beta}], 
\end{equation} 
where $E_{H}[\rho] \equiv \frac{1}{2} \iint \frac{\rho({\bf r})\rho({\bf r'})}{|{\bf r} - {\bf r'}|}d{\bf r}d{\bf r'}$ is the Hartree energy. Spin-unpolarized (spin-restricted) TAO-DFT can be formulated 
by imposing the constraints of $\psi_{i\alpha}({\bf r})$ = $\psi_{i\beta}({\bf r})$ and $f_{i\alpha}$ = $f_{i\beta}$ to spin-polarized TAO-DFT.

\subsection{Density functional approximations} 

As the exact $E_{xc}[\rho_{\alpha},\rho_{\beta}]$ and $E_{\theta}[\rho_{\alpha},\rho_{\beta}]$ (i.e., the essential ingredients of spin-polarized TAO-DFT) remain unknown, DFAs for both of them 
(denoted as TAO-DFAs) are necessary for practical applications. Consequently, the performance of TAO-DFAs depends on the accuracy of DFAs and the choice of the fictitious temperature 
$\theta$. Note that $E_{xc}^{\text{DFA}}[\rho_{\alpha},\rho_{\beta}]$ can be readily obtained from that of KS-DFA, and $E_{\theta}^{\text {DFA}}[\rho_{\alpha},\rho_{\beta}]$ can be obtained with 
the knowledge of $A_{s}^{\text{DFA},\theta}[\rho_{\alpha},\rho_{\beta}]$ as follows: 
\begin{equation}\label{eq:tao11} 
\begin{split} 
E_{\theta}^{\text{DFA}}[\rho_{\alpha},\rho_{\beta}] 
\equiv&\ A_{s}^{\text{DFA},\theta=0}[\rho_{\alpha},\rho_{\beta}] - A_{s}^{\text{DFA},\theta}[\rho_{\alpha},\rho_{\beta}] \\ 
=&\ \frac{1}{2} (A_{s}^{\text{DFA},\theta=0}[2\rho_{\alpha}] + A_{s}^{\text{DFA},\theta=0}[2\rho_{\beta}]) - \frac{1}{2} (A_{s}^{\text{DFA},\theta}[2\rho_{\alpha}] + A_{s}^{\text{DFA},\theta}[2\rho_{\beta}]), 
\end{split} 
\end{equation} 
where $A_{s}^{\text{DFA},\theta}[\rho_{\alpha},\rho_{\beta}]$ is expressed in terms of $A_{s}^{\text{DFA},\theta}[\rho]$ (in its spin-unpolarized form) based on the spin-scaling relation of 
$A_{s}^{\theta}[\rho_{\alpha},\rho_{\beta}]$ \cite{spin-scaling}. Note that $E_{\theta=0}^{\text {DFA}}[\rho_{\alpha},\rho_{\beta}]=0$ (i.e., an exact property of $E_{\theta}[\rho_{\alpha},\rho_{\beta}]$) 
is ensured by Eq.\ (\ref{eq:tao11}). Accordingly, TAO-DFAs at $\theta = 0$ reduce to KS-DFAs.

\subsection{Strong static correlation from TAO-DFAs} 

In 2012, we developed TAO-LDA \cite{TAO1}, employing the LDA XC functional $E_{xc}^{\text{LDA}}[\rho_{\alpha},\rho_{\beta}]$ \cite{LDAX,LDAC} and 
$E_{\theta}^{\text{LDA}}[\rho_{\alpha},\rho_{\beta}]$ (given by Eq.\ (\ref{eq:tao11}) with $A_{s}^{\text{LDA},\theta}[\rho]$, the LDA for $A_{s}^{\theta}[\rho]$ (see Appendix A of Ref.\ \cite{As} and 
Eq.\ (37) of Ref.\ \cite{TAO1})) in TAO-DFT. Even at the simplest LDA level, TAO-LDA was shown to provide a reasonably accurate treatment of static correlation via the entropy contribution 
$E_{S}^{\theta}[\{f_{i\alpha} \}, \{f_{i\beta} \}]$ (see Eq.\ (\ref{eq:tao9})), when the distribution of TAO orbital occupation numbers (TOONs) $\{f_{i\sigma}\}$ (related to the chosen $\theta$) is 
close to the distribution of NOONs. However, this implies that a 
$\theta$ related to the distribution of NOONs should be employed to properly describe strong static correlation effects. For simplicity, an optimal value of $\theta$ = 7 mhartree 
was previously defined for TAO-LDA, based on physical arguments and numerical investigations. TAO-LDA (with $\theta$ = 7 mhartree) was shown to consistently outperform KS-LDA for 
multi-reference systems (due to the appropriate treatment of static correlation), while performing comparably to KS-LDA for single-reference systems (i.e., in the absence of strong static 
correlation effects). 

To improve the accuracy of TAO-LDA for single-reference systems, in 2014, we developed TAO-GGAs \cite{TAO2}, adopting the GGA XC functionals 
$E_{xc}^{\text{GGA}}[\rho_{\alpha},\rho_{\beta}]$ and $E_{\theta}^{\text{GEA}}[\rho_{\alpha},\rho_{\beta}]$ (given by Eq.\ (\ref{eq:tao11}) with $A_{s}^{\text{GEA},\theta}[\rho]$, the gradient 
expansion approximation (GEA) for $A_{s}^{\theta}[\rho]$ (see Appendices A and B of Ref.\ \cite{As})) in TAO-DFT. As TAO-GGAs should improve upon TAO-LDA mainly for the properties 
governed by short-range XC effects (due to the more accurate treatment of on-top hole density) \cite{Parr,DFTreview1,DFTreview2,DFTreview3,Perdew09}, and the orbital energy gaps of 
TAO-LDA and TAO-GGAs should be similar \cite{LCAC,EB}, the optimal $\theta$ values for TAO-LDA and TAO-GGAs should remain similar (when the same physical arguments and numerical 
investigations are adopted to define the optimal $\theta$ values). Therefore, we adopted an optimal value of $\theta$ = 7 mhartree for both TAO-LDA and TAO-GGAs. While 
$E_{\theta}^{\text{GEA}}[\rho_{\alpha},\rho_{\beta}]$ should be more accurate than $E_{\theta}^{\text{LDA}}[\rho_{\alpha},\rho_{\beta}]$ for the nearly uniform electron gas, for a small value of 
$\theta$ (i.e., 7 mhartree), their difference was found to be much smaller than the difference between two different XC energy functionals. Unsurprisingly, since 
$E_{\theta=0}^{\text{LDA}}=E_{\theta=0}^{\text{GEA}}=0$, the difference between $E_{\theta}^{\text{LDA}}$ and $E_{\theta}^{\text{GEA}}$ should remain small for a small value of $\theta$ 
(i.e., 7 mhartree). Accordingly, $E_{\theta}^{\text{LDA}}[\rho_{\alpha},\rho_{\beta}]$ may also be adopted for TAO-GGAs. 

While TAO-DFAs (i.e., TAO-LDA and TAO-GGAs) outperform KS-DFAs for multi-reference systems, they perform similarly to KS-DFAs for single-reference systems. 
As mentioned previously, hybrid functionals in KS-DFT, which provide an improved description of nonlocal exchange effects, are generally superior to KS-DFAs in performance for a broad 
range of applications \cite{LCAC,EB}. Therefore, a possible hybrid functional in TAO-DFT is expected to outperform TAO-DFAs for a wide variety of single-reference systems. 
In the following section, we define the exact exchange in TAO-DFT, and propose the corresponding global and range-separated hybrid schemes in TAO-DFT.

\section{Hybrid Schemes in TAO-DFT} 

\subsection{Exact exchange} 

In KS-DFT, the exact exchange $E_{x}[\rho_{\alpha},\rho_{\beta}]$ is defined as the HF exchange energy of the occupied KS orbitals 
$\{\phi_{i\sigma}({\bf r})\}$ \cite{Parr,DFTreview1,DFTreview2,DFTreview3}: 
\begin{equation}\label{eq:ks-hf} 
\begin{split} 
E_{x}[\rho_{\alpha},\rho_{\beta}] 
\equiv&\ E_{x}^{\text{HF}}[\{\phi_{i\alpha} \}, \{\phi_{i\beta} \}]  \\ 
=&\ - \frac{1}{2} \sum_{\sigma}^{\alpha,\beta} \sum_{i,j=1}^{N_{\sigma}} 
\iint \frac{\phi_{i\sigma}^{*}({\bf r}_{1}) \phi_{j\sigma}^{*}({\bf r}_{2}) \phi_{j\sigma}({\bf r}_{1}) \phi_{i\sigma}({\bf r}_{2})}{r_{12}} d{\bf r}_{1} d{\bf r}_{2}  \\ 
=&\ - \frac{1}{2} \sum_{\sigma}^{\alpha,\beta} \iint \frac{|\gamma_{\sigma}^{\text{KS}}({\bf r}_{1},{\bf r}_{2})|^{2}}{r_{12}} d{\bf r}_{1} d{\bf r}_{2}, 
\end{split} 
\end{equation} 
where $r_{12} = |{\bf r}_{1} - {\bf r}_{2}|$ is the interelectronic distance. Here, 
$\gamma_{\sigma}^{\text{KS}}({\bf r}_{1},{\bf r}_{2}) = \sum_{i=1}^{N_{\sigma}} \phi_{i\sigma}^{*}({\bf r}_{1}) \phi_{i\sigma}({\bf r}_{2})$ is the $\sigma$-spin 1-RDM in KS-DFT, and its diagonal 
element $\gamma_{\sigma}^{\text{KS}}({\bf r},{\bf r}) = \sum_{i=1}^{N_{\sigma}} |\phi_{i\sigma}({\bf r})|^{2} = \rho_{\sigma}({\bf r})$ is the $\sigma$-spin density in KS-DFT. 

In TAO-DFT, the exact exchange $F_{x}^{\theta}[\rho_{\alpha},\rho_{\beta}]$ can be defined as 
the HF exchange free energy of the TAO orbitals $\{\psi_{i\sigma}({\bf r})\}$ and their occupation numbers $\{f_{i\sigma} \}$ at the fictitious temperature $\theta$: 
\begin{equation}\label{eq:tao-hf} 
\begin{split} 
F_{x}^{\theta}[\rho_{\alpha},\rho_{\beta}] 
\equiv&\ F_{x}^{\text{HF},\theta}[\{f_{i\alpha}, \psi_{i\alpha} \}, \{f_{i\beta}, \psi_{i\beta} \}]  \\ 
=&\ - \frac{1}{2} \sum_{\sigma}^{\alpha,\beta} \sum_{i,j=1}^{\infty} f_{i\sigma} f_{j\sigma} 
\iint \frac{\psi_{i\sigma}^{*}({\bf r}_{1}) \psi_{j\sigma}^{*}({\bf r}_{2}) \psi_{j\sigma}({\bf r}_{1}) \psi_{i\sigma}({\bf r}_{2})}{r_{12}} d{\bf r}_{1} d{\bf r}_{2}  \\ 
=&\ - \frac{1}{2} \sum_{\sigma}^{\alpha,\beta} \iint \frac{|\gamma_{\sigma}^{\text{TAO}}({\bf r}_{1},{\bf r}_{2})|^{2}}{r_{12}} d{\bf r}_{1} d{\bf r}_{2}. 
\end{split} 
\end{equation} 
Here, 
\begin{equation}\label{eq:tao-sdm} 
\gamma_{\sigma}^{\text{TAO}}({\bf r}_{1},{\bf r}_{2}) = \sum_{i=1}^{\infty} f_{i\sigma} \psi_{i\sigma}^{*}({\bf r}_{1}) \psi_{i\sigma}({\bf r}_{2}) 
\end{equation} 
is the $\sigma$-spin 1-RDM in TAO-DFT, and its diagonal element 
\begin{equation} 
\gamma_{\sigma}^{\text{TAO}}({\bf r},{\bf r}) = \sum_{i=1}^{\infty} f_{i\sigma} |\psi_{i\sigma}({\bf r})|^{2} = \sum_{i=1}^{\infty} \rho_{i\sigma}({\bf r}) = \rho_{\sigma}({\bf r}) 
\end{equation} 
is the $\sigma$-spin density in TAO-DFT, where $\rho_{i\sigma}({\bf r}) \equiv f_{i\sigma} |\psi_{i\sigma}({\bf r})|^{2}$ is the $i$-th $\sigma$-spin orbital density in TAO-DFT. 
Note that the TAO orbitals $\{\psi_{i\sigma}({\bf r})\}$ and their occupation numbers $\{f_{i\sigma}\}$ are the eigenfunctions and eigenvalues, respectively, of 
$\gamma_{\sigma}^{\text{TAO}}({\bf r}_{1},{\bf r}_{2})$: 
\begin{equation}\label{eq:tao-sdm-decomp} 
\begin{split} 
\int \gamma_{\sigma}^{\text{TAO}}({\bf r}_{1},{\bf r}_{2}) \psi_{i\sigma}({\bf r}_{1}) d{\bf r}_{1} 
=&\ \sum_{j=1}^{\infty} f_{j\sigma} \psi_{j\sigma}({\bf r}_{2}) \int \psi_{j\sigma}^{*}({\bf r}_{1}) \psi_{i\sigma}({\bf r}_{1}) d{\bf r}_{1}  \\ 
=&\ \sum_{j=1}^{\infty} f_{j\sigma} \psi_{j\sigma}({\bf r}_{2}) \delta_{ij} 
= f_{i\sigma} \psi_{i\sigma}({\bf r}_{2}), 
\end{split} 
\end{equation} 
where $\delta_{ij}$ is the Kronecker delta function. 
At $\theta = 0$, TAO-DFT is the same as KS-DFT, and hence, Eq.\ (\ref{eq:tao-hf}) is reduced to Eq.\ (\ref{eq:ks-hf}). 

To justify the use of $F_{x}^{\theta}[\rho_{\alpha},\rho_{\beta}] \equiv F_{x}^{\text{HF},\theta}[\{f_{i\alpha}, \psi_{i\alpha} \}, \{f_{i\beta}, \psi_{i\beta} \}]$ (given by Eq.\ (\ref{eq:tao-hf})) 
as the definition of exact exchange in TAO-DFT, here, we comment on the self-interaction energy associated with the exact exchange in TAO-DFT. 
On the basis of \Cref{eq:tao3,eq:tao6}, the Hartree energy can be expressed as 
\begin{equation}\label{eq:tao-h} 
\begin{split} 
E_{H}[\rho] 
\equiv&\ \frac{1}{2} \iint \frac{\rho({\bf r}_{1})\rho({\bf r}_{2})}{r_{12}}d{\bf r}_{1} d{\bf r}_{2} 
= \frac{1}{2} \sum_{\sigma}^{\alpha,\beta} \sum_{\sigma'}^{\alpha,\beta} \iint \frac{\rho_{\sigma}({\bf r}_{1})\rho_{\sigma'}({\bf r}_{2})}{r_{12}}d{\bf r}_{1} d{\bf r}_{2}  \\ 
=&\ \frac{1}{2} \sum_{\sigma}^{\alpha,\beta} \sum_{\sigma'}^{\alpha,\beta} \sum_{i,j=1}^{\infty} 
f_{i\sigma} f_{j\sigma'} \iint \frac{|\psi_{i\sigma}({\bf r}_1)|^2 |\psi_{j\sigma'}({\bf r}_2)|^2}{r_{12}}d{\bf r}_{1} d{\bf r}_{2}. 
\end{split} 
\end{equation} 
Accordingly, the self-Hartree energy \cite{PZ-SIC}, 
\begin{equation}\label{eq:tao-she} 
\text{self-Hartree energy} 
\equiv \sum_{\sigma}^{\alpha,\beta} \sum_{i=1}^{\infty} E_{H}[\rho_{i\sigma}] 
= \frac{1}{2} \sum_{\sigma}^{\alpha,\beta} \sum_{i=1}^{\infty} f_{i\sigma}^{2} \iint \frac{|\psi_{i\sigma}({\bf r}_1)|^2 |\psi_{i\sigma}({\bf r}_2)|^2}{r_{12}}d{\bf r}_{1} d{\bf r}_{2}, 
\end{equation} 
which is the sum of the ($\sigma=\sigma'$ and $i=j$) terms in Eq.\ (\ref{eq:tao-h}), can be exactly cancelled by the self-exchange energy \cite{PZ-SIC}, 
\begin{equation}\label{eq:tao-see} 
\begin{split} 
\text{self-exchange energy} 
\equiv&\ \sum_{\sigma}^{\alpha,\beta} \sum_{i=1}^{\infty} F_{x}^{\theta}[\rho_{i\sigma},0] 
= - \frac{1}{2} \sum_{\sigma}^{\alpha,\beta} \sum_{i=1}^{\infty} f_{i\sigma}^{2} \iint \frac{|\psi_{i\sigma}({\bf r}_1)|^2 |\psi_{i\sigma}({\bf r}_2)|^2}{r_{12}} d{\bf r}_{1} d{\bf r}_{2}  \\ 
=&\ - \sum_{\sigma}^{\alpha,\beta} \sum_{i=1}^{\infty} E_{H}[\rho_{i\sigma}], 
\end{split} 
\end{equation} 
which is the sum of the ($i=j$) terms in Eq.\ (\ref{eq:tao-hf}), on an orbital-by-orbital basis (i.e., each term in Eq.\ (\ref{eq:tao-she}) can be exactly cancelled by a term in 
Eq.\ (\ref{eq:tao-see})). Therefore, complete cancellation of the self-interaction in the Hartree energy would require the full exact exchange (given by Eq.\ (\ref{eq:tao-hf})) 
in TAO-DFT. By contrast, such perfect cancellation may not be achieved by the HF exchange (given by Eq.\ (\ref{eq:ks-hf})) with the KS orbitals being replaced by the 
TAO orbitals. Besides, the self-Hartree energy in TAO-DFT is unlikely to be exactly cancelled by the self-XC energy, i.e., 
$\sum_{\sigma}^{\alpha,\beta} \sum_{i=1}^{\infty} E_{xc}^{\text{DFA}}[\rho_{i\sigma},0]$, associated with the DFA XC functional $E_{xc}^{\text{DFA}}[\rho_{\alpha},\rho_{\beta}]$, 
implying that the SIEs associated with TAO-DFAs may remain pronounced for both single- and multi-reference systems! 

From Eq.\ (\ref{eq:ks-hf}), $E_{x}^{\text{HF}}[\{\phi_{i\alpha} \}, \{\phi_{i\beta} \}]$ (i.e., the exact exchange in KS-DFT) can be expressed as 
\begin{equation}\label{eq:tao12} 
\begin{split} 
E_{x}^{\text{HF}}[\{\phi_{i\alpha} \}, \{\phi_{i\beta} \}] 
=&\ E_{x}[\rho_{\alpha},\rho_{\beta}]  \\ 
=&\ F_{x}^{\theta}[\rho_{\alpha},\rho_{\beta}] + (E_{x}[\rho_{\alpha},\rho_{\beta}] - F_{x}^{\theta}[\rho_{\alpha},\rho_{\beta}])  \\ 
=&\ F_{x}^{\text{HF},\theta}[\{f_{i\alpha}, \psi_{i\alpha} \}, \{f_{i\beta}, \psi_{i\beta} \}] + E_{x,\theta}[\rho_{\alpha},\rho_{\beta}], 
\end{split} 
\end{equation} 
the sum of $F_{x}^{\text{HF},\theta}[\{f_{i\alpha}, \psi_{i\alpha} \}, \{f_{i\beta}, \psi_{i\beta} \}]$ (i.e., the exact exchange in TAO-DFT) and 
$E_{x,\theta}[\rho_{\alpha},\rho_{\beta}] \equiv E_{x}[\rho_{\alpha},\rho_{\beta}] - F_{x}^{\theta}[\rho_{\alpha},\rho_{\beta}] 
= F_{x}^{\theta=0}[\rho_{\alpha},\rho_{\beta}] - F_{x}^{\theta}[\rho_{\alpha},\rho_{\beta}]$ (i.e., the difference between the exchange free energy at zero temperature 
and that at the fictitious temperature $\theta$). 
Subsequently, a DFA can be made for $E_{x,\theta}[\rho_{\alpha},\rho_{\beta}]$ as follows: 
\begin{equation}\label{eq:ex_dfa} 
\begin{split} 
E_{x,\theta}^{\text{DFA}}[\rho_{\alpha},\rho_{\beta}] \equiv F_{x}^{\text{DFA},\theta=0}[\rho_{\alpha},\rho_{\beta}] - F_{x}^{\text{DFA},\theta}[\rho_{\alpha},\rho_{\beta}], 
\end{split} 
\end{equation} 
where $F_{x}^{\text{DFA},\theta}[\rho_{\alpha},\rho_{\beta}]$ is the DFA for $F_{x}^{\theta}[\rho_{\alpha},\rho_{\beta}]$. Note that 
$E_{x,\theta=0}^{\text{DFA}}[\rho_{\alpha},\rho_{\beta}]=0$ (i.e., an exact property of $E_{x,\theta}[\rho_{\alpha},\rho_{\beta}]$) can be readily achieved by Eq.\ (\ref{eq:ex_dfa}). 
Besides, from the spin-scaling relation of $F_{x}^{\theta}[\rho_{\alpha},\rho_{\beta}]$ \cite{spin-scaling}, $E_{x,\theta}^{\text{DFA}}[\rho_{\alpha},\rho_{\beta}]$ can be 
conveniently expressed in terms of $F_{x}^{\text{DFA},\theta}[\rho]$ (in its spin-unpolarized form): 
\begin{equation}\label{eq:exs_dfa} 
E_{x,\theta}^{\text{DFA}}[\rho_{\alpha},\rho_{\beta}] 
= \frac{1}{2} (F_{x}^{\text{DFA},\theta=0}[2\rho_{\alpha}] + F_{x}^{\text{DFA},\theta=0}[2\rho_{\beta}]) - \frac{1}{2} (F_{x}^{\text{DFA},\theta}[2\rho_{\alpha}] + F_{x}^{\text{DFA},\theta}[2\rho_{\beta}]). 
\end{equation} 
From Eqs.\ (\ref{eq:tao12}) and (\ref{eq:ex_dfa}), the exact exchange in KS-DFT is approximately given by 
\begin{equation}\label{eq:tao13} 
E_{x}[\rho_{\alpha},\rho_{\beta}] = E_{x}^{\text{HF}}[\{\phi_{i\alpha} \}, \{\phi_{i\beta} \}] 
\approx F_{x}^{\text{HF},\theta}[\{f_{i\alpha}, \psi_{i\alpha} \}, \{f_{i\beta}, \psi_{i\beta} \}] + E_{x,\theta}^{\text{DFA}}[\rho_{\alpha},\rho_{\beta}], 
\end{equation} 
the sum of the exact exchange in TAO-DFT and $E_{x,\theta}^{\text{DFA}}[\rho_{\alpha},\rho_{\beta}]$. 
Note that the approximation becomes exact, when the exact $E_{x,\theta}^{\text{DFA}}[\rho_{\alpha},\rho_{\beta}]$ is employed. 

While the exact exchange in TAO-DFT is free of the SIE, the scheme is not expected to perform satisfactorily for most systems, due to the lack of correlation energy 
$E_{c}[\rho_{\alpha},\rho_{\beta}]$. Besides, it is well known that the exact exchange is incompatible with the DFA correlation in KS-DFT, implying that 
TAO-DFT with the exact exchange and DFA correlation would not perform well for single-reference systems (i.e., in the absence of strong static correlation effects). 
Therefore, similar to the hybrid schemes in KS-DFT, it may be useful to incorporate the exact exchange with the DFA XC functional in TAO-DFT. 
In the following subsections, the global and range-separated hybrid schemes in TAO-DFT are proposed.

\subsection{Global hybrid scheme} 

In KS-DFT, a global hybrid (GH) functional \cite{hybrid1,hybrid2,B3LYP,DFA0,DFA0a,PBE0,PBE0a,SCAN0} is generally expressed as 
\begin{equation}\label{eq:ks-gh} 
E_{xc}^{\text{KS-GH}} = a_{x} E_{x}^{\text{HF}}[\{\phi_{i\alpha} \}, \{\phi_{i\beta} \}] + (1 - a_{x}) E_{x}^{\text{DFA}}[\rho_{\alpha},\rho_{\beta}] + E_{c}^{\text{DFA}}[\rho_{\alpha},\rho_{\beta}], 
\end{equation} 
where $E_{x}^{\text{HF}}$ is the HF exchange energy (given by Eq.\ (\ref{eq:ks-hf})), $E_{x}^{\text{DFA}}$ is the DFA exchange energy, and $E_{c}^{\text{DFA}}$ is the DFA correlation energy. 
The fraction of HF exchange $a_{x}$, typically ranging from 0.2 to 0.25 for thermochemistry and from 0.4 to 0.6 for kinetics, can be determined by empirical fitting or physical arguments. 

After substituting Eq.\ (\ref{eq:tao13}) into Eq.\ (\ref{eq:ks-gh}), the corresponding global hybrid functional in TAO-DFT can be defined as 
\begin{equation}\label{eq:tao-gh} 
\begin{split} 
E_{xc}^{\text{TAO-GH}} 
=&\ a_{x} \bigg\lbrace F_{x}^{\text{HF},\theta}[\{f_{i\alpha}, \psi_{i\alpha} \}, \{f_{i\beta}, \psi_{i\beta} \}] + E_{x,\theta}^{\text{DFA}}[\rho_{\alpha},\rho_{\beta}] \bigg\rbrace  \\ 
+&\ (1 - a_{x}) E_{x}^{\text{DFA}}[\rho_{\alpha},\rho_{\beta}] + E_{c}^{\text{DFA}}[\rho_{\alpha},\rho_{\beta}], 
\end{split} 
\end{equation} 
and the resulting ground-state energy is evaluated by 
\begin{equation}\label{eq:tao-gh-e} 
E^{\text{TAO-GH}} = A_{s}^{\theta}[\{f_{i\alpha}, \psi_{i\alpha} \}, \{f_{i\beta}, \psi_{i\beta} \}] + \int \rho({\bf r}) v_{ext}({\bf r}) d{\bf r} + E_{H}[\rho] 
+ E_{xc}^{\text{TAO-GH}} + E_{\theta}^{\text{DFA}}[\rho_{\alpha},\rho_{\beta}]. 
\end{equation} 
While an evaluation of the functional derivative of $F_{x}^{\text{HF},\theta}[\{f_{i\alpha}, \psi_{i\alpha} \}, \{f_{i\beta}, \psi_{i\beta} \}]$ (i.e., an explicit functional of the TAO orbitals and their 
occupation numbers) with respect to the density $\rho_{\sigma}$ (see Eq.\ (\ref{eq:tao2})) can be achieved with the finite-temperature exact-exchange and related schemes \cite{FTEXX}, 
the resulting scheme can be computationally demanding. To reduce the computational complexity, in this work, the electronic energy for a global hybrid functional in TAO-DFT is minimized 
with respect to the 1-RDM $\gamma_{\sigma}^{\text{TAO}}$ (as is usual in the finite-temperature HF (FT-HF) and related 
schemes \cite{FTHF1,FTHF2,FTHF3,FTHF4,FTHF5,FTHF6,FTHF7,FTHF8,FTHF9}). The resulting self-consistent equations for the $\sigma$-spin electrons can be expressed as 
\begin{equation}\label{eq:ghtao1} 
\bigg\lbrace -\frac{1}{2} {\bf \nabla}^2 \ + \ v_{s,\sigma}^{loc}({\bf r}) \bigg\rbrace \psi_{i\sigma}({\bf r}) 
- a_{x} \sum_{j=1}^{\infty} f_{j\sigma} \int \frac{\psi_{j\sigma}^{*}({\bf r'}) \psi_{i\sigma}({\bf r'})}{|{\bf r} - {\bf r'}|} \psi_{j\sigma}({\bf r}) d{\bf r'} 
= \epsilon_{i\sigma} \psi_{i\sigma}({\bf r}), 
\end{equation} 
where 
\begin{equation}\label{eq:ghtao2} 
\begin{split} 
v_{s,\sigma}^{loc}({\bf r}) 
=&\ v_{ext}({\bf r}) + \int \frac{\rho({\bf r'})}{|{\bf r} - {\bf r'}|}d{\bf r'} + \frac{\delta E_{\theta}^{\text{DFA}}[\rho_{\alpha},\rho_{\beta}]}{\delta \rho_{\sigma}({\bf r})}  \\ 
+&\ (1 - a_{x}) \frac{\delta E_{x}^{\text{DFA}}[\rho_{\alpha},\rho_{\beta}]}{\delta \rho_{\sigma}({\bf r})} 
+ \frac{\delta E_{c}^{\text{DFA}}[\rho_{\alpha},\rho_{\beta}]}{\delta \rho_{\sigma}({\bf r})} 
+ a_{x} \frac{\delta E_{x,\theta}^{\text{DFA}}[\rho_{\alpha},\rho_{\beta}]}{\delta \rho_{\sigma}({\bf r})}. 
\end{split} 
\end{equation} 
is the local part of the effective potential. The two sets (one for each spin function) of self-consistent equations, \Cref{eq:ghtao1,eq:ghtao2,eq:tao3,eq:tao4,eq:tao5}, 
for $\rho_{\alpha}({\bf r})$ and $\rho_{\beta}({\bf r})$, respectively, are coupled with the ground-state density (given by Eq.\ (\ref{eq:tao6})). 

Note that $E_{xc}^{\text{TAO-GH}}$ reduces to $E_{xc}^{\text{DFA}}$ (i.e., the DFA XC functional) for $a_{x} = 0$, and reduces to 
$F_{x}^{\text{HF},\theta} + E_{x,\theta}^{\text{DFA}} + E_{c}^{\text{DFA}}$ (i.e., the exact exchange in TAO-DFT, the DFA for $E_{x,\theta}$, and the DFA correlation functional) for $a_{x} = 1$. 
At $\theta = 0$, TAO-DFT with $E_{xc}^{\text{TAO-GH}}$ is the same as KS-DFT with $E_{xc}^{\text{KS-GH}}$. 

On the other hand, if the constraints of $a_{x} = 1$ and $E_{c}^{\text{DFA}} = E_{\theta}^{\text{DFA}} = E_{x,\theta}^{\text{DFA}} = 0$ are imposed to the global hybrid scheme in TAO-DFT, 
the resulting scheme resembles the FT-HF scheme. Therefore, the computational cost of the global hybrid scheme in TAO-DFT is similar to that of the global hybrid scheme in KS-DFT or 
the FT-HF scheme.

\subsection{Range-separated hybrid scheme} 

In KS-DFT, a range-separated hybrid (RSH) functional \cite{RSH,wB97X,wB97X-2} is generally given by 
\begin{equation}\label{eq:ks-rsh} 
E_{xc}^{\text{KS-RSH}} = E_{x}^{\text{HF}}(\text{I})[\{\phi_{i\alpha} \}, \{\phi_{i\beta} \}] + E_{x}^{\text{DFA}}(\bar{\text{I}})[\rho_{\alpha},\rho_{\beta}] + E_{c}^{\text{DFA}}[\rho_{\alpha},\rho_{\beta}], 
\end{equation} 
where $E_{x}^{\text{HF}}(\text{I})$ is the HF exchange energy of an interelectronic repulsion operator $\text{I}(r_{12})$, 
\begin{equation}\label{eq:ks-ihf} 
E_{x}^{\text{HF}}(\text{I})[\{\phi_{i\alpha} \}, \{\phi_{i\beta} \}] = - \frac{1}{2} \sum_{\sigma}^{\alpha,\beta} \sum_{i,j=1}^{N_{\sigma}} 
\iint \text{I}(r_{12}) \phi_{i\sigma}^{*}({\bf r}_{1}) \phi_{j\sigma}^{*}({\bf r}_{2}) \phi_{j\sigma}({\bf r}_{1}) \phi_{i\sigma}({\bf r}_{2}) d{\bf r}_{1} d{\bf r}_{2}, 
\end{equation} 
and $E_{x}^{\text{DFA}}(\bar{\text{I}})$ is the DFA exchange energy of the complementary operator $\bar{\text{I}}(r_{12}) \equiv 1/r_{12} - \text{I}(r_{12})$. 
Similar to the previous trick, we replace the Coulomb operator $1/r_{12}$ in Eq.\ (\ref{eq:tao13}) by the operator $\text{I}(r_{12})$, yielding the following expression: 
\begin{equation}\label{eq:tao14} 
E_{x}^{\text{HF}}(\text{I})[\{\phi_{i\alpha} \}, \{\phi_{i\beta} \}] 
\approx F_{x}^{\text{HF},\theta}(\text{I})[\{f_{i\alpha}, \psi_{i\alpha} \}, \{f_{i\beta}, \psi_{i\beta} \}] + E_{x,\theta}^{\text{DFA}}(\text{I})[\rho_{\alpha},\rho_{\beta}], 
\end{equation} 
where 
\begin{equation}\label{eq:tao-ihf} 
\begin{split} 
F_{x}^{\text{HF},\theta}(\text{I})[\{f_{i\alpha}, \psi_{i\alpha} \}, \{f_{i\beta}, \psi_{i\beta} \}] 
=&\ - \frac{1}{2} \sum_{\sigma}^{\alpha,\beta} \sum_{i,j=1}^{\infty} f_{i\sigma} f_{j\sigma}  \\ 
\times&\ \iint \text{I}(r_{12}) \psi_{i\sigma}^{*}({\bf r}_{1}) \psi_{j\sigma}^{*}({\bf r}_{2}) \psi_{j\sigma}({\bf r}_{1}) \psi_{i\sigma}({\bf r}_{2}) d{\bf r}_{1} d{\bf r}_{2} 
\end{split} 
\end{equation} 
is the HF exchange free energy of the operator $\text{I}(r_{12})$ at the fictitious temperature $\theta$, and 
\begin{equation}\label{eq:exi_dfa} 
E_{x,\theta}^{\text{DFA}}(\text{I})[\rho_{\alpha},\rho_{\beta}] \equiv F_{x}^{\text{DFA},\theta=0}(\text{I})[\rho_{\alpha},\rho_{\beta}] 
- F_{x}^{\text{DFA},\theta}(\text{I})[\rho_{\alpha},\rho_{\beta}] 
\end{equation} 
is the difference between the DFA exchange free energy of the operator $\text{I}(r_{12})$ at zero temperature and that at the fictitious temperature $\theta$. 
Note that the approximation (see Eq.\ (\ref{eq:tao14})) becomes exact, when the exact $E_{x,\theta}^{\text{DFA}}(\text{I})[\rho_{\alpha},\rho_{\beta}]$ is employed. 

After substituting Eq.\ (\ref{eq:tao14}) into Eq.\ (\ref{eq:ks-rsh}), the corresponding range-separated hybrid functional in TAO-DFT can be defined as 
\begin{equation}\label{eq:tao-rsh} 
\begin{split} 
E_{xc}^{\text{TAO-RSH}} 
=&\ \bigg\lbrace F_{x}^{\text{HF},\theta}(\text{I})[\{f_{i\alpha}, \psi_{i\alpha} \}, \{f_{i\beta}, \psi_{i\beta} \}] + E_{x,\theta}^{\text{DFA}}(\text{I})[\rho_{\alpha},\rho_{\beta}] \bigg\rbrace  \\ 
+&\ E_{x}^{\text{DFA}}(\bar{\text{I}})[\rho_{\alpha},\rho_{\beta}] + E_{c}^{\text{DFA}}[\rho_{\alpha},\rho_{\beta}], 
\end{split} 
\end{equation} 
For $\text{I}(r_{12}) = a_{x}/r_{12}$, $E_{xc}^{\text{TAO-RSH}}$ reduces to $E_{xc}^{\text{TAO-GH}}$. 
However, for a general operator $\text{I}(r_{12})$ (e.g., the erf \cite{op}, erfgau \cite{erfgau}, or terf \cite{terf} operator), while $F_{x}^{\text{HF},\theta}(\text{I})$ is defined, and 
$E_{x}^{\text{DFA}}(\bar{\text{I}})$ and $E_{c}^{\text{DFA}}$ are available from those of the range-separated hybrid scheme in KS-DFT, $F_{x}^{\text{DFA},\theta}(\text{I})$ (and hence, 
$E_{x,\theta}^{\text{DFA}}(\text{I})$) is mostly unavailable, and needs to be developed for practical applications. Therefore, in this work, while the range-separated hybrid scheme 
in TAO-DFT is proposed, our numerical results are only available for the global hybrid scheme in TAO-DFT.

\section{Global Hybrid Functionals in TAO-DFT} 

\subsection{Definition of the optimal $\theta$ values} 

As previously mentioned, the fictitious temperature $\theta$ in TAO-DFT should be chosen so that the distribution of TOONs is close to that of NOONs \cite{TAO1,TAO2}. In this situation, 
the strong static correlation effects can be properly described by the entropy contribution. For single-reference systems, as the exact NOONs are close to either 0 or 1, the optimal $\theta$ 
should be sufficiently small. However, for multi-reference systems, as the distribution of NOONs can be diverse (due to the varying strength of static correlation), the optimal $\theta$ can 
span a wide range of values. Therefore, for a global hybrid functional in TAO-DFT, it is impossible to adopt a $\theta$ that is optimal for both single- and multi-reference systems. 
Nevertheless, it remains useful to define an optimal $\theta$ value for a global hybrid functional in TAO-DFT to provide an explicit description of orbital occupations. 

To be consistent with the previous definition of the optimal $\theta$ value for TAO-DFAs, in this work, the same physical arguments and numerical investigations are adopted to define 
the optimal $\theta$ value for a global hybrid functional in TAO-DFT. Specifically, the performance of various global hybrid functionals in TAO-DFT 
(with $\theta$ = 0, 5, 10, 15, 20, 25, 30, 35, 40, 45, and 50 mhartree) is examined for the following single-reference systems: 
\begin{itemize} 
\item the reaction energies of the 30 chemical reactions in the NHTBH38/04 and HTBH38/04 sets \cite{BH}, 
\item the 166 equilibrium geometries of the equilibrium experimental test set (EXTS) \cite{EXTS}. 
\end{itemize} 
The optimal $\theta$ value for a global hybrid functional in TAO-DFT is defined as the largest $\theta$ value for which the performance of the global hybrid functional in TAO-DFT 
(with this $\theta$) and the corresponding global hybrid functional in KS-DFT (i.e., the $\theta = 0$ case) is similar for the aforementioned systems. 

For the choice of global hybrid functionals, we adopt the following four popular functionals (see Eq.\ (\ref{eq:tao-gh})): 
\begin{itemize} 
\item B3LYP \cite{hybrid2,B3LYP}: 
$a_{x} = 1/5$, $E_{x}^{\text{DFA}} = 0.10\ E_{x}^{\text{LDA}} + 0.90\ E_{x}^{\text{B88}}$, $E_{c}^{\text{DFA}} = 0.19\ E_{c}^{\text{VWN1RPA}} + 0.81\ E_{c}^{\text{LYP}}$, 
\item B3LYP-D3 \cite{DFT-D3}: 
B3LYP with the -D3 dispersion corrections ($s_{r,6} = 1.261$ and $s_{8} = 1.703$), 
\item PBE0 \cite{PBE0,PBE0a}: 
$a_{x} = 1/4$, $E_{x}^{\text{DFA}} = E_{x}^{\text{PBE}}$, $E_{c}^{\text{DFA}} = E_{c}^{\text{PBE}}$, 
\item BHHLYP \cite{hybrid1}: 
$a_{x} = 1/2$, $E_{x}^{\text{DFA}} = E_{x}^{\text{B88}}$, $E_{c}^{\text{DFA}} = E_{c}^{\text{LYP}}$, 
\end{itemize} 
where $E_{x}^{\text{LDA}}$ is the LDA exchange energy \cite{LDAX}, $E_{x}^{\text{B88}}$ is the B88 exchange energy \cite{B88}, $E_{x}^{\text{PBE}}$ is the PBE exchange energy \cite{PBE}, 
$E_{c}^{\text{VWN1RPA}}$ is the VWN formula 1 RPA local correlation energy \cite{VWN}, $E_{c}^{\text{LYP}}$ is the LYP correlation energy \cite{LYP}, and $E_{c}^{\text{PBE}}$ is the PBE 
correlation energy \cite{PBE}. Note that $s_{r,6}$ and $s_{8}$ are the parameters controlling the strength of the -D3 dispersion corrections (see Eq.\ (3) of Ref.\ \cite{DFT-D3}). 

Besides, we adopt the following $\theta$-dependent energy functionals (see Eqs.\ (\ref{eq:tao-gh}) and (\ref{eq:tao-gh-e})): 
\begin{itemize} 
\item $E_{\theta}^{\text{DFA}} = E_{\theta}^{\text{LDA}}$: 
given by Eq.\ (\ref{eq:tao11}) with $A_{s}^{\text{LDA},\theta}[\rho]$ \cite{As} (also see Eq.\ (37) of Ref.\ \cite{TAO1}). 
\item $E_{x,\theta}^{\text{DFA}} = E_{x,\theta}^{\text{LDA}}$: 
given by Eq.\ (\ref{eq:exs_dfa}) with $F_{x}^{\text{LDA},\theta}[\rho]$ \cite{Fx} (also see Eq.\ (\ref{eq:fx_lda})). 
\end{itemize} 
For completeness of this work, $F_{x}^{\text{LDA},\theta}[\rho]$ (in its spin-unpolarized form), which is the LDA for $F_{x}^{\theta}[\rho]$, is explicitly given here, 
\begin{equation}\label{eq:fx_lda} 
F_{x}^{\text{LDA},\theta}[\rho] = \int f_{x}^{\text{LDA},\theta}({\bf r}) d{\bf r}, 
\end{equation} 
where $f_{x}^{\text{LDA},\theta}({\bf r}) \equiv -(3 / \pi)^{1/3} \rho^{4/3}({\bf r}) g(t)$, $t \equiv 2\theta / (3 \pi^2 \rho({\bf r}))^{2/3}$, and $g(t)$ is a parametrized function: 
\begin{equation} 
g(t) = \frac{0.75 + 3.04363\ t^2 - 0.092270\ t^3 + 1.70350\ t^4}{1 + 8.31051\ t^2 + 5.1105\ t^4} \times \text{tanh}(1/t). 
\end{equation} 
The $\theta = 0$ case, $F_{x}^{\text{LDA},\theta=0}[\rho]$, is the same as the LDA exchange energy functional \cite{LDAX}, 
\begin{equation} 
F_{x}^{\text{LDA},\theta=0}[\rho] = E_{x}^{\text{LDA}}[\rho] = -\frac{3}{4}\left(\frac{3}{\pi}\right)^{1/3} \int \rho^{4/3}({\bf r}) d{\bf r}. 
\end{equation} 
For consistency, in this work, we evaluate the functional derivative of $F_{x}^{\text{LDA},\theta}[\rho]$ based on Eq.\ (\ref{eq:fx_lda}), instead of adopting the independent parametrization 
given by Eq.\ (3.3) of Ref.\ \cite{Fx}. 

The B3LYP, B3LYP-D3, PBE0, and BHHLYP global hybrid functionals (together with $E_{\theta}^{\text{LDA}}$ and $E_{x,\theta}^{\text{LDA}}$) in TAO-DFT are denoted as TAO-B3LYP, 
TAO-B3LYP-D3, TAO-PBE0, and TAO-BHHLYP, respectively, which reduce to KS-B3LYP, KS-B3LYP-D3, KS-PBE0, and KS-BHHLYP, respectively (i.e., the corresponding global hybrid 
functionals in KS-DFT) at $\theta = 0$. 

All calculations are performed with a development version of \textsf{Q-Chem 4.3} \cite{QChem}. 
Spin-restricted theory is employed for singlet states and spin-unrestricted theory for others, unless noted otherwise. For the interaction energies of the weakly bound systems, 
the counterpoise correction \cite{CP} is adopted to reduce the basis set superposition error (BSSE). Results are calculated using the 6-311++G(3df,3pd) basis set with the fine grid 
EML(75,302), consisting of 75 Euler-Maclaurin radial grid points \cite{EM} and 302 Lebedev angular grid points \cite{L}, unless noted otherwise. 
The error for each entry is defined as (error = theoretical value $-$ reference value). The notation adopted for characterizing statistical errors is as follows: 
mean signed errors (MSEs), mean absolute errors (MAEs), root-mean-square (rms) errors, maximum negative errors (Max($-$)), and maximum positive errors (Max(+)). 

\begin{figure} 
\includegraphics[scale=0.83]{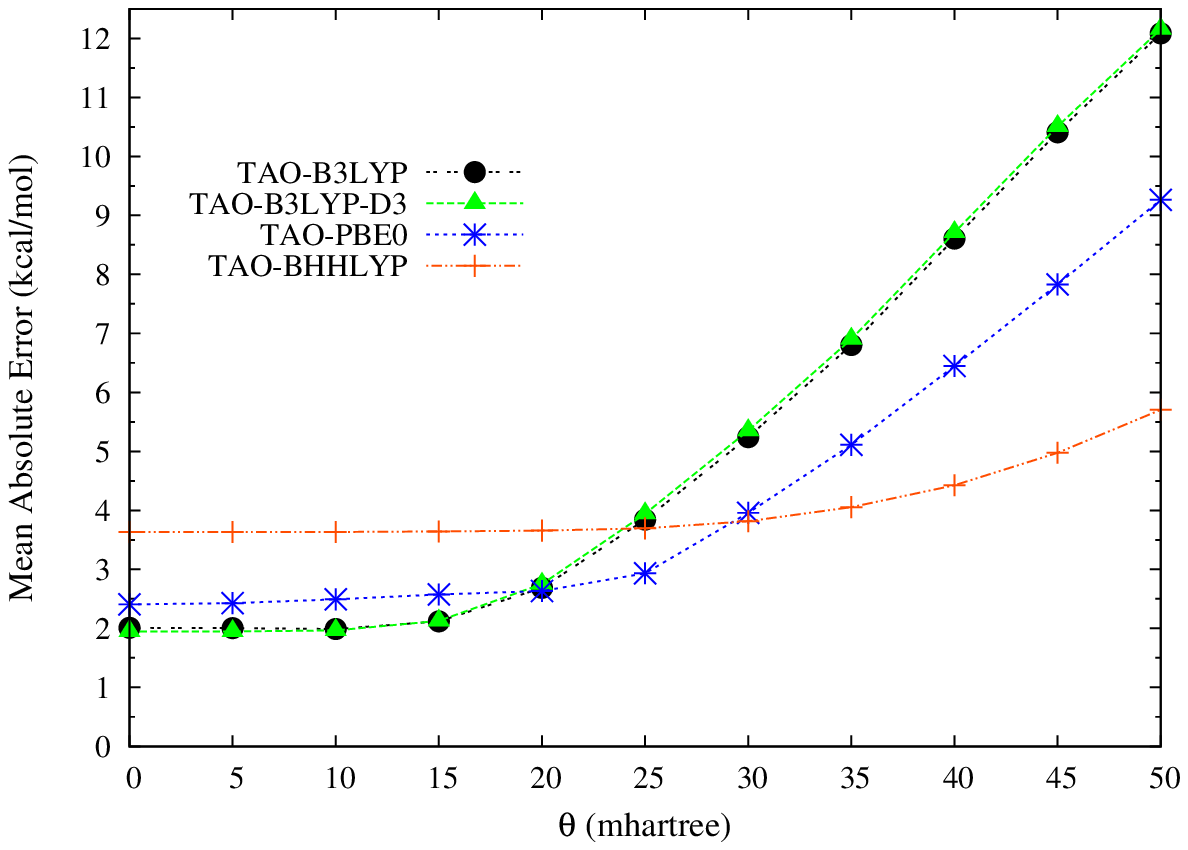} 
\caption{\label{fig:mae_re} 
Mean absolute errors of the reaction energies of the 30 chemical reactions in the NHTBH38/04 and HTBH38/04 sets \cite{BH}, calculated using 
TAO-B3LYP, TAO-B3LYP-D3, TAO-PBE0, and TAO-BHHLYP (with various $\theta$). The $\theta = 0$ cases correspond to 
KS-B3LYP, KS-B3LYP-D3, KS-PBE0, and KS-BHHLYP, respectively.} 
\end{figure}

The reaction energies of the 30 chemical reactions with different barrier heights for the forward and backward directions in the NHTBH38/04 and HTBH38/04 sets \cite{BH} are adopted 
to assess the performance of TAO-B3LYP, TAO-B3LYP-D3, TAO-PBE0, and TAO-BHHLYP (with various $\theta$ values). As shown in \Cref{fig:mae_re}, the global hybrid functionals in 
TAO-DFT (with sufficiently small $\theta$ values) perform similarly to the corresponding global hybrid functionals in KS-DFT (i.e., the $\theta = 0$ cases). Unsurprisingly, these systems 
do not have significant amounts of static correlation, and hence, the exact NOONs should be close to either 0 or 1, which can be properly simulated by the TOONs of TAO-B3LYP, 
TAO-B3LYP-D3, TAO-PBE0, and TAO-BHHLYP (with sufficiently small $\theta$ values). 

\begin{figure} 
\includegraphics[scale=0.83]{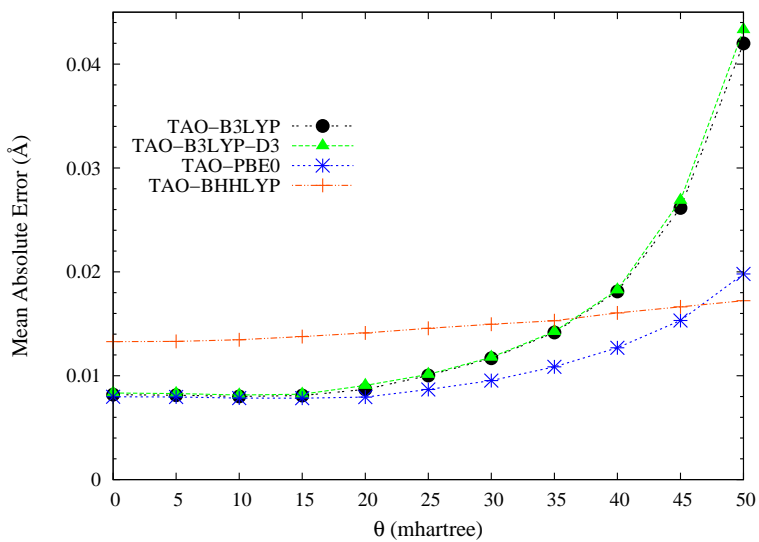} 
\caption{\label{fig:mae_exts} 
Mean absolute errors of the 166 bond lengths in the EXTS set \cite{EXTS}, calculated using 
TAO-B3LYP, TAO-B3LYP-D3, TAO-PBE0, and TAO-BHHLYP (with various $\theta$). The $\theta = 0$ cases correspond to 
KS-B3LYP, KS-B3LYP-D3, KS-PBE0, and KS-BHHLYP, respectively.} 
\end{figure}

An accurate and efficient prediction of molecular geometries can be essential for practical applications. Geometry optimizations for TAO-B3LYP, TAO-B3LYP-D3, TAO-PBE0, and 
TAO-BHHLYP (with various $\theta$ values) are performed using analytical nuclear gradients on the equilibrium experimental test set (EXTS) \cite{EXTS}, which contains 166 symmetry 
unique experimental bond lengths for small to medium size molecules. As shown in \Cref{fig:mae_exts}, the global hybrid functionals in TAO-DFT (with sufficiently small $\theta$ values) 
have similar performance to the corresponding global hybrid functionals in KS-DFT (i.e., the $\theta = 0$ cases). As the ground states of these molecules near their equilibrium geometries 
do not exhibit significant multi-reference character, the exact NOONs are close to either 0 or 1, which can be well described by the TOONs of the global hybrid functionals in TAO-DFT 
(with sufficiently small $\theta$ values). 

In this work, the optimal $\theta$ value for a global hybrid functional in TAO-DFT is defined as the largest $\theta$ value for which the difference between the MAE of the global hybrid 
functional in TAO-DFT (with this $\theta$) and that of the corresponding global hybrid functional in KS-DFT (i.e., the $\theta$ = 0 case) is less than 0.5 kcal/mol for the 30 reaction energies, 
and less than 0.003 {\AA} for the 166 bond lengths. On the basis of our numerical investigations, the optimal $\theta$ value is estimated to be 15 mhartree for TAO-B3LYP and 
TAO-B3LYP-D3, 20 mhartree for TAO-PBE0, and 35 mhartree for TAO-BHHLYP. 
Although only four global hybrid functionals in TAO-DFT are examined in this work, some common characteristics are summarized as follows. 
Since the dispersion corrections have no effects on the TAO orbitals and their occupation numbers, the optimal $\theta$ value for a global hybrid functional with and without the 
dispersion corrections in TAO-DFT is the same. In addition, as previously mentioned, the choice of DFA functionals (e.g., $E_{x}^{\text{DFA}}$, $E_{c}^{\text{DFA}}$, 
$E_{\theta}^{\text{DFA}}$, and $E_{x,\theta}^{\text{DFA}}$) has insignificant effects on the optimal $\theta$ values \cite{TAO2}. Accordingly, we expect that the optimal $\theta$ value 
for a global hybrid functional in TAO-DFT should be mainly dependent on the fraction of exact exchange $a_{x}$. A global hybrid functional with a larger fraction of exact exchange gives 
larger orbital energy gaps \cite{LCAC,EB}, requiring a larger $\theta$ value to yield a similar distribution of TOONs. 

Here, based on a simple linear interpolation between 
the optimal $\theta$ = 7 mhartree for TAO-DFAs \cite{TAO1,TAO2} ($a_{x} = 0$) and the optimal $\theta$ = 20 mhartree for TAO-PBE0 ($a_{x} = 1/4$), the optimal $\theta$ (in mhatree) 
\begin{equation}\label{eq:opt_theta} 
\theta = 7 + 52\ a_{x} 
\end{equation} 
for a global hybrid functional in TAO-DFT (see Eq.\ (\ref{eq:tao-gh})) is expressed as a linear function of the fraction of exact exchange $a_{x}$. 
As shown in \Cref{table:opt_theta}, the optimal $\theta$ value, given by Eq.\ (\ref{eq:opt_theta}), is 17.4 mhartree for TAO-B3LYP and TAO-B3LYP-D3, 20 mhartree for TAO-PBE0, and 
33 mhartree for TAO-BHHLYP, matching well with the aforementioned optimal $\theta$ values. Therefore, the optimal $\theta$ value for a global hybrid functional with 0--50\% exact 
exchange (i.e., most of the existing global hybrid functionals) in TAO-DFT should be reliably given by Eq.\ (\ref{eq:opt_theta}), while the optimal $\theta$ value for a global hybrid 
functional with 50--100\% exact exchange in TAO-DFT may also be reasonably given by Eq.\ (\ref{eq:opt_theta}). 

The 30 reaction energies (see \Cref{table:reall}) and 166 bond lengths (see \Cref{table:EXTS}) calculated using TAO-B3LYP, TAO-B3LYP-D3, TAO-PBE0, and TAO-BHHLYP (with the 
optimal $\theta$ values given in \Cref{table:opt_theta}) are indeed similar to those calculated using KS-B3LYP, KS-B3LYP-D3, KS-PBE0, and KS-BHHLYP, respectively 
(see the supplementary material). 
In addition, relative to TAO-DFAs (see Tables II and III of Ref.\ \cite{TAO2}), the global hybrid functionals in TAO-DFT are superior in performance for the 30 reaction energies and 
166 bond lengths.

\subsection{Results and discussion for the test sets} 

Here, we examine the performance of TAO-B3LYP, TAO-B3LYP-D3, TAO-PBE0, and TAO-BHHLYP (with the optimal $\theta$ values given in \Cref{table:opt_theta}, unless noted otherwise) 
on various test sets, including both single- and multi-reference systems. The results are compared with those obtained from KS-B3LYP, KS-B3LYP-D3, KS-PBE0, and KS-BHHLYP (i.e., 
the corresponding global hybrid functionals in KS-DFT). 

\subsubsection{$\omega$B97 training set} 

The $\omega$B97 training set \cite{wB97X} contains different types of databases, such as 
\begin{itemize} 
\item the 223 atomization energies (AEs) of the G3/99 set \cite{G399}, 
\item the 40 ionization potentials (IPs), 25 electron affinities (EAs), and 8 proton affinities (PAs) of the G2-1 set \cite{G21}, 
\item the 76 barrier heights (BHs) of the NHTBH38/04 and HTBH38/04 sets \cite{BH}, 
\item the 22 noncovalent interactions of the S22 set \cite{S22}. 
\end{itemize} 
Since these systems do not exhibit significant static correlation, TAO-B3LYP, TAO-B3LYP-D3, TAO-PBE0, and TAO-BHHLYP perform comparably to KS-B3LYP, KS-B3LYP-D3, KS-PBE0, 
and KS-BHHLYP, respectively (see \Cref{table:training}) (see the supplementary material). In particular, TAO-B3LYP ($a_{x} = 1/5$) performs well for thermochemistry, and 
TAO-BHHLYP ($a_{x} = 1/2$) 
performs well for kinetics. For the noncovalent interactions of the S22 set, the dispersion corrected functionals, KS-B3LYP-D3 and TAO-B3LYP-D3, perform better than the other 
functionals, suggesting that the DFT-D schemes can be adopted in both KS-DFT and TAO-DFT for an accurate description of noncovalent interactions. 
Besides, due to the improved treatment of nonlocal exchange effects, TAO-B3LYP, TAO-B3LYP-D3, TAO-PBE0, and TAO-BHHLYP are shown to significantly outperform 
TAO-DFAs (see Table I of Ref.\ \cite{TAO2}) for the 223 AEs of the G3/99 set and the 76 BHs of the NHTBH38/04 and HTBH38/04 sets. 

\subsubsection{Dissociation of H$_2$ and N$_2$} 

Owing to the presence of strong static correlation effects, the dissociation of molecular hydrogen H$_2$ (a single-bond breaking system) remains very challenging for KS-DFT. On the 
basis of the symmetry constraint, the spin-restricted and spin-unrestricted dissociation energy curves of H$_2$ calculated using the exact theory, should be identical. Accordingly, the 
difference between the spin-restricted and spin-unrestricted dissociation limits calculated using an approximate electronic structure method, can be taken as a quantitative measure of 
the SCE of the method \cite{SciYang,SCE}. Conventional LDA, GGA, hybrid, and double-hybrid functionals in spin-restricted KS-DFT have been shown to yield very large SCEs for 
the dissociation of H$_2$, owing to the inappropriate treatment of static correlation. By contrast, spin-restricted TAO-LDA and TAO-GGAs (with a $\theta$ between 30 and 50 mhartree) 
are able to dissociate H$_2$ correctly (yielding vanishingly small SCEs) to the respective spin-unrestricted dissociation limits, which is closely related to that the distribution of TOONs 
(related to the chosen $\theta$) matches reasonably well with that of NOONs \cite{TAO1,TAO2}. 

\begin{figure} 
\subfigure 
{\includegraphics[scale=0.603]{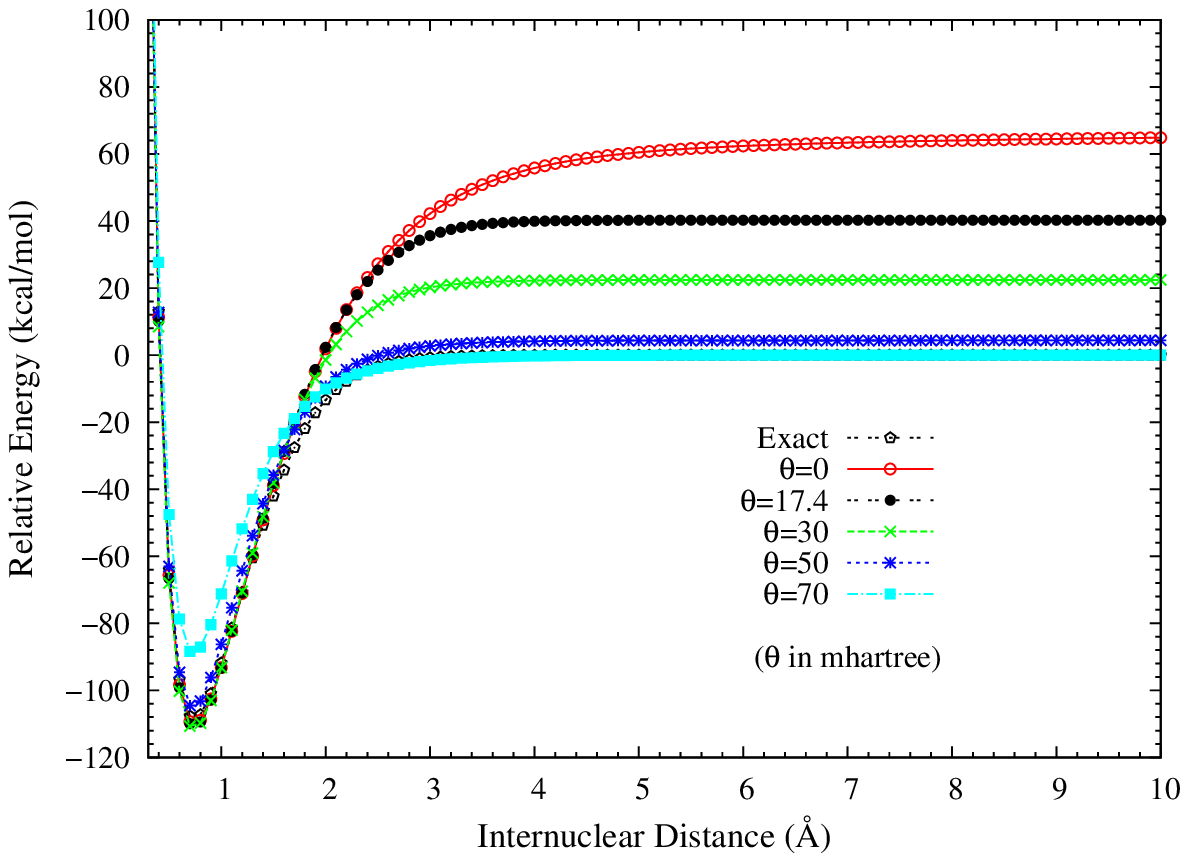}(a)} 
\subfigure 
{\includegraphics[scale=0.603]{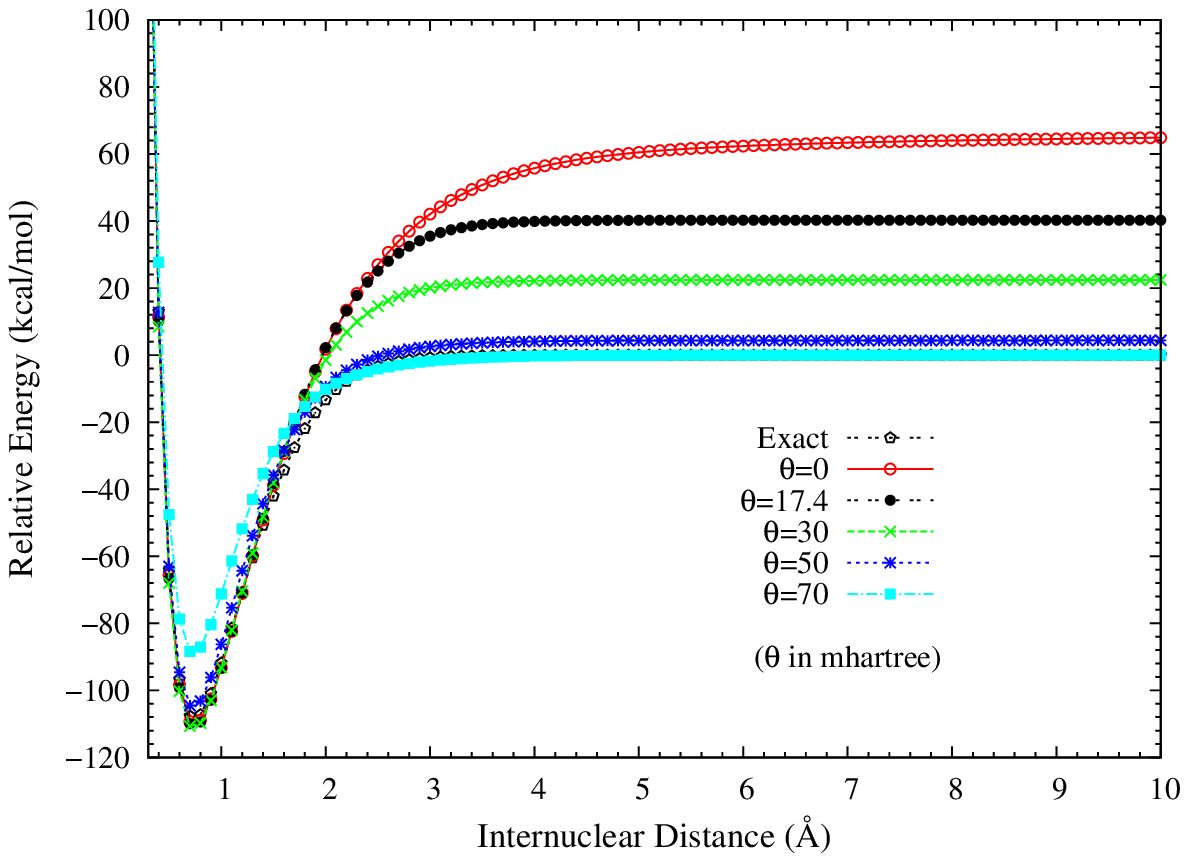}(b)} 
\subfigure 
{\includegraphics[scale=0.603]{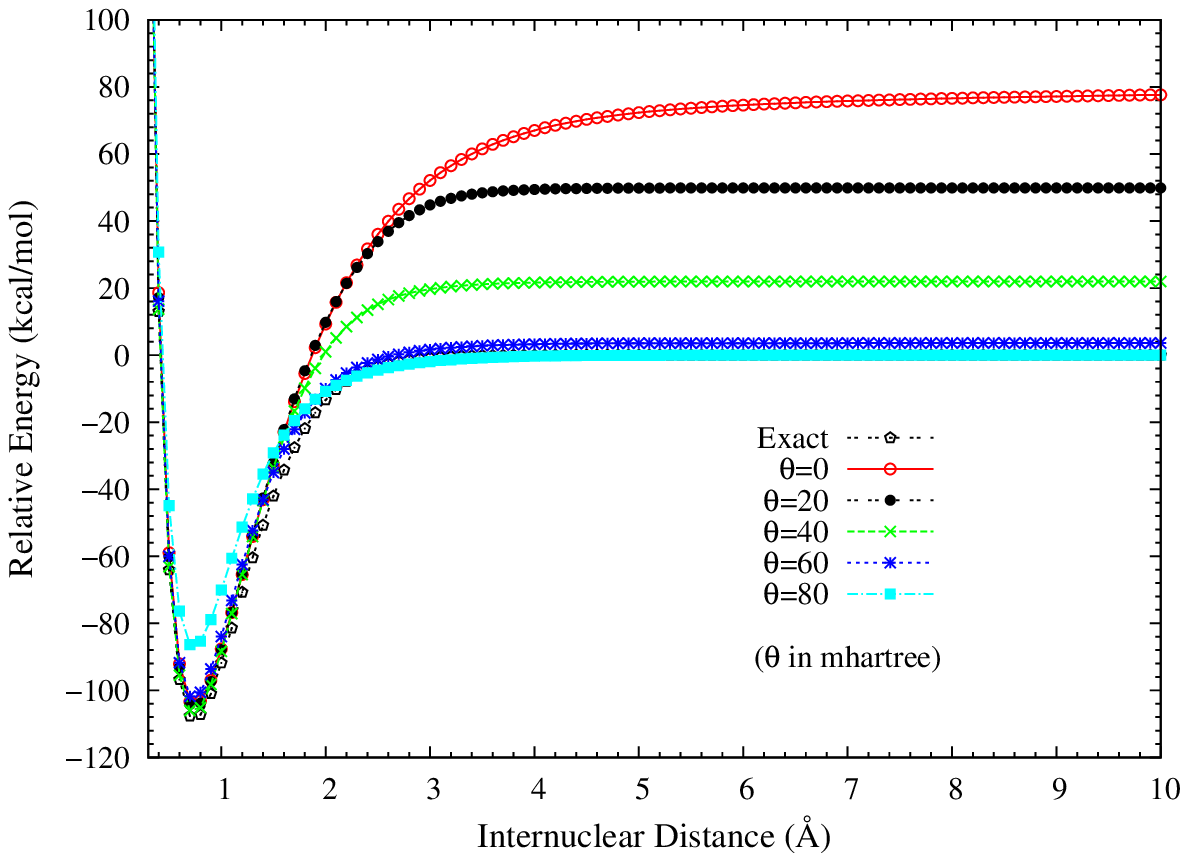}(c)} 
\subfigure 
{\includegraphics[scale=0.603]{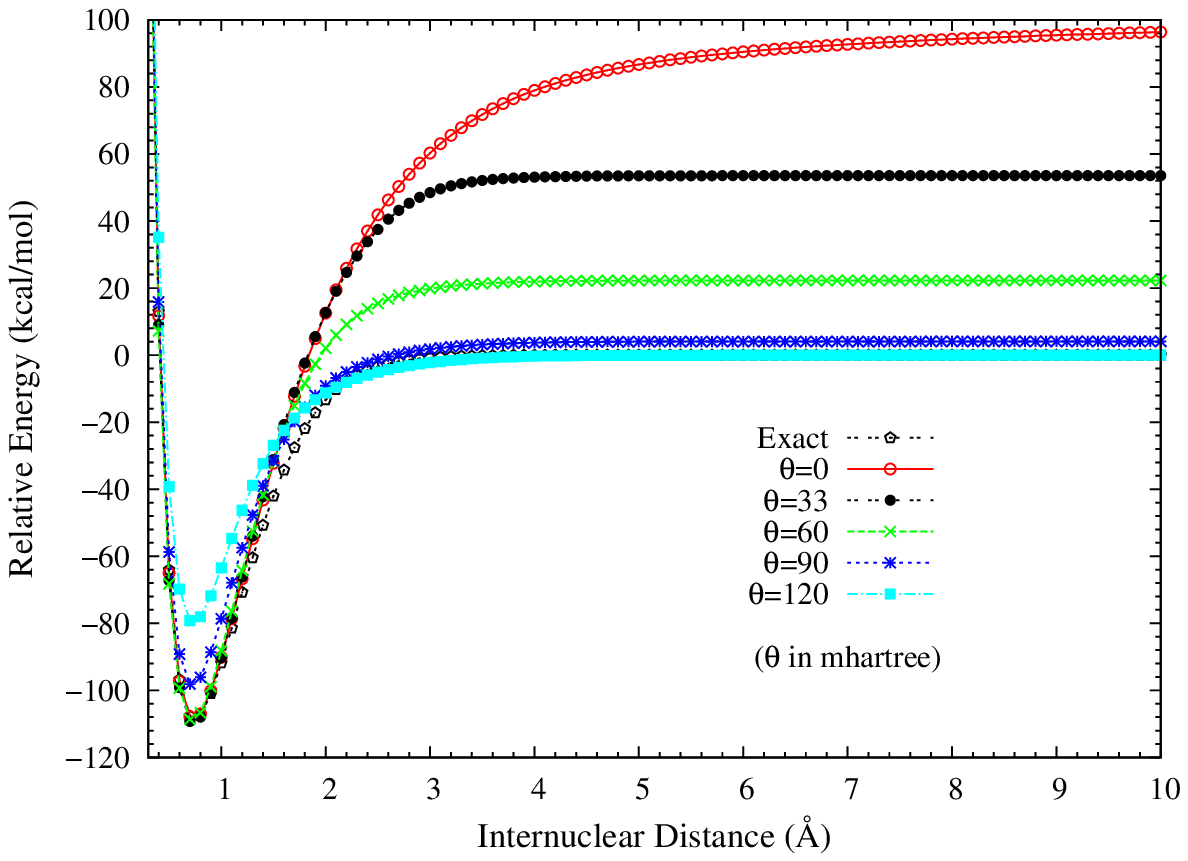}(d)} 
\caption{\label{fig:h2} 
Potential energy curves (in relative energy) for the ground state of H$_2$, calculated using spin-restricted 
(a) TAO-B3LYP, (b) TAO-B3LYP-D3, (c) TAO-PBE0, and (d) TAO-BHHLYP (with various $\theta$). 
The $\theta = 0$ cases correspond to spin-restricted 
(a) KS-B3LYP, (b) KS-B3LYP-D3, (c) KS-PBE0, and (d) KS-BHHLYP, respectively. 
The exact curve is calculated using the CCSD theory. 
The zeros of energy are set at the respective spin-unrestricted dissociation limits.} 
\end{figure}

To assess the performance of the present method upon the SCE problems, the potential energy curves (in relative energy) for the ground state of H$_2$ are calculated using 
spin-restricted TAO-B3LYP, TAO-B3LYP-D3, TAO-PBE0, and TAO-BHHLYP with various $\theta$ values (see \Cref{fig:h2}), 
where the zeros of energy are set at the respective spin-unrestricted dissociation limits. The results are compared with the exact curve calculated using 
the coupled-cluster theory with iterative singles and doubles (CCSD) \cite{CCSD}, which is equivalent to the FCI method for any two-electron system \cite{H2_NOON}. 

Near the equilibrium bond length of H$_2$, where the single-reference character is predominant, the global hybrid functionals in TAO-DFT (with the optimal $\theta$ values given 
in \Cref{table:opt_theta}) perform similarly to the corresponding global hybrid functionals in KS-DFT (i.e., the $\theta = 0$ cases), matching reasonably well with the exact curve. 
However, at the dissociation limit, where the multi-reference character becomes pronounced, they have noticeable SCEs. By contrast, spin-restricted 
TAO-B3LYP and TAO-B3LYP-D3 (with a $\theta$ between 50 and 70 mhartree), TAO-PBE0 (with a $\theta$ between 60 and 80 mhartree), and 
TAO-BHHLYP (with a $\theta$ between 90 and 120 mhartree) can properly dissociate H$_2$ (yielding vanishingly small SCEs) to the respective spin-unrestricted dissociation limits. 

\begin{figure} 
\subfigure 
{\includegraphics[scale=0.603]{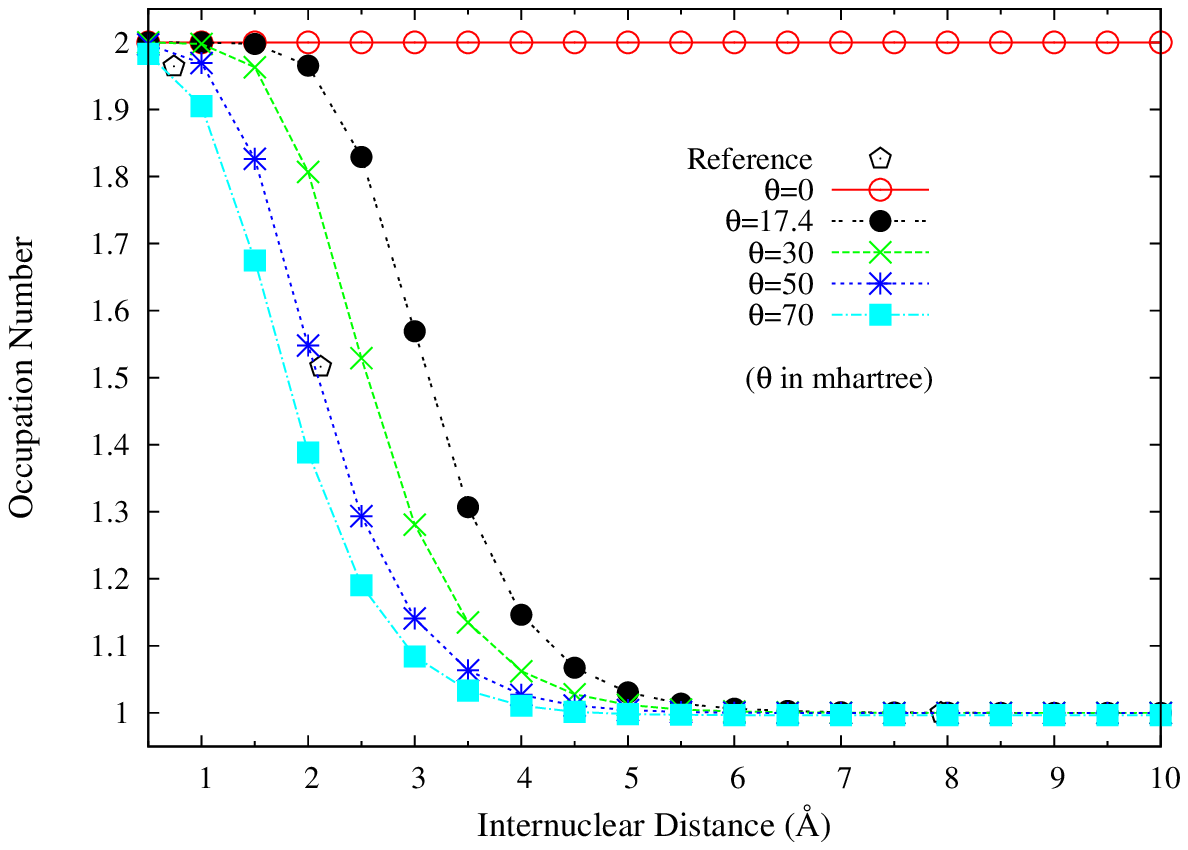}(a)} 
\subfigure 
{\includegraphics[scale=0.603]{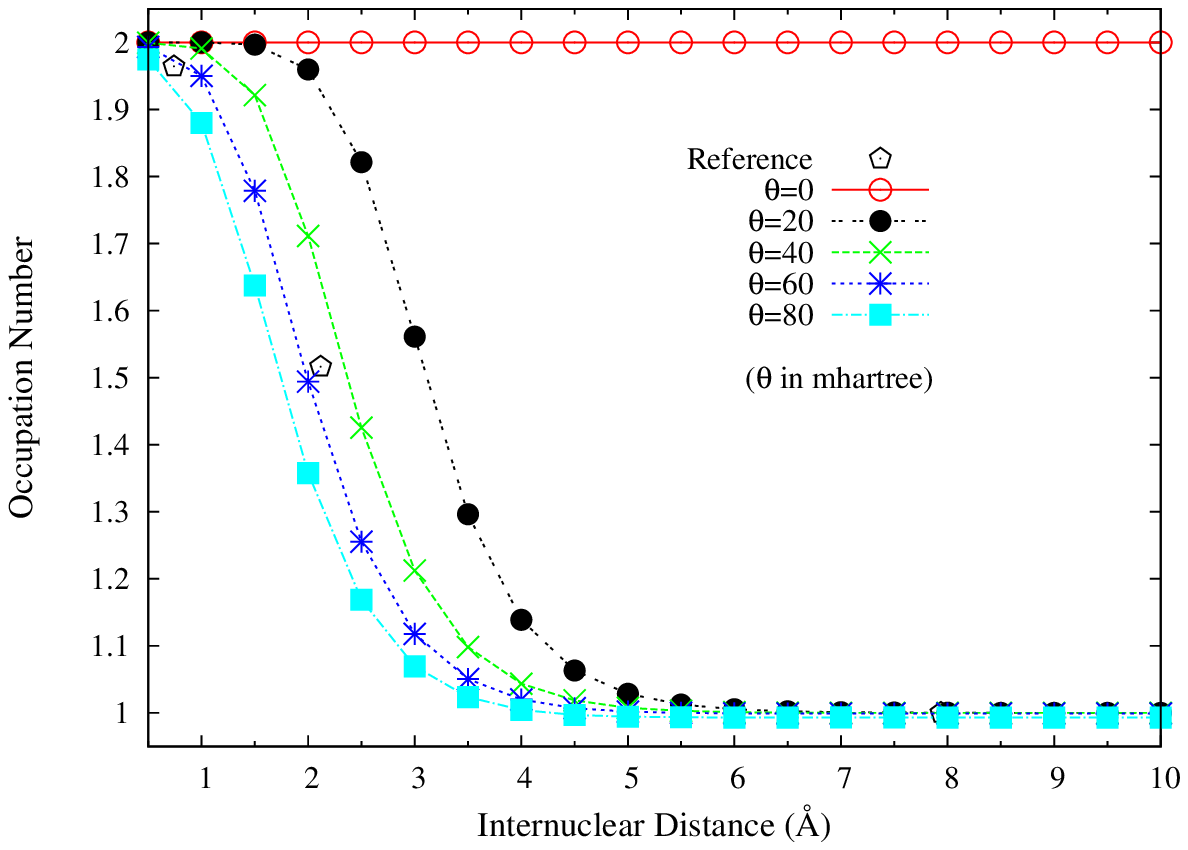}(b)} 
\subfigure 
{\includegraphics[scale=0.603]{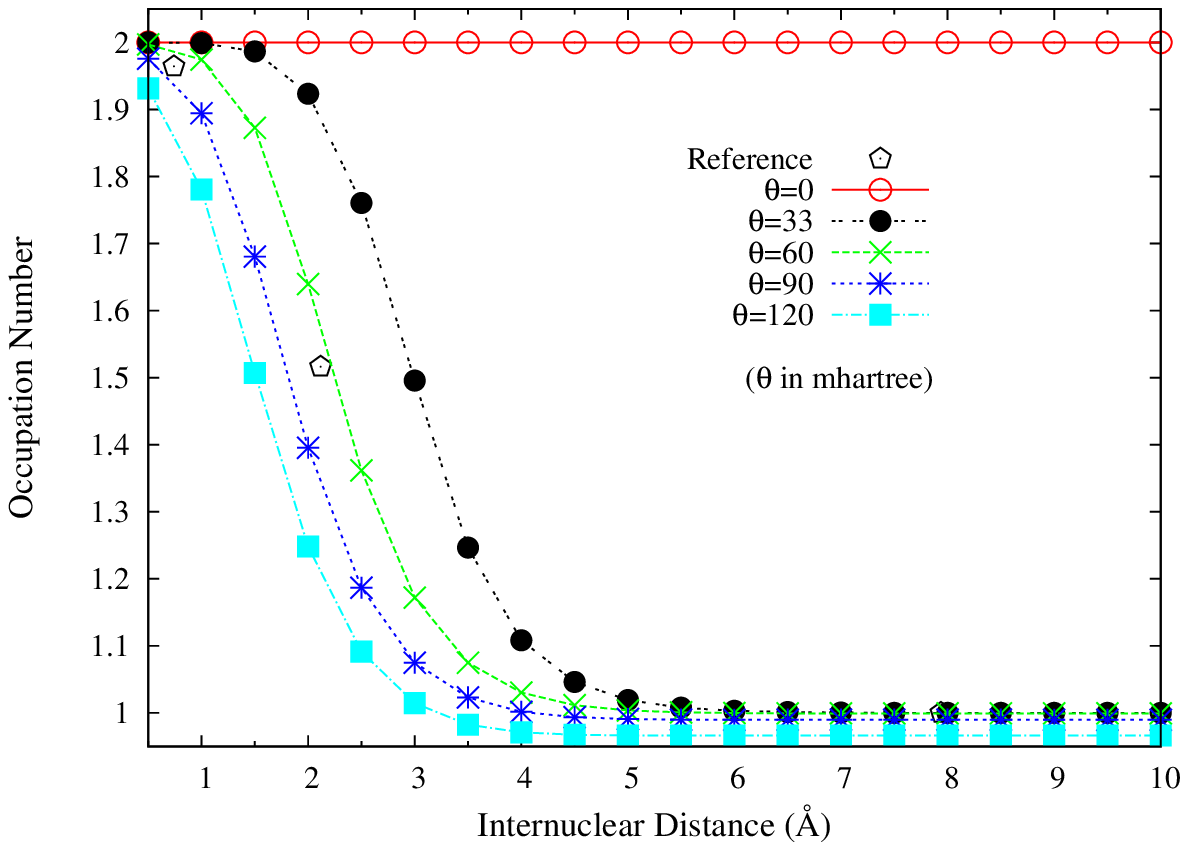}(c)} 
\caption{\label{fig:h2noon} 
Occupation numbers of the $1\sigma_g$ orbital for the ground state of H$_2$ as a function of the internuclear distance $R$, calculated using spin-restricted 
(a) TAO-B3LYP/TAO-B3LYP-D3, (b) TAO-PBE0, and (c) TAO-BHHLYP (with various $\theta$). 
The $\theta = 0$ cases correspond to spin-restricted 
(a) KS-B3LYP/KS-B3LYP-D3, (b) KS-PBE0, and (c) KS-BHHLYP, respectively. 
The reference data are the FCI NOONs \cite{H2_NOON}.} 
\end{figure}

To examine if this is related to the distribution of TOONs, we plot the occupation numbers of the $1\sigma_g$ orbital for the ground state of H$_2$ as a function of the internuclear 
distance $R$, calculated using spin-restricted TAO-B3LYP/TAO-B3LYP-D3, TAO-PBE0, and TAO-BHHLYP with various $\theta$ values (see \Cref{fig:h2noon}), 
where the reference data are the FCI NOONs \cite{H2_NOON}. The FCI NOON is 1.9643 at $R$ = 0.741 {\AA} (i.e., at the equilibrium geometry), 
1.5162 at $R$ = 2.117 {\AA}, and 1.0000 at $R$ = 7.938 {\AA}. As shown, the $1\sigma_g$ orbital occupation numbers of spin-restricted TAO-B3LYP/TAO-B3LYP-D3 (with a $\theta$ 
between 50 and 70 mhartree), TAO-PBE0 (with a $\theta$ between 60 and 80 mhartree), and TAO-BHHLYP (with a $\theta$ between 90 and 120 mhartree) match reasonably well 
with the FCI NOONs, which is closely related to the vanishingly small SCEs of these global hybrid functionals in TAO-DFT (with the same $\theta$ values). This highlights the 
importance of adopting a $\theta$ related to the distribution of NOONs in TAO-DFT. 

\begin{figure} 
\subfigure 
{\includegraphics[scale=0.603]{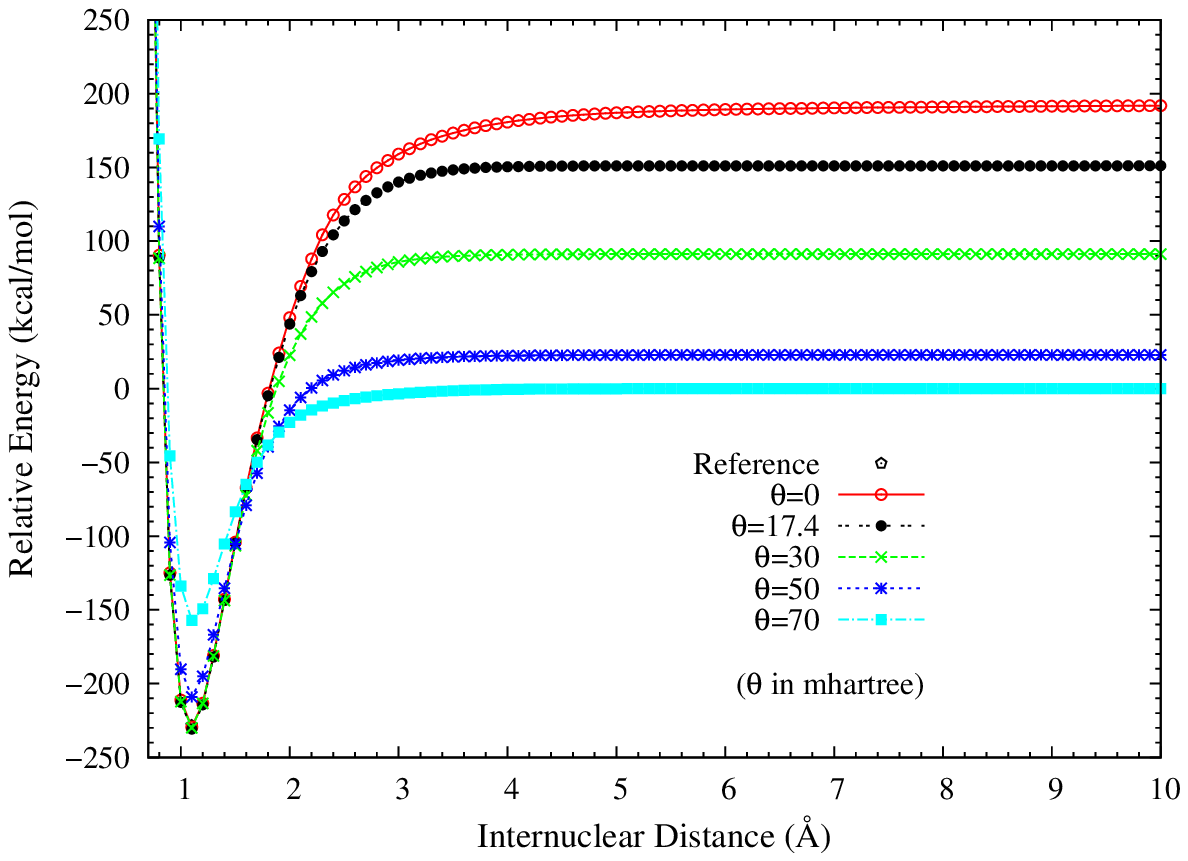}(a)} 
\subfigure 
{\includegraphics[scale=0.603]{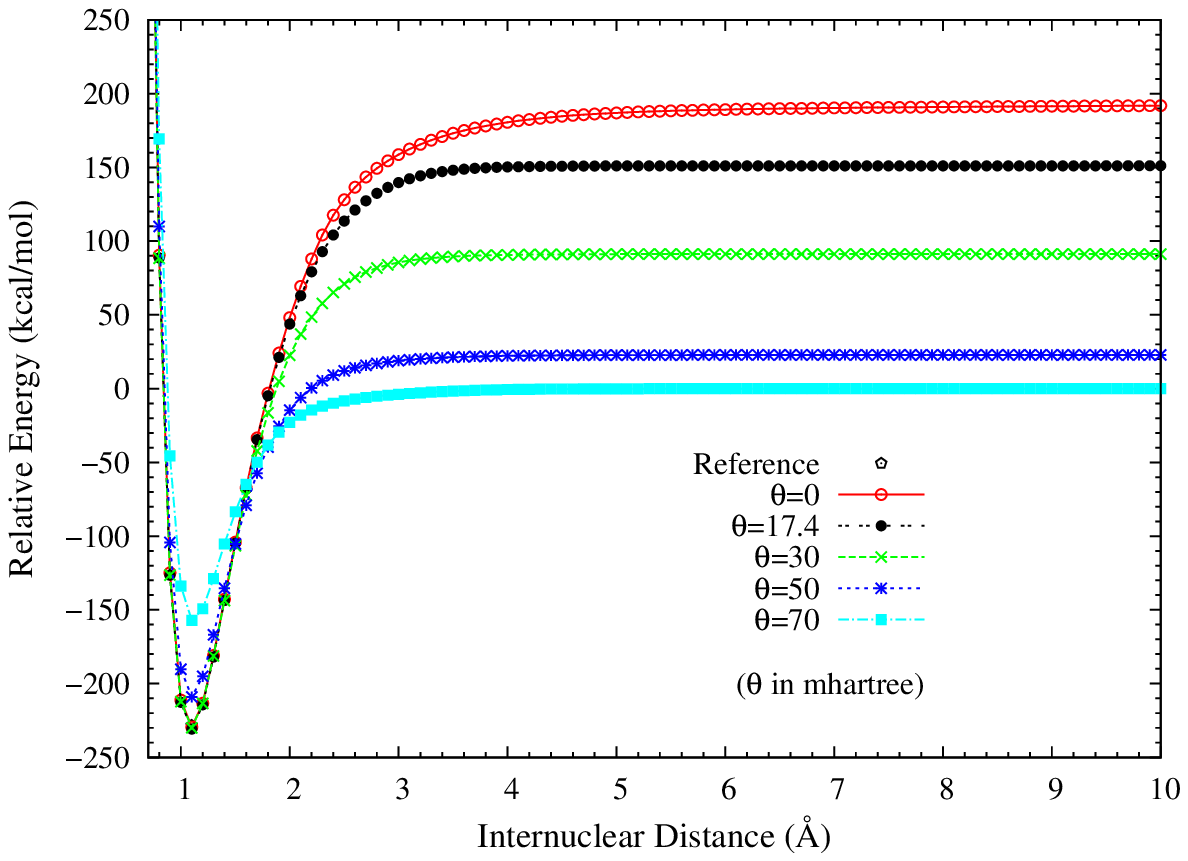}(b)} 
\subfigure 
{\includegraphics[scale=0.603]{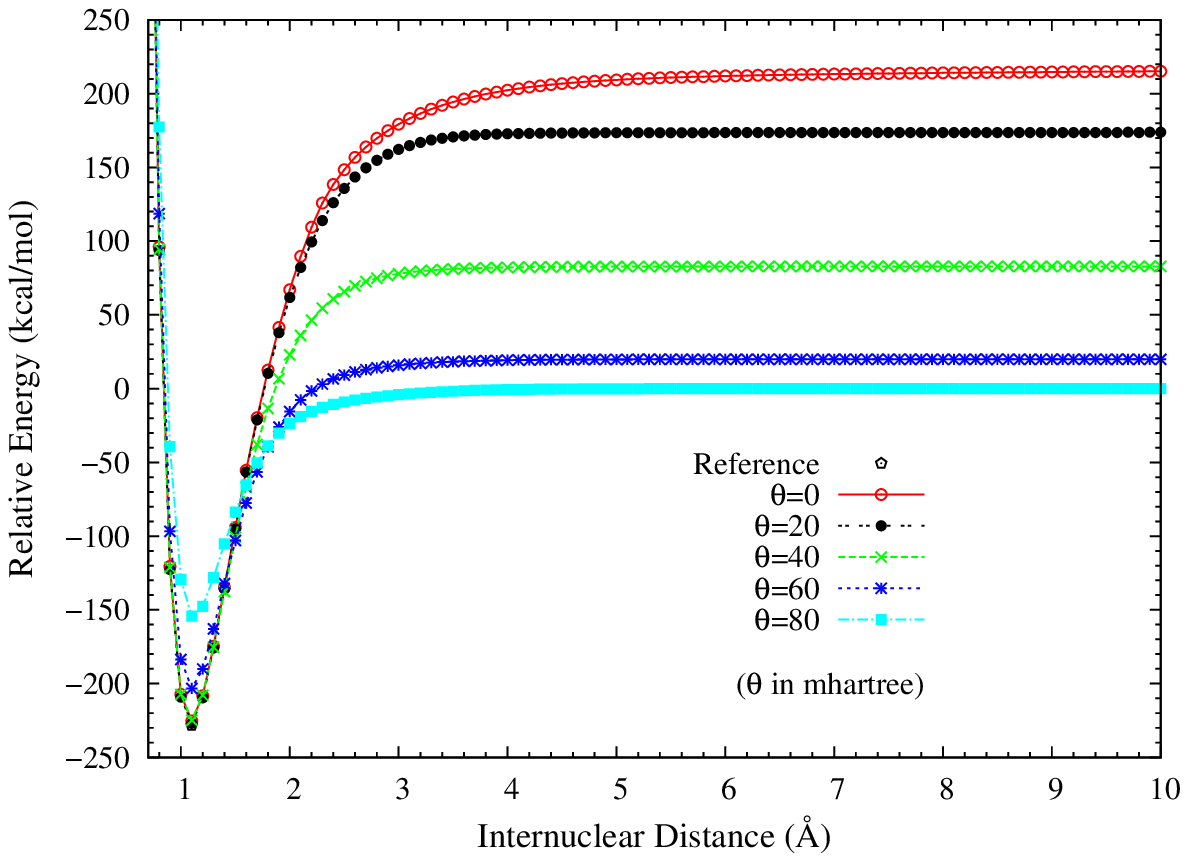}(c)} 
\subfigure 
{\includegraphics[scale=0.603]{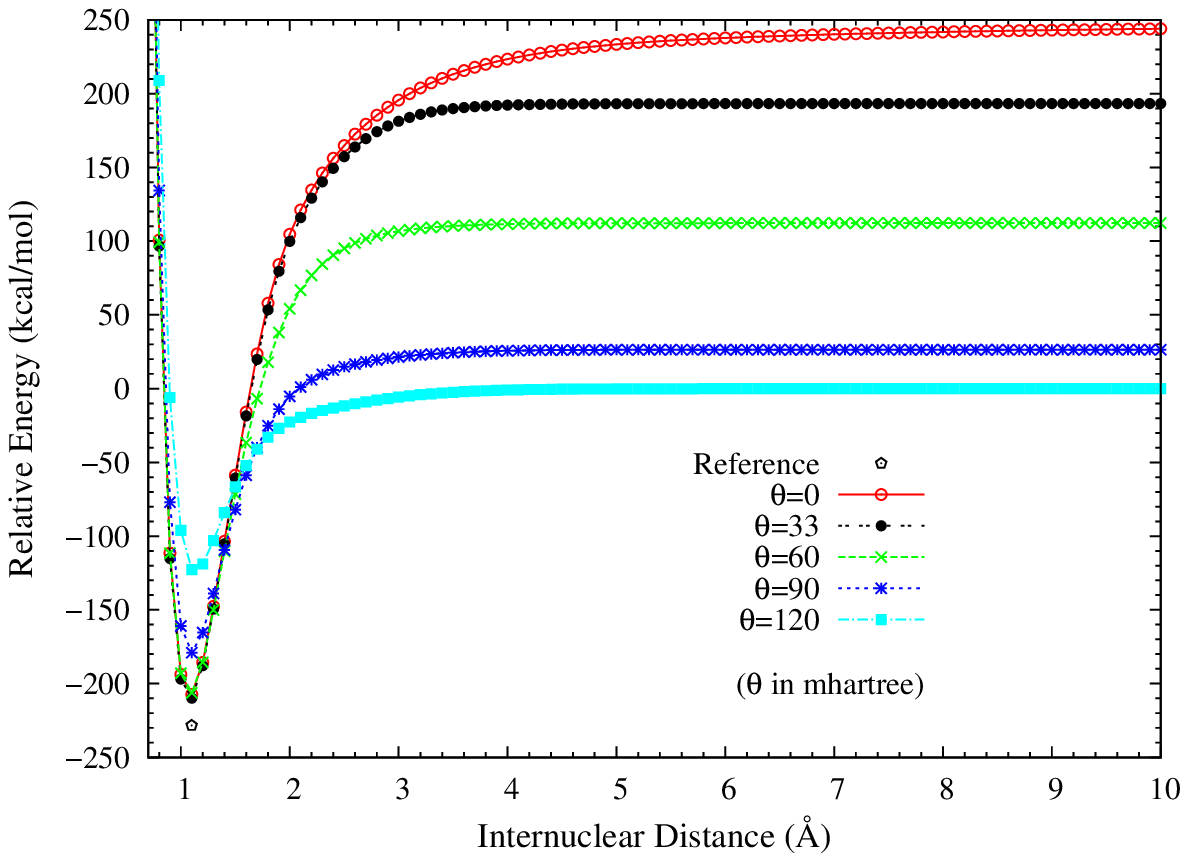}(d)} 
\caption{\label{fig:n2} 
Potential energy curves (in relative energy) for the ground state of N$_2$, calculated using spin-restricted 
(a) TAO-B3LYP, (b) TAO-B3LYP-D3, (c) TAO-PBE0, and (d) TAO-BHHLYP (with various $\theta$). 
The $\theta = 0$ cases correspond to spin-restricted 
(a) KS-B3LYP, (b) KS-B3LYP-D3, (c) KS-PBE0, and (d) KS-BHHLYP, respectively. 
The reference data ($-$228.3 (kcal/mol) at $R$ = 1.098 {\AA} (i.e., at the equilibrium geometry)) are the experimental results \cite{N2_GeomBE}. 
The zeros of energy are set at the respective spin-unrestricted dissociation limits.} 
\end{figure}

\begin{figure} 
\subfigure 
{\includegraphics[scale=0.603]{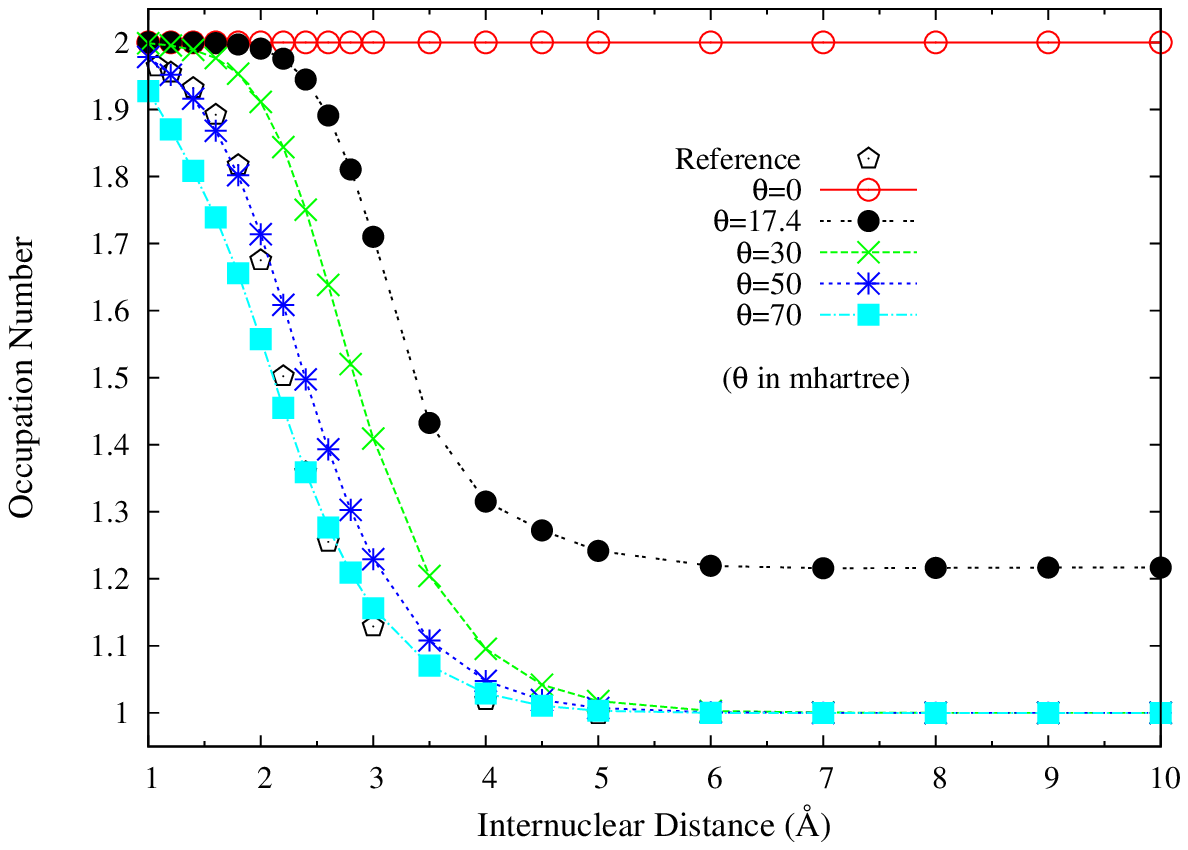}(a)} 
\subfigure 
{\includegraphics[scale=0.603]{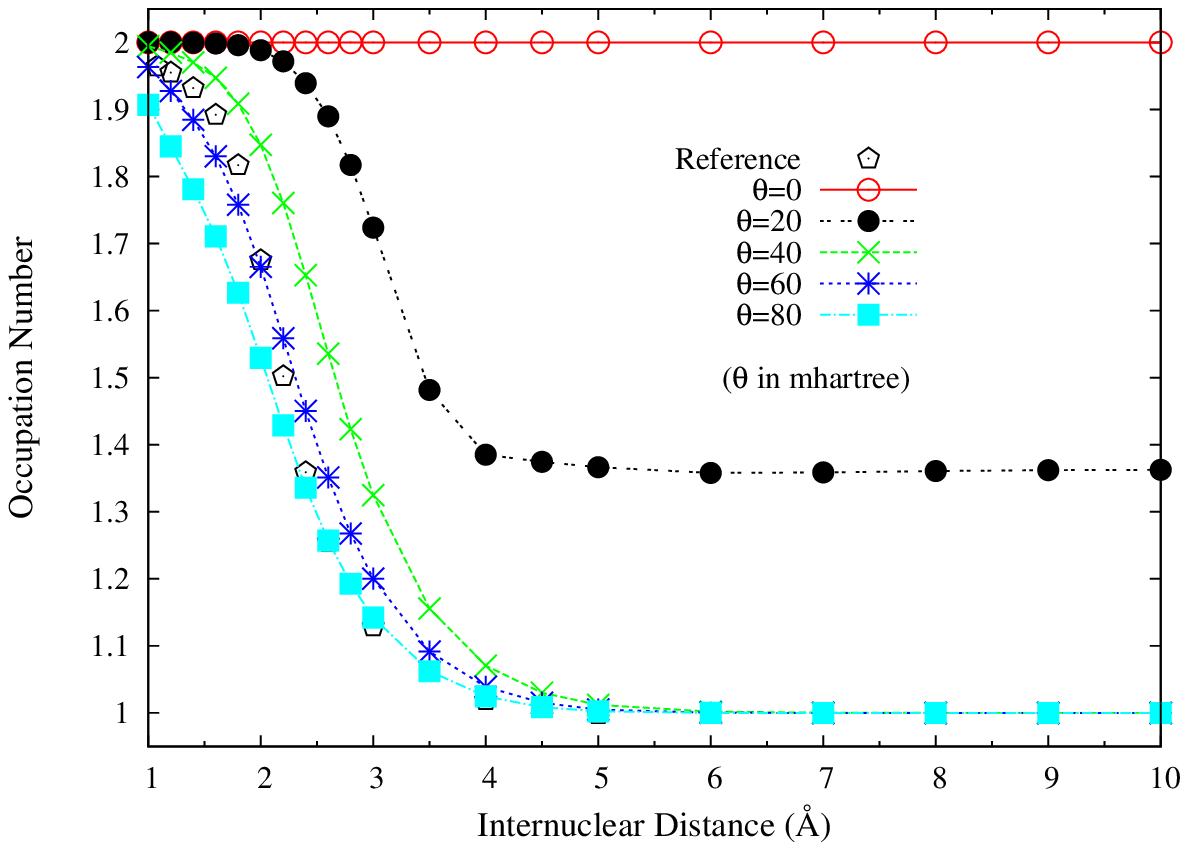}(b)} 
\subfigure 
{\includegraphics[scale=0.603]{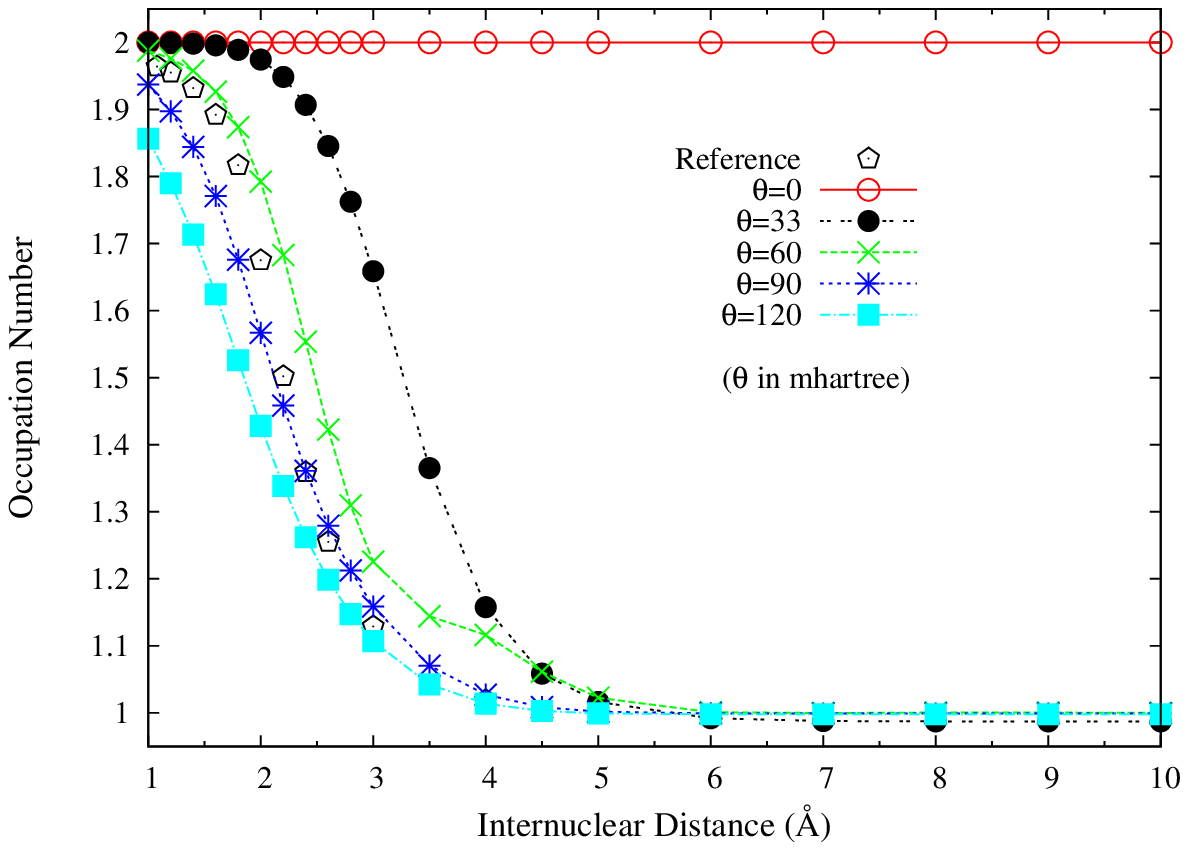}(c)} 
\caption{\label{fig:n2noon_g} 
Occupation numbers of the $3\sigma_g$ orbital for the ground state of N$_2$ as a function of the internuclear distance $R$, calculated using spin-restricted 
(a) TAO-B3LYP/TAO-B3LYP-D3, (b) TAO-PBE0, and (c) TAO-BHHLYP (with various $\theta$). 
The $\theta = 0$ cases correspond to spin-restricted 
(a) KS-B3LYP/KS-B3LYP-D3, (b) KS-PBE0, and (c) KS-BHHLYP, respectively. 
The reference data are the NOONs of MRCI method \cite{N2_NOON}.} 
\end{figure}

\begin{figure} 
\subfigure 
{\includegraphics[scale=0.603]{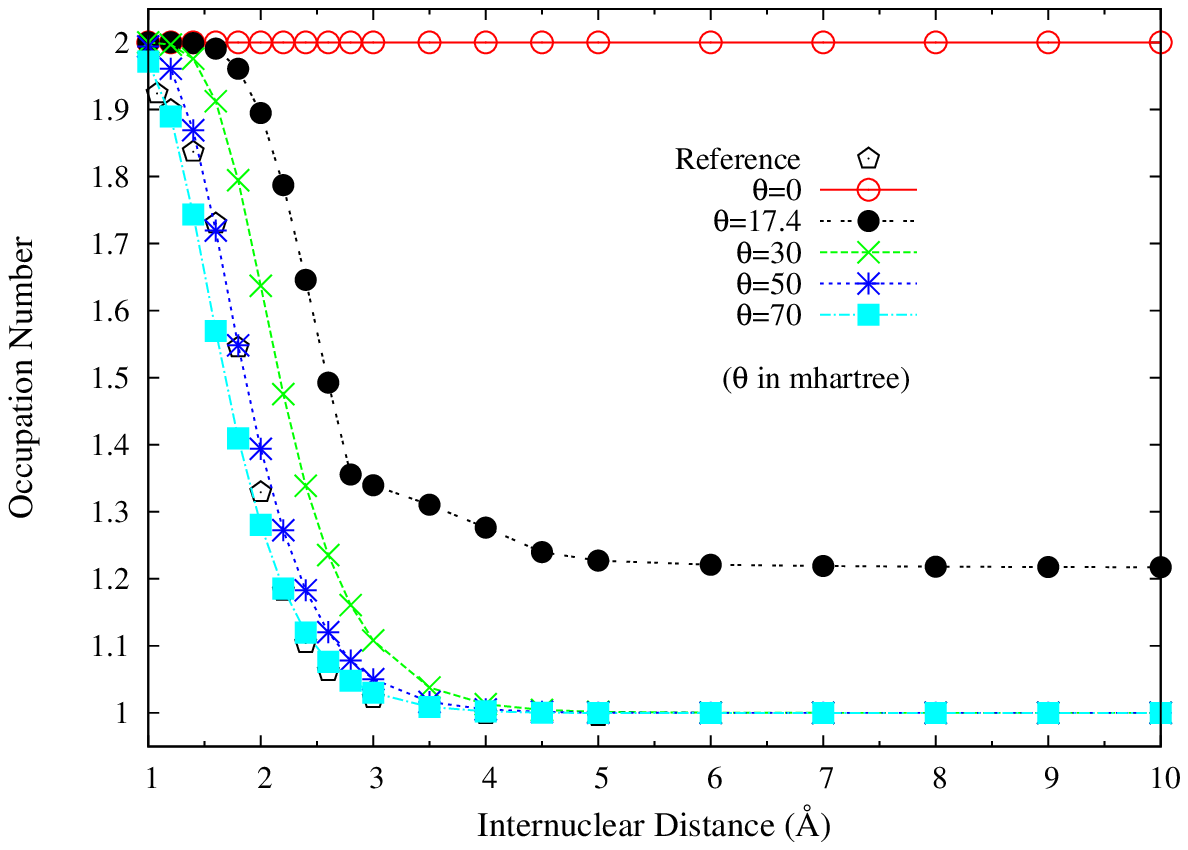}(a)} 
\subfigure 
{\includegraphics[scale=0.603]{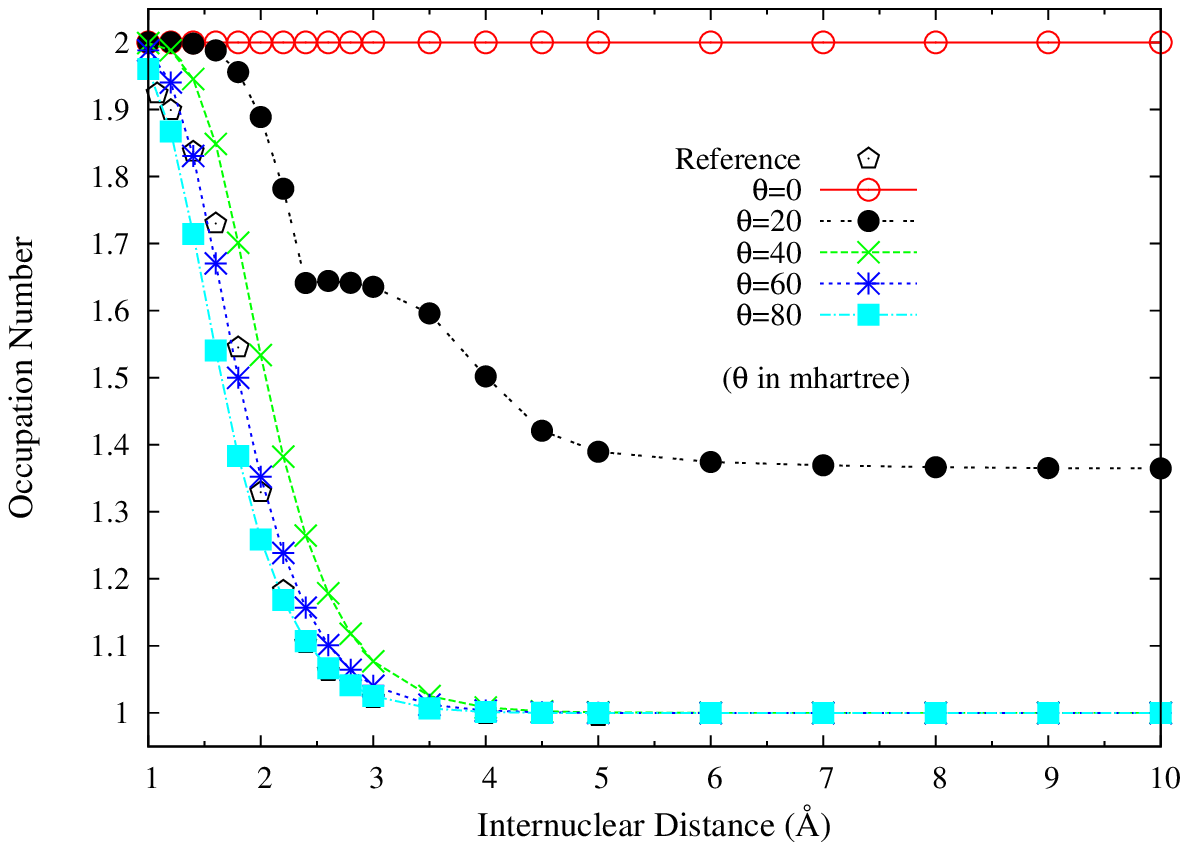}(b)} 
\subfigure 
{\includegraphics[scale=0.603]{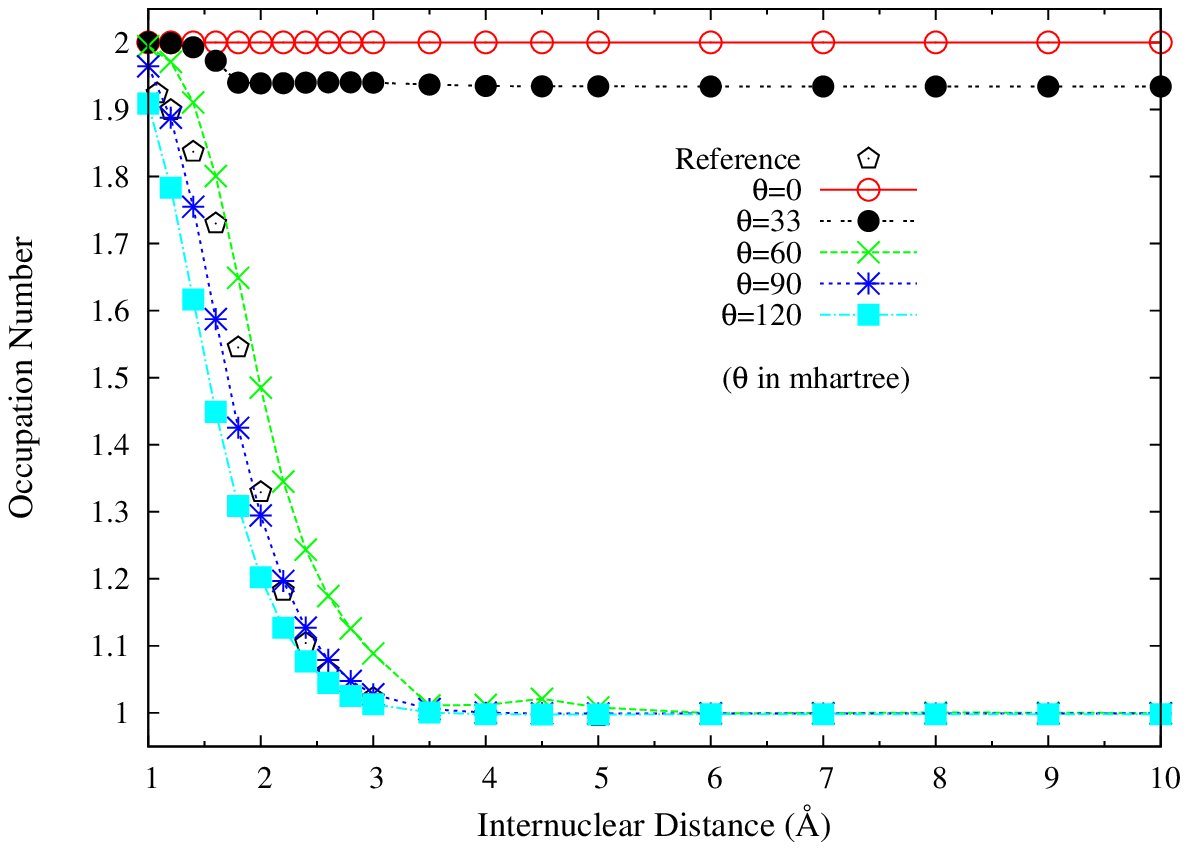}(c)} 
\caption{\label{fig:n2noon_pi} 
Occupation numbers of the $1\pi_{ux}$ orbital for the ground state of N$_2$ as a function of the internuclear distance $R$, calculated using spin-restricted 
(a) TAO-B3LYP/TAO-B3LYP-D3, (b) TAO-PBE0, and (c) TAO-BHHLYP (with various $\theta$). 
The $\theta = 0$ cases correspond to spin-restricted 
(a) KS-B3LYP/KS-B3LYP-D3, (b) KS-PBE0, and (c) KS-BHHLYP, respectively. 
The reference data are the NOONs of MRCI method \cite{N2_NOON}.} 
\end{figure}

Similar results are also found for N$_2$ dissociation (a triple-bond breaking system), where experimental results are also presented \cite{N2_GeomBE}. As shown in \Cref{fig:n2}, 
spin-restricted TAO-B3LYP and TAO-B3LYP-D3 (with a $\theta$ between 50 and 70 mhartree), TAO-PBE0 (with a $\theta$ between 60 and 80 mhartree), and 
TAO-BHHLYP (with a $\theta$ between 90 and 120 mhartree) can dissociate N$_2$ adequately (yielding very small SCEs) to the respective spin-unrestricted dissociation limits, 
which is closely correlated with the fact that the occupation numbers of the $3\sigma_g$ (see \Cref{fig:n2noon_g}) and $1\pi_{ux}$ (see \Cref{fig:n2noon_pi}) orbitals 
for the ground state of N$_2$ as functions of the internuclear distance $R$, calculated using these global hybrid functionals in TAO-DFT (with the same $\theta$ values), 
match reasonably well with the corresponding NOONs of multi-reference configuration interaction (MRCI) method (i.e., the reference data) \cite{N2_NOON}. 
By contrast, while the global hybrid functionals in TAO-DFT (with the optimal $\theta$ values given in \Cref{table:opt_theta}) perform comparably to the corresponding global hybrid 
functionals in KS-DFT (i.e., the $\theta = 0$ cases) near the equilibrium bond length of N$_2$, they yield considerable SCEs at the dissociation limit (as the TOONs do not match well 
with the accurate MRCI NOONs). This again shows the significance of adopting a $\theta$ related to the distribution of NOONs in TAO-DFT. 

\subsubsection{Twisted ethylene} 

The torsion of ethylene (C$_2$H$_4$) remains very difficult for KS-DFT due to the presence of strong static correlation effects. 
The $\pi$ (1b$_2$) and $\pi^*$ (2b$_2$) orbitals in ethylene should be degenerate when the HCCH torsion angle is 90$^{\circ}$. 
However, spin-restricted KS-DFT cannot properly describe such degeneracy, yielding a torsion potential with an unphysical cusp and a too high barrier. 

\begin{figure} 
\subfigure 
{\includegraphics[scale=0.603]{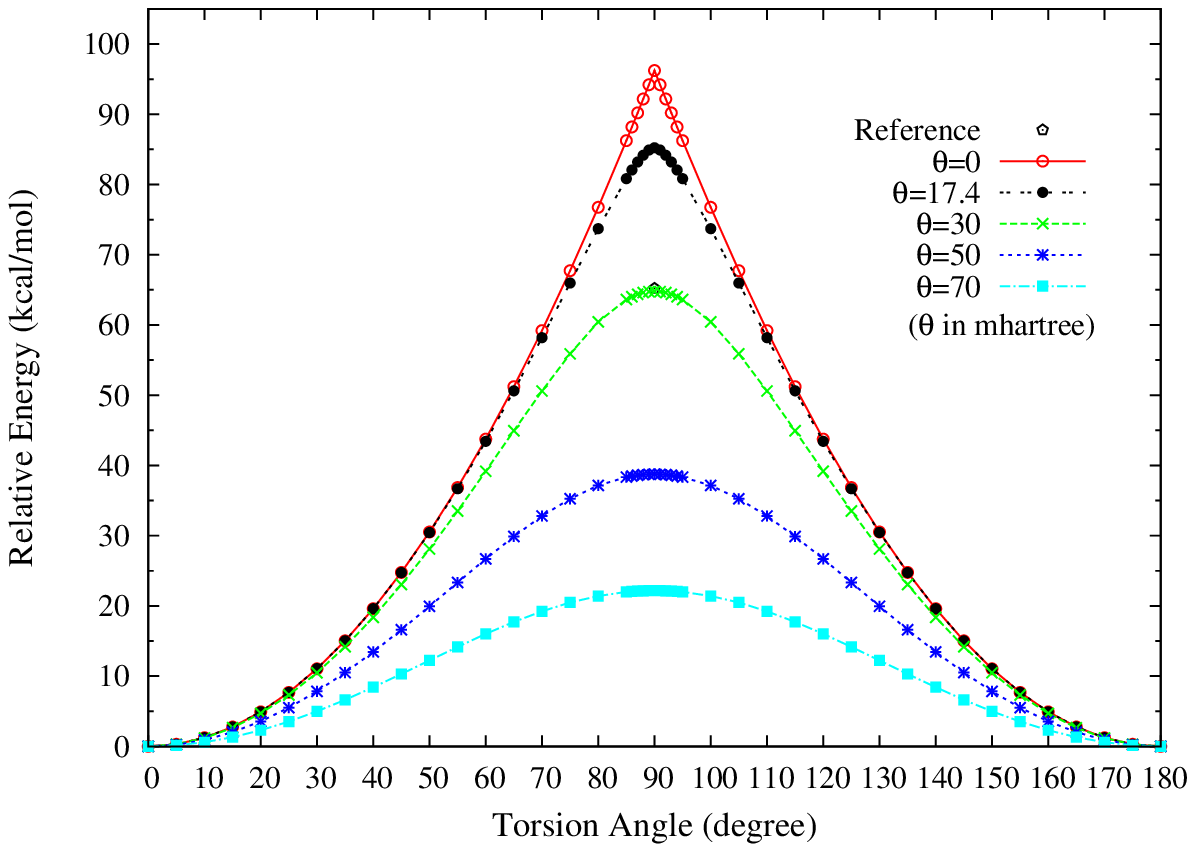}(a)} 
\subfigure 
{\includegraphics[scale=0.603]{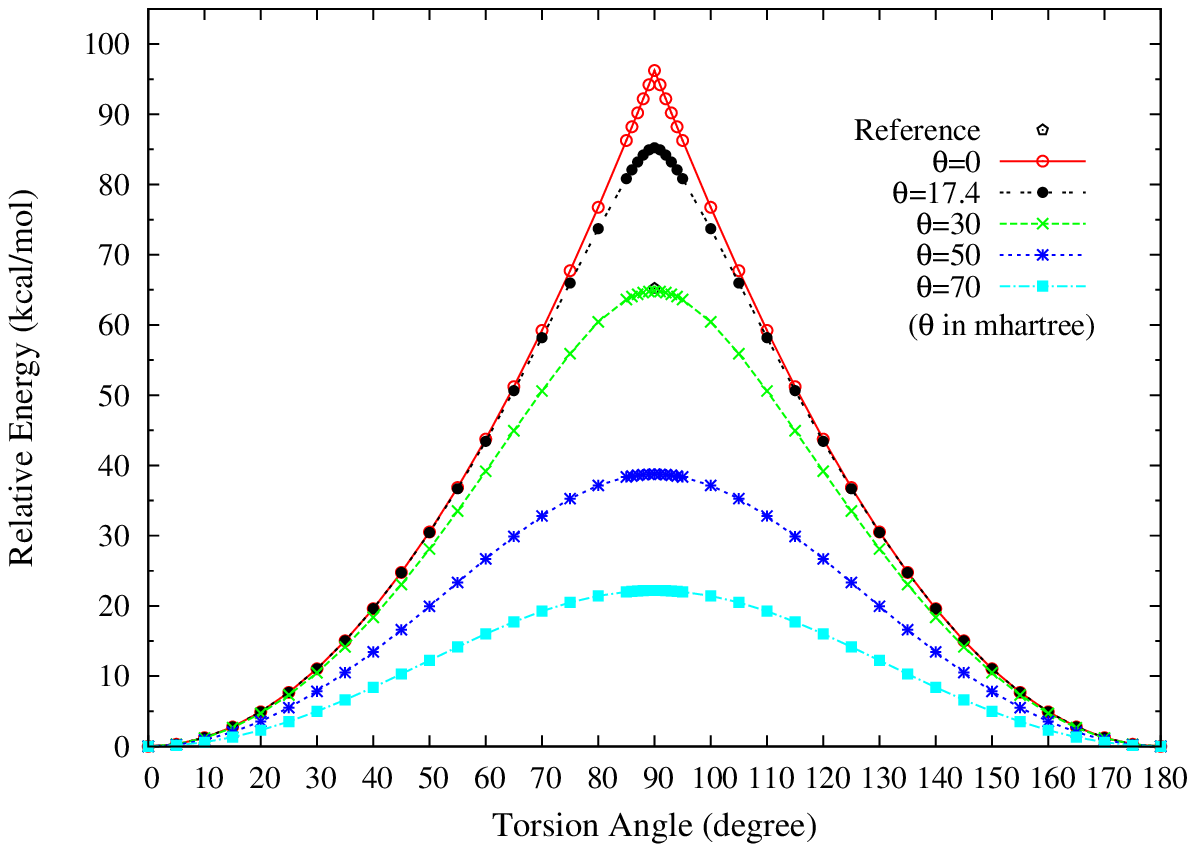}(b)} 
\subfigure 
{\includegraphics[scale=0.603]{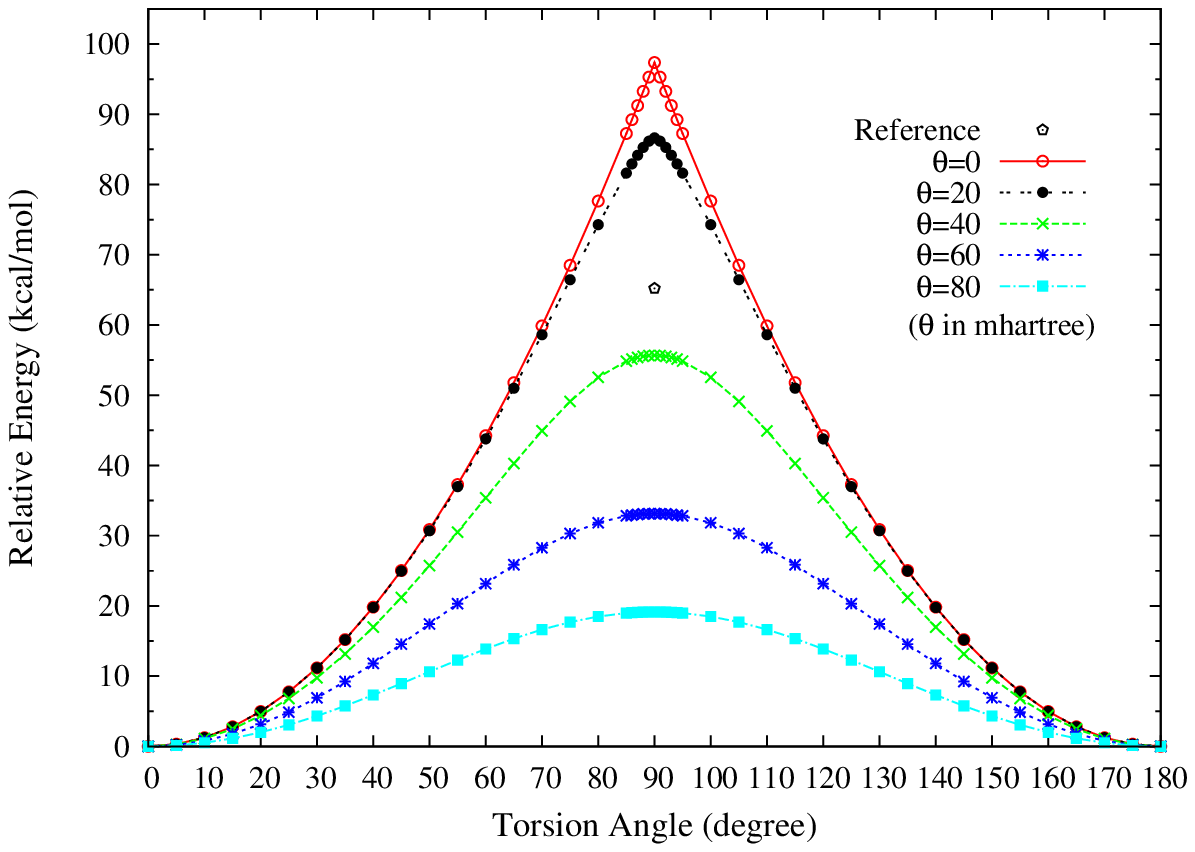}(c)} 
\subfigure 
{\includegraphics[scale=0.603]{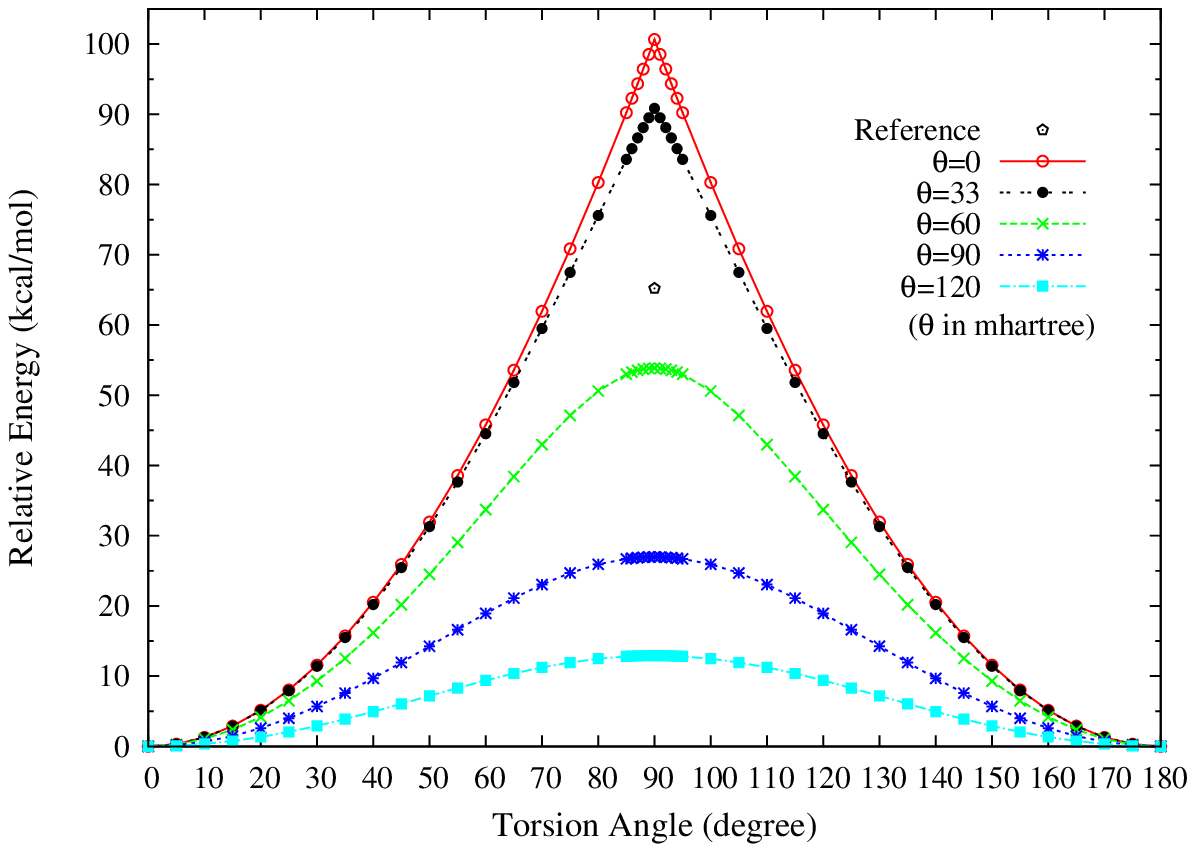}(d)} 
\caption{\label{fig:c2h4} 
Torsion potential energy curves (in relative energy) for the ground state of twisted ethylene as a function of the HCCH torsion angle, calculated using spin-restricted 
(a) TAO-B3LYP, (b) TAO-B3LYP-D3, (c) TAO-PBE0, and (d) TAO-BHHLYP (with various $\theta$). 
The $\theta = 0$ cases correspond to spin-restricted 
(a) KS-B3LYP, (b) KS-B3LYP-D3, (c) KS-PBE0, and (d) KS-BHHLYP, respectively. 
The reference data are the CASPT2 results \cite{C2H4_CASPT2}. 
The zeros of energy are set at the respective minimum energies.} 
\end{figure}

To investigate if spin-restricted TAO-DFT alleviates these problems, we plot 
the torsion potential energy curves (in relative energy) for the ground state of twisted ethylene as a function of the HCCH torsion angle, calculated using 
spin-restricted TAO-B3LYP, TAO-B3LYP-D3, TAO-PBE0, and TAO-BHHLYP with various $\theta$ values (see \Cref{fig:c2h4}), 
where the zeros of energy are set at the respective minimum energies. The experimental geometry of C$_2$H$_4$ ($R_{\text{CC}}$ = 1.339 {\AA}, 
$R_{\text{CH}}$ = 1.086 {\AA}, and $\angle_{\text{HCH}}$ = 117.6$^{\circ}$) \cite{C2H4_Geom} is adopted in the calculations. 
Spin-restricted TAO-B3LYP and TAO-B3LYP-D3 (with $\theta$ = 30 mhartree), TAO-PBE0 (with $\theta$ = 40 mhartree), and TAO-BHHLYP (with $\theta$ = 60 mhartree) 
can remove the unphysical cusp, and the corresponding torsion barriers are close to the torsion barrier of complete-active-space second-order perturbation theory (CASPT2), that is, 
65.2 (kcal/mol) \cite{C2H4_CASPT2}. We note, however, that the torsion barrier of TAO-DFT can be too low for a very large $\theta$ value, and too high for a very small $\theta$ value. 
While the global hybrid functionals in TAO-DFT (with the optimal $\theta$ values given in \Cref{table:opt_theta}) consistently outperform the corresponding global hybrid functionals in 
KS-DFT (i.e., the $\theta = 0$ cases), the predicted torsion barriers remain too high. Therefore, this indicates a limited applicability of TAO-DFT in its present form, showing the importance 
of finding an efficient way to estimate the appropriate $\theta$ value. 

\begin{figure} 
\subfigure 
{\includegraphics[scale=0.603]{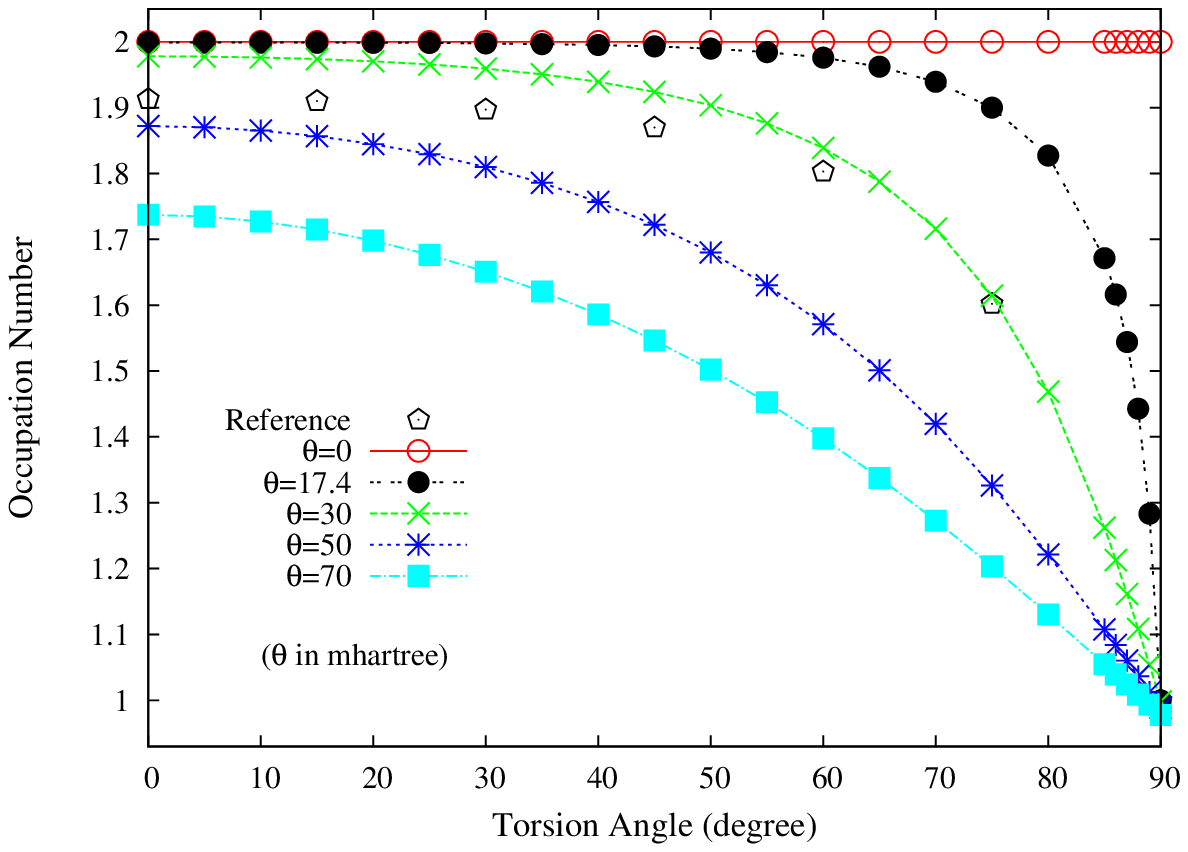}(a)} 
\subfigure 
{\includegraphics[scale=0.603]{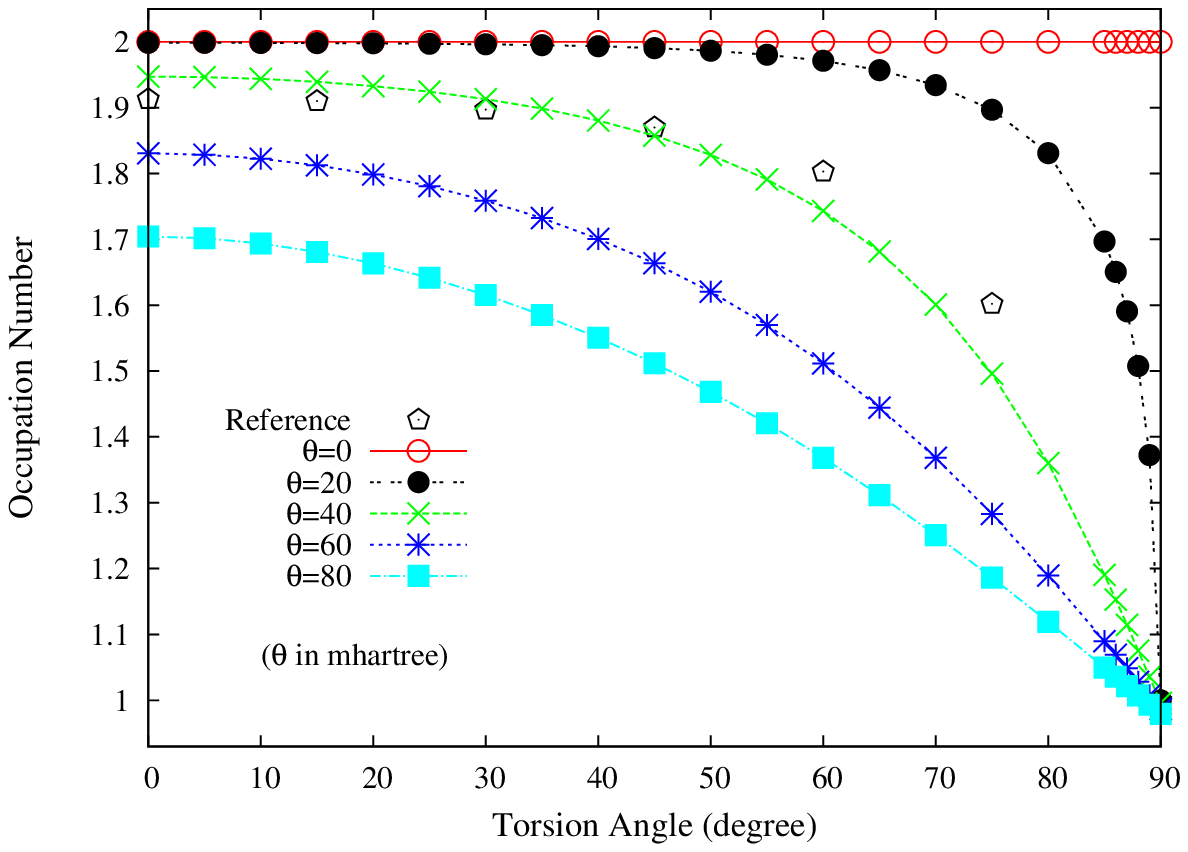}(b)} 
\subfigure 
{\includegraphics[scale=0.603]{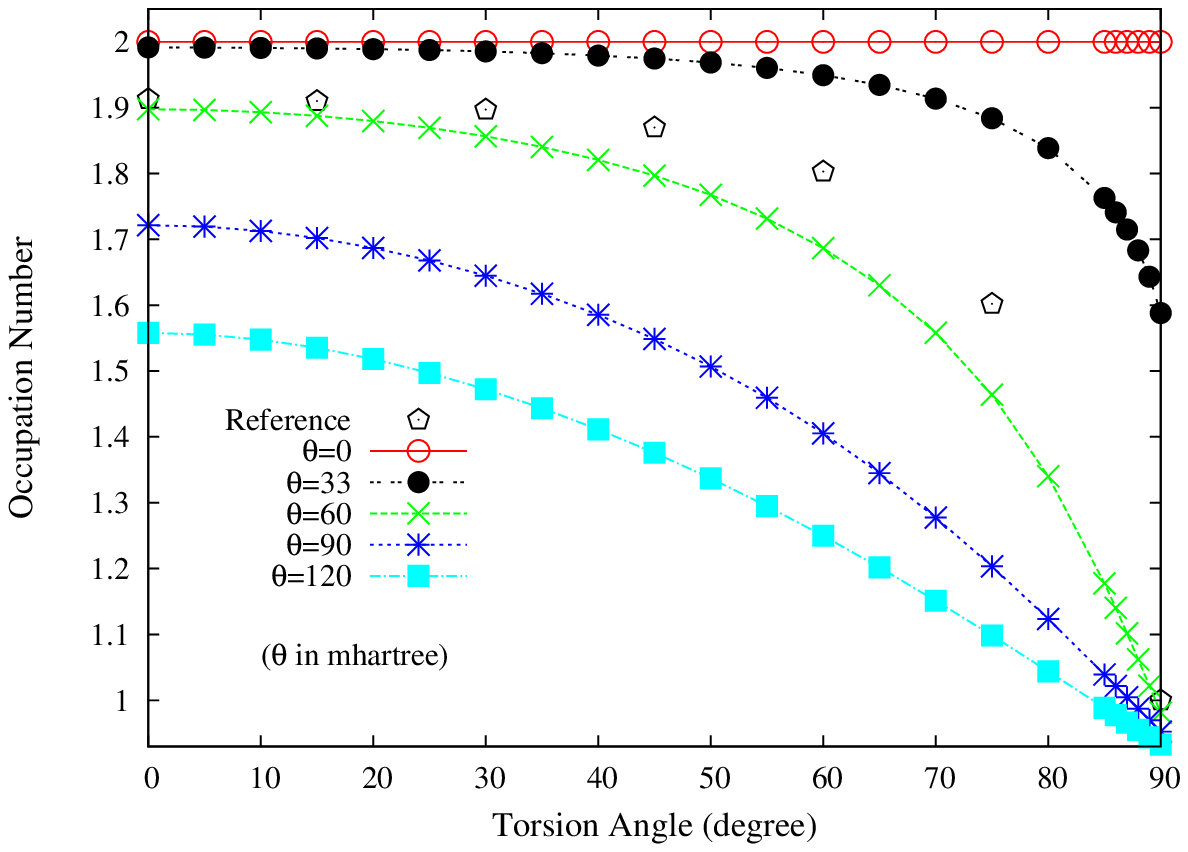}(c)} 
\caption{\label{fig:c2h4noon} 
Occupation numbers of the $\pi$ (1b$_2$) orbital for the ground state of twisted ethylene as a function of the HCCH torsion angle, calculated using spin-restricted 
(a) TAO-B3LYP/TAO-B3LYP-D3, (b) TAO-PBE0, and (c) TAO-BHHLYP (with various $\theta$). 
The $\theta = 0$ cases correspond to spin-restricted 
(a) KS-B3LYP/KS-B3LYP-D3, (b) KS-PBE0, and (c) KS-BHHLYP, respectively. 
The reference data are the half-projected NOONs of CASSCF method (HPNO-CAS) \cite{C2H4_NOON}.} 
\end{figure}

To assess if this is also related to the distribution of TOONs, we plot the occupation numbers of the $\pi$ (1b$_2$) orbital for the ground state of twisted ethylene as a function of the 
HCCH torsion angle, calculated using spin-restricted 
TAO-B3LYP/TAO-B3LYP-D3, TAO-PBE0, and TAO-BHHLYP with various $\theta$ values (see \Cref{fig:c2h4noon}), where the reference data are 
the half-projected NOONs of complete-active-space self-consistent field (CASSCF) method \cite{C2H4_NOON}. As shown, the $\pi$ (1b$_2$) orbital occupation numbers of spin-restricted 
TAO-B3LYP/TAO-B3LYP-D3 (with $\theta$ = 30 mhartree), TAO-PBE0 (with $\theta$ = 40 mhartree), and TAO-BHHLYP (with $\theta$ = 60 mhartree) match reasonably well with the 
accurate NOONs, which is closely related to the accurate torsion potential energy curves obtained from these global hybrid functionals in TAO-DFT (with the same $\theta$ values). 
Note that the $\pi$ (1b$_2$) orbital occupation numbers of spin-restricted TAO-BHHLYP (with a $\theta$ between 0 and 33 mhartree) are not correctly reduced to unity (singly occupied) 
near 90$^{\circ}$, yielding an unphysical cusp in the torsion potential. Again, this highlights the importance of adopting a $\theta$ related to the distribution of NOONs in TAO-DFT. 

\subsubsection{Electronic properties of linear acenes} 

\begin{figure} 
\includegraphics[scale=0.38]{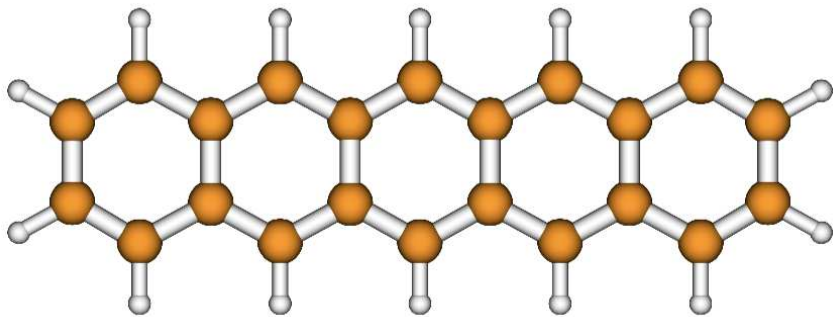} 
\caption{\label{fig:pentacene} 
Pentacene, containing 5 linearly fused benzene rings, is designated as 5-acene.} 
\end{figure}

Recently, linear $n$-acenes (C$_{4n+2}$H$_{2n+4}$), containing $n$ linearly fused benzene rings (see \Cref{fig:pentacene}), have attracted considerable interest in the research community 
owing to their promising electronic properties \cite{2-acene,3-acene,4-acene,5-acene,aceneTH,aceneIP,aceneChan,acene_IPEAFG,aceneJiang,aceneEA,aceneHajgato,aceneHajgato2,
aceneMazziotti,GNRs-DMRG,entropy,MR2,TAO1,TAO2,TAO3,TAO4,TAO5,TAO6}. The electronic properties of $n$-acenes have been found to be highly dependent on the chain lengths. 
Although there has been a keen interest in $n$-acenes, it remains very challenging to study the electronic properties of long-chain $n$-acenes from both experimental and theoretical 
approaches. On the experimental side, the synthetic procedures have been extremely difficult, and have not succeeded in synthesizing long-chain $n$-acenes, which may be attributed to 
their highly reactive nature. Consequently, the experimental singlet-triplet energy gaps (ST gaps) of $n$-acenes are only available up to pentacene \cite{2-acene,3-acene,4-acene,5-acene}. 
On the theoretical side, since $n$-acenes belong to conjugated $\pi$-orbital systems, high-level {\it ab initio} multi-reference methods, such as the density matrix renormalization group 
(DMRG) algorithm \cite{aceneChan,GNRs-DMRG}, the variational two-electron reduced density matrix (2-RDM) method \cite{aceneMazziotti,MR2}, and other high-level 
methods \cite{aceneIP,aceneEA,aceneHajgato,aceneHajgato2}, are typically required to capture the essential strong static correlation effects. Nevertheless, as the number of electrons in 
$n$-acene, $26n + 16$, quickly increases with the increase of $n$, there have been very scarce studies on the electronic properties of long-chain $n$-acenes using multi-reference methods 
due to their prohibitively high cost. 

On the other hand, despite their computational efficiency, conventional LDA, GGA, hybrid, and double-hybrid functionals in KS-DFT can perform very poorly for systems with strong static 
correlation effects \cite{SciYang,SCE,MR1,TAO1,TAO2}, and hence, their predicted electronic properties of $n$-acenes can be problematic \cite{TAO1,TAO2,TAO3,aceneChan,GNRs-DMRG}. 
By contrast, TAO-LDA and TAO-GGAs (with $\theta$ = 7 mhartree) were recently applied to study the electronic properties of $n$-acenes \cite{TAO1,TAO2,TAO3}, and the predicted electronic 
properties were shown to be in good agreement with the existing experimental and high-level {\it ab initio} data. 

To examine how global hybrid functionals in TAO-DFT improve upon the corresponding global hybrid functionals in KS-DFT here, spin-unrestricted calculations, employing TAO-B3LYP, 
TAO-B3LYP-D3, TAO-PBE0, and TAO-BHHLYP (with the optimal $\theta$ values given in \Cref{table:opt_theta}), are performed using the 6-31G(d) basis set (up to 30-acene), for the 
lowest singlet and triplet energies on the respective geometries that were fully optimized at the same level of theory. The ST gap of $n$-acene is calculated as $(E_{\text{T}} - E_{\text{S}})$, 
the energy difference between the lowest triplet (T) and singlet (S) states of $n$-acene. The results are compared with those calculated using the corresponding global hybrid functionals 
in spin-unrestricted KS-DFT. Besides, to compare with the ST gaps obtained from high-level {\it ab initio} methods, the DMRG data are taken from Ref.\ \cite{aceneChan}, and the 
CCSD(T)/CBS data (calculated using the CCSD theory with perturbative treatment of triple substitutions at the complete basis set limit) are taken from Ref.\ \cite{aceneHajgato2}. 

\begin{figure} 
\includegraphics[scale=0.83]{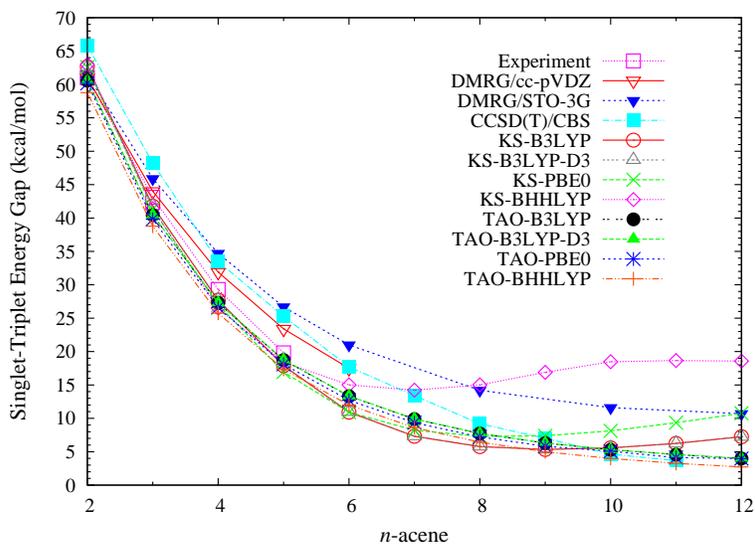} 
\caption{\label{fig:stg} 
Singlet-triplet energy gap as a function of the acene length, calculated using various hybrid functionals 
in spin-unrestricted KS-DFT and TAO-DFT (with the optimal $\theta$ values given in \Cref{table:opt_theta}). 
The experimental data (uncorrected for zero-point vibrations, thermal vibrations, etc.) are taken from Refs.\ \cite{2-acene,3-acene,4-acene,5-acene}, 
the DMRG data are taken from Ref.\ \cite{aceneChan}, and the CCSD(T)/CBS data are taken from Ref.\ \cite{aceneHajgato2}.} 
\end{figure}

\begin{figure} 
\includegraphics[scale=0.83]{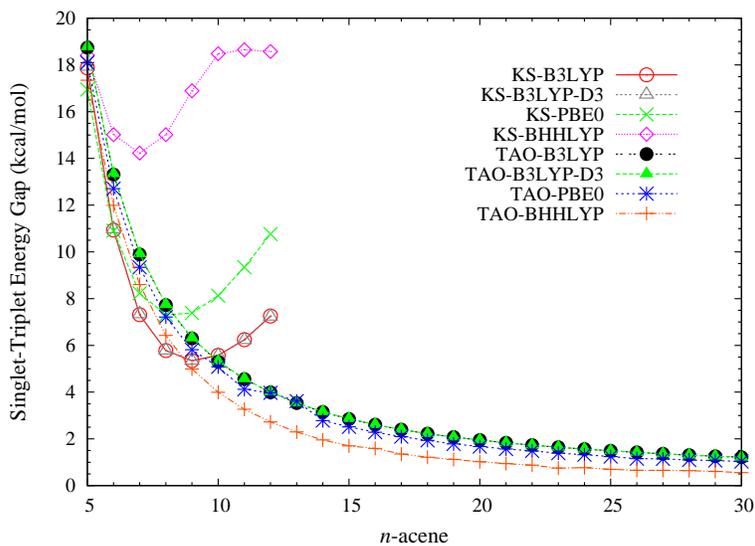} 
\caption{\label{fig:stg2} 
Same as \Cref{fig:stg}, but for the larger acenes.} 
\end{figure}

As shown in \Cref{fig:stg,fig:stg2}, in contrast to the accurate DMRG and CCSD(T)/CBS data, the ST gaps calculated using spin-unrestricted KS-DFT, unexpectedly increase beyond 
9-acene for KS-B3LYP and KS-B3LYP-D3, 8-acene for KS-PBE0, and 7-acene for KS-BHHLYP, due to unphysical symmetry-breaking effects (see the supplementary material). By contrast, 
the ST gaps calculated 
using spin-unrestricted TAO-B3LYP, TAO-B3LYP-D3, and TAO-PBE0 decrease monotonically as the size of the acene increases, which are in good agreement with the existing 
experimental \cite{2-acene,3-acene,4-acene,5-acene} and high-level {\it ab initio} \cite{aceneChan,aceneHajgato2} data. While the ST gaps calculated using spin-unrestricted TAO-BHHLYP 
unexpectedly increase beyond 23-acene, the deviation remains very small (within 0.02 kcal/mol). Similar to previous findings \cite{TAO1,TAO2,TAO3,aceneChan,GNRs-DMRG,MR2}, 
the ground states of $n$-acenes are singlets for all the chain lengths investigated. 

The spin-restricted and spin-unrestricted energies for the lowest singlet state of $n$-acene, calculated using the exact theory, should be identical due to the symmetry constraint. 
To examine this property, spin-restricted TAO-DFT calculations are also performed for the lowest singlet energies on the respective geometries that were fully optimized at the same level. 
For TAO-B3LYP/TAO-B3LYP-D3, the spin-unrestricted and spin-restricted calculations are found to essentially yield the same energy value for the lowest singlet state of $n$-acene 
(i.e., no unphysical symmetry-breaking effects). For TAO-PBE0 and TAO-BHHLYP, while symmetry-breaking effects occur, the maximum deviation between the spin-unrestricted and 
spin-restricted energy values remains small (within 0.5 kcal/mol). 

\begin{figure} 
\includegraphics[scale=0.83]{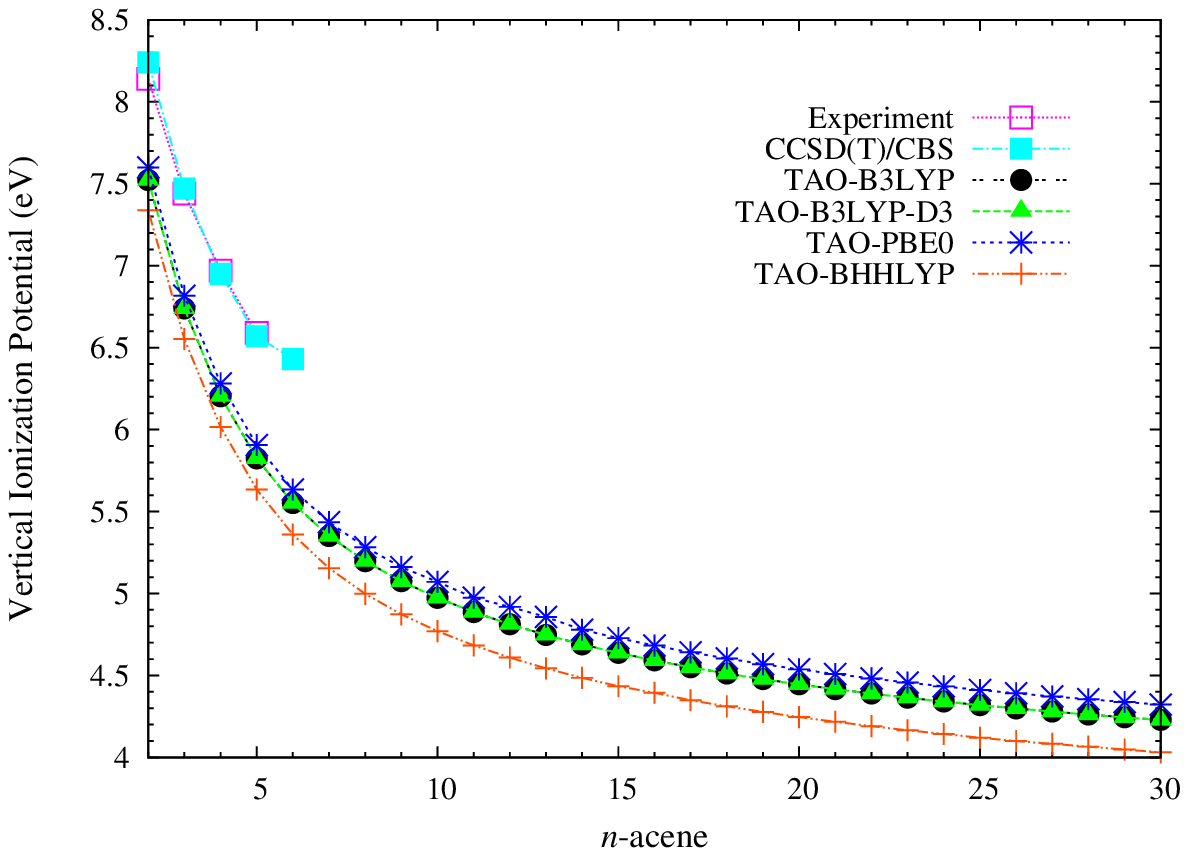} 
\caption{\label{fig:ip} 
Vertical ionization potential for the lowest singlet state of $n$-acene as a function of the acene length, calculated using various hybrid functionals 
in spin-unrestricted TAO-DFT (with the optimal $\theta$ values given in \Cref{table:opt_theta}). 
The experimental data are taken from the compilation in Ref.\ \cite{acene_IPEAFG}, and the CCSD(T)/CBS data are taken from Ref.\ \cite{aceneIP}.} 
\end{figure}

\begin{figure} 
\includegraphics[scale=0.83]{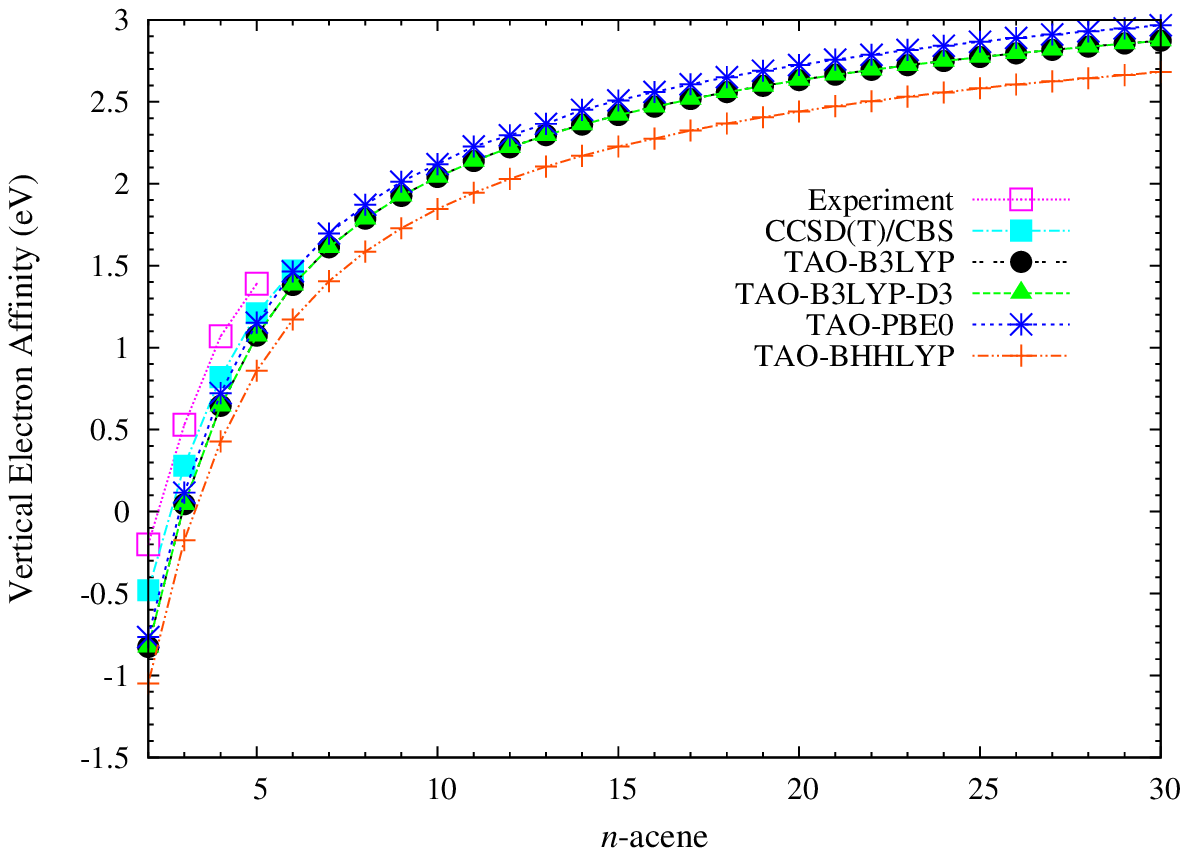} 
\caption{\label{fig:ea} 
Vertical electron affinity for the lowest singlet state of $n$-acene as a function of the acene length, calculated using various hybrid functionals 
in spin-unrestricted TAO-DFT (with the optimal $\theta$ values given in \Cref{table:opt_theta}). 
The experimental data are taken from the compilation in Ref.\ \cite{acene_IPEAFG}, and the CCSD(T)/CBS data are taken from Ref.\ \cite{aceneEA}.} 
\end{figure}

\begin{figure} 
\includegraphics[scale=0.83]{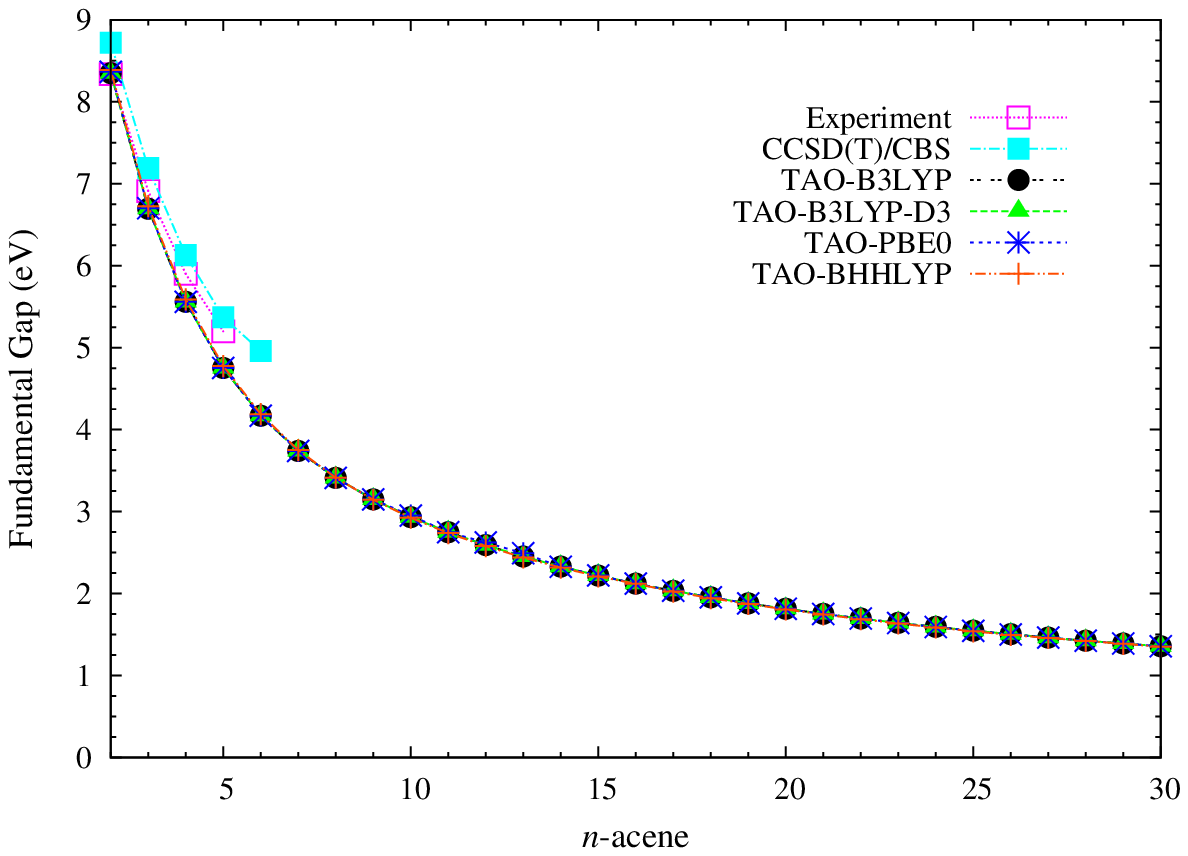} 
\caption{\label{fig:fg} 
Fundamental gap for the lowest singlet state of $n$-acene as a function of the acene length, calculated using various hybrid functionals 
in spin-unrestricted TAO-DFT (with the optimal $\theta$ values given in \Cref{table:opt_theta}). 
The experimental data are taken from the compilation in Ref.\ \cite{acene_IPEAFG}, and the CCSD(T)/CBS data are taken from Refs.\ \cite{aceneIP,aceneEA}.} 
\end{figure}

At the optimized geometry of the lowest singlet state (i.e., the ground state) of $n$-acene (containing $N$ electrons), the vertical ionization potential $\text{IP}_{v} = {E}_{N-1} - {E}_{N}$, 
vertical electron affinity $\text{EA}_{v} = {E}_{N} - {E}_{N+1}$, and fundamental gap $E_{g} = \text{IP}_{v} - \text{EA}_{v} = {E}_{N+1} + {E}_{N-1} - 2{E}_{N}$ are calculated using multiple 
energy-difference methods, where ${E}_{N}$ is the total energy of the $N$-electron system. With increasing chain length, $\text{IP}_{v}$ (see \Cref{fig:ip}) monotonically decreases, 
$\text{EA}_{v}$ (see \Cref{fig:ea}) monotonically increases, and hence $E_{g}$ (see \Cref{fig:fg}) monotonically decreases. The calculated $\text{IP}_{v}$, $\text{EA}_{v}$, and $E_{g}$ 
values are in good agreement with the available experimental \cite{acene_IPEAFG} and high-level {\it ab initio} \cite{aceneIP,aceneEA} data. Similar to our previous findings \cite{TAO2}, 
$E_{g}$ is rather insensitive to the choice of the XC functionals in TAO-DFT. 

\begin{figure} 
\includegraphics[scale=0.83]{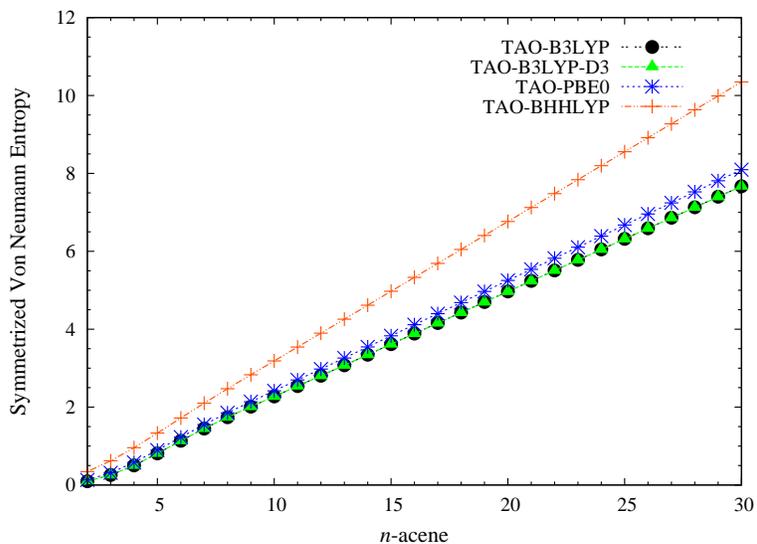} 
\caption{\label{fig:entropy} 
Symmetrized von Neumann entropy for the lowest singlet state of $n$-acene as a function of the acene length, calculated using various hybrid functionals 
in spin-restricted TAO-DFT (with the optimal $\theta$ values given in \Cref{table:opt_theta}).} 
\end{figure}

Since the TOONs are closely related to the NOONs, to investigate the possible polyradical character of $n$-acene, we compute the symmetrized von Neumann entropy 
(e.g., see Eq.\ (9) of Ref.\ \cite{entropy}) 
\begin{equation}\label{eq:svn} 
S_{\text{vN}} = -\frac{1}{2} \sum_{\sigma}^{\alpha,\beta} \sum_{i=1}^{\infty} \bigg\lbrace f_{i\sigma}\ \text{ln}(f_{i\sigma}) + (1-f_{i\sigma})\ \text{ln}(1-f_{i\sigma}) \bigg\rbrace   
\end{equation} 
for the lowest singlet state of $n$-acene as a function of the chain length, using spin-restricted TAO-DFT. Note that 
$S_{\text{vN}} = -\frac{1}{2 \theta} E_{S}^{\theta}[\{f_{i\alpha} \}, \{f_{i\beta} \}]$, which can be readily obtained in TAO-DFT, provides insignificant contributions for single-reference systems, 
and quickly increases with the number of fractionally occupied orbitals (i.e., active orbitals) for multi-reference systems. 
As shown in \Cref{fig:entropy}, $S_{\text{vN}}$ increases monotonically with the chain length. 

\begin{figure} 
\subfigure 
{\includegraphics[scale=0.603]{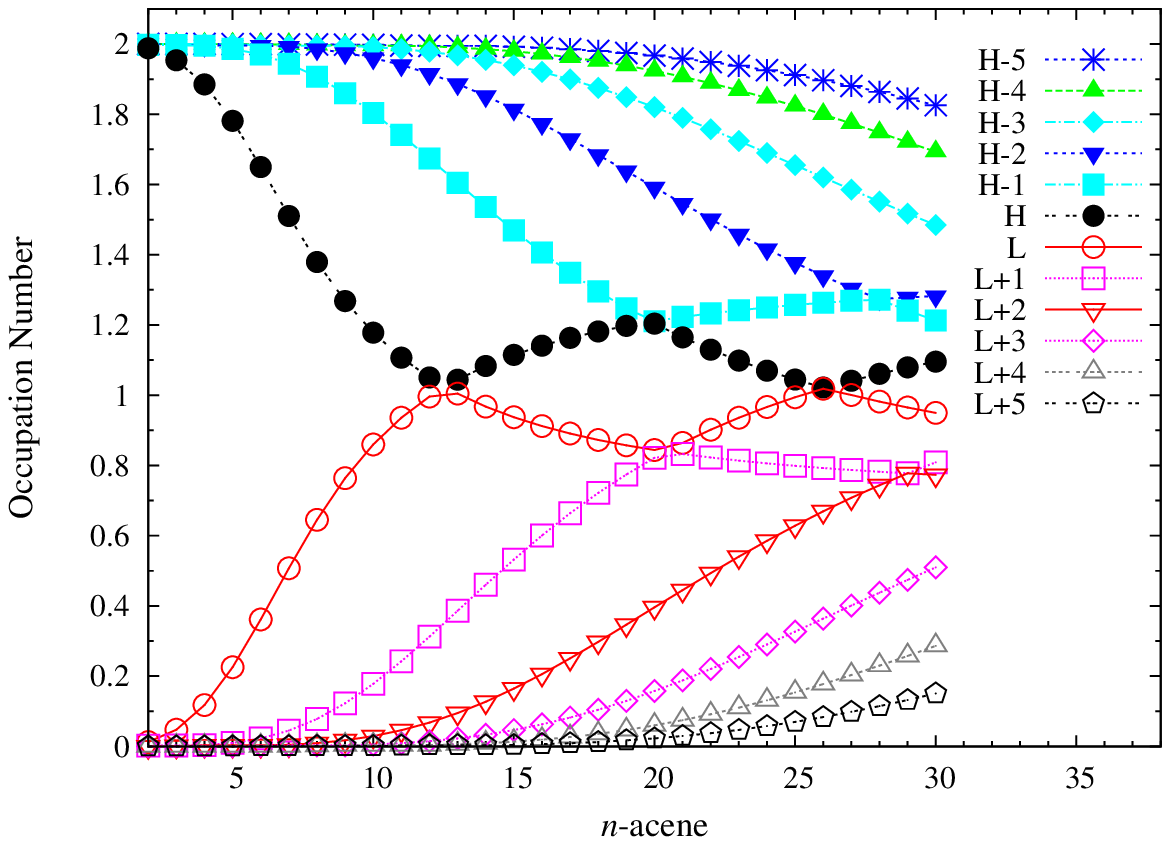}(a)} 
\subfigure 
{\includegraphics[scale=0.603]{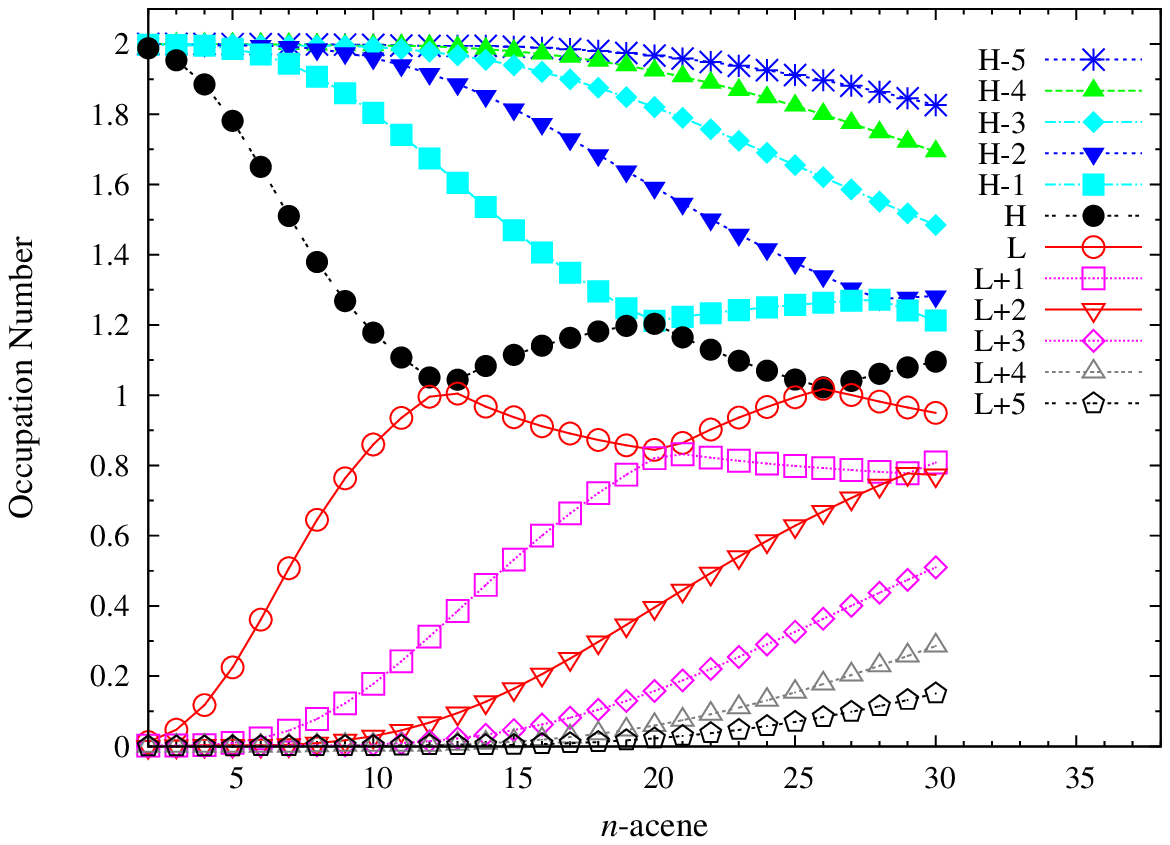}(b)} 
\subfigure 
{\includegraphics[scale=0.603]{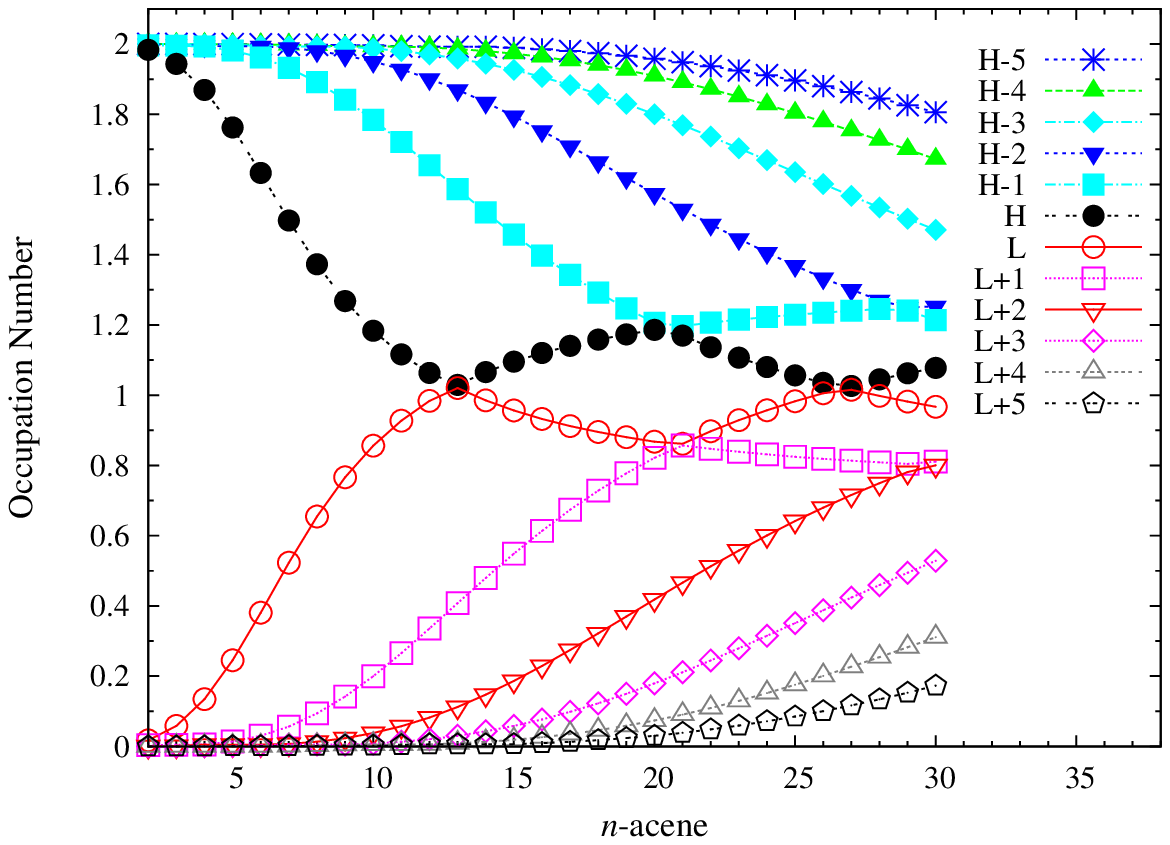}(c)} 
\subfigure 
{\includegraphics[scale=0.603]{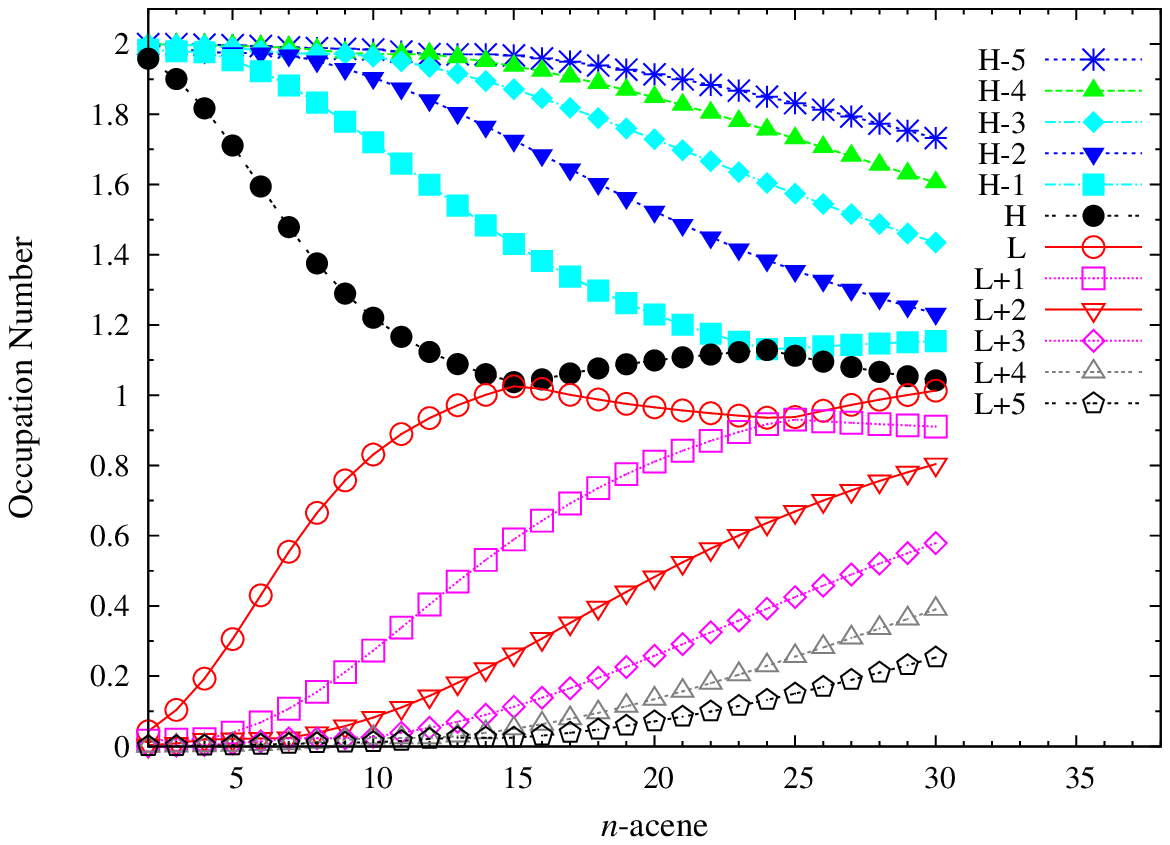}(d)} 
\caption{\label{fig:acenenoon} 
Active orbital occupation numbers (HOMO-5, ..., HOMO-1, HOMO, LUMO, LUMO+1, ..., and LUMO+5) for the lowest singlet state of $n$-acene as a function of the acene length, 
calculated using spin-restricted 
(a) TAO-B3LYP (with $\theta$ = 17.4 mhartree), (b) TAO-B3LYP-D3 (with $\theta$ = 17.4 mhartree), (c) TAO-PBE0 (with $\theta$ = 20 mhartree), and (d) TAO-BHHLYP (with $\theta$ = 33 mhartree).} 
\end{figure}

To understand the reasons of increasing $S_{\text{vN}}$ with the chain length, we plot the active orbital occupation numbers for the lowest singlet state of $n$-acene as a function of the 
chain length, calculated using spin-restricted TAO-B3LYP, TAO-B3LYP-D3, TAO-PBE0, and TAO-BHHLYP (see \Cref{fig:acenenoon}). 
Here, the highest occupied molecular orbital (HOMO) is the $(N/2)$-th orbital, and the lowest unoccupied molecular orbital (LUMO) 
is the $(N/2+1)$-th orbital, with $N$ being the number of electrons in $n$-acene. For brevity, HOMO, HOMO$-$1, ..., and HOMO$-$5, are denoted as H, H$-$1, ..., and H$-$5, respectively, 
while LUMO, LUMO+1, ..., and LUMO+5, are denoted as L, L+1, ..., and L+5, respectively. As shown, the number of fractionally occupied orbitals increases with the increase of chain length, 
supporting the previous findings that longer acenes should possess increasing polyradical character \cite{aceneChan,aceneJiang,GNRs-DMRG,entropy,TAO1,TAO2,TAO3}. However, 
in contrast to some previous studies \cite{aceneChan,entropy}, the active orbital occupation numbers display a curve crossing behavior in the approach to unity (singly occupied) with 
the increase of chain length. For examples, the orbital with HOMO (LUMO) character in short acenes may become the LUMO (HOMO) in long acenes. This curve crossing behavior was 
first observed from our TAO-LDA calculations \cite{TAO1,TAO3}, and was recently confirmed by highly accurate 2-RDM calculations \cite{MR2}. 
This is a very encouraging result, showing the value of TAO-DFT.

\section{Conclusions} 

In summary, we have proposed the global and range-separated hybrid schemes in TAO-DFT, incorporating the exact exchange into TAO-DFAs. For a global hybrid functional in TAO-DFT, 
a linear relationship between the optimal fictitious temperature $\theta$ and the fraction of exact exchange $a_{x}$ has been established. Global hybrid functionals in TAO-DFT (with the 
optimal $\theta$ values) have been shown to consistently improve upon the corresponding global hybrid functionals in KS-DFT for multi-reference systems, while performing similarly to the 
corresponding global hybrid functionals in KS-DFT for single-reference systems. In addition, the inclusion of dispersion corrections in hybrid TAO-DFT has been shown to yield an efficient 
and reasonably accurate description of noncovalent interactions. Relative to TAO-DFAs, global hybrid functionals in TAO-DFT are generally superior in performance for a broad range of 
applications, such as thermochemistry, kinetics, reaction energies, and optimized geometries. Owing to the computational efficiency, four global hybrid functionals in TAO-DFT (with the 
optimal $\theta$ values) have been applied to study the electronic properties of linear acenes, including the ST gaps, vertical ionization potentials, vertical electron affinities, fundamental 
gaps, symmetrized von Neumann entropy, and active orbital occupation numbers. The ground states of acenes have been found to be singlets for all the cases examined. With increasing 
acene length, the ST gaps, vertical ionization potentials, and fundamental gaps decrease monotonically, while the vertical electron affinities and symmetrized von Neumann entropy 
increase monotonically. Long acenes should possess singlet polyradical character in their ground states. 

Nonetheless, for a few multi-reference systems (e.g., the dissociation of H$_2$ and N$_2$, twisted ethylene, etc.), global hybrid functionals in TAO-DFT (with the optimal $\theta$ values) 
may not provide a sufficient amount of static correlation energy. Since a $\theta$ related to the distribution of NOONs should improve the performance of global hybrid 
functionals in TAO-DFT for a wide variety of systems, work in this direction is in progress. Besides, as the development of a possible range-separated hybrid functional in TAO-DFT would 
require $E_{x,\theta}^{\text{DFA}}(\text{I})$ (see Eq.\ (\ref{eq:exi_dfa})), which is mostly unavailable, we plan to pursue this in the future.

\section*{supplementary material} 

See supplementary material for further numerical results.

\begin{acknowledgements} 

This work was supported by the Ministry of Science and Technology of Taiwan (Grant No.\ MOST104-2628-M-002-011-MY3), National Taiwan University (Grant No.\ NTU-CDP-105R7818), 
the Center for Quantum Science and Engineering at NTU (Subproject Nos.:\ NTU-ERP-105R891401 and NTU-ERP-105R891403), and the National Center for Theoretical Sciences of Taiwan. 
We are grateful to the Computer and Information Networking Center at NTU for the support of high-performance computing facilities. 

\end{acknowledgements} 

\bibliographystyle{jcp}

\newpage 

\newpage 

\newpage 

\newpage 

\newpage 

\newpage 

\newpage 

\newpage 

\newpage 

\newpage 

\newpage 

\newpage 

\newpage 

\newpage 

\newpage 

\newpage 

\newpage 

\newpage 
\begin{table*} 
\begin{center} 
\caption{\label{table:opt_theta} 
Optimal fictitious temperature $\theta$ (in mhartree), given by Eq.\ (\ref{eq:opt_theta}), for TAO-B3LYP, TAO-B3LYP-D3, TAO-PBE0, and TAO-BHHLYP, 
where $a_{x}$ is the fraction of exact exchange.} 
\begin{ruledtabular} 
\begin{tabular}{lcccc} 
& TAO-B3LYP & TAO-B3LYP-D3 & TAO-PBE0 & TAO-BHHLYP \\ 
\hline 
$a_{x}$   &  1/5    &  1/5    &  1/4  &  1/2  \\ 
$\theta$  &  17.4  &  17.4  &  20   &  33  \\ 
\end{tabular} 
\end{ruledtabular} 
\end{center} 
\end{table*} 

\newpage 
\begin{table*} 
\begin{center} 
\caption{\label{table:reall} 
Statistical errors (in kcal/mol) of the reaction energies of the 30 chemical reactions in the NHTBH38/04 and HTBH38/04 sets \cite{BH}, calculated using 
TAO-B3LYP, TAO-B3LYP-D3, TAO-PBE0, and TAO-BHHLYP (with the optimal $\theta$ values given in \Cref{table:opt_theta}). The $\theta = 0$ cases correspond to 
KS-B3LYP, KS-B3LYP-D3, KS-PBE0, and KS-BHHLYP, respectively.} 
\begin{ruledtabular} 
\begin{tabular}{lrrrrrrrr} 
& \multicolumn{3}{r}{KS-DFT} & \multicolumn{4}{r}{TAO-DFT}  \\ 
\cline{2-5} 
\cline{6-9} 
& B3LYP & B3LYP-D3 & PBE0 & BHHLYP & B3LYP & B3LYP-D3 & PBE0 & BHHLYP \\ 
\hline 
MSE&       	-0.23 &	-0.27 &	-0.03 &	-1.25   &	-0.66 &	-0.70 &	-0.41  &	-1.76 \\ 
MAE&	         2.01 &	1.95 	 &	2.41 	 &	3.63 	   &	2.33 	 &	2.36 	 &	2.63 	  &	3.95 	\\ 
rms&                2.66 	&	2.61 	 &	3.35 	 &	4.72 	   &	3.05 	 &	3.07 	 &	3.69 	  &	5.00 	\\ 
Max($-$)& 	-7.38 &	-7.41 &	-7.11  &	-14.00 &	-8.44 &	-8.46 &	-8.40  &	-14.21 \\ 
Max($+$)&	4.46 	&	4.13 	 &	10.20 &	7.63 	   &	4.34 	 &	4.01 	 &	10.52 &	6.55 	\\ 
\end{tabular} 
\end{ruledtabular} 
\end{center} 
\end{table*} 

\newpage 
\begin{table*} 
\begin{center} 
\caption{\label{table:EXTS} 
Statistical errors (in {\AA}) of the 166 bond lengths in the EXTS set \cite{EXTS}, calculated using 
TAO-B3LYP, TAO-B3LYP-D3, TAO-PBE0, and TAO-BHHLYP (with the optimal $\theta$ values given in \Cref{table:opt_theta}). The $\theta = 0$ cases correspond to 
KS-B3LYP, KS-B3LYP-D3, KS-PBE0, and KS-BHHLYP, respectively.} 
\begin{ruledtabular} 
\begin{tabular}{lrrrrrrrr} 
& \multicolumn{3}{r}{KS-DFT} & \multicolumn{4}{r}{TAO-DFT}  \\ 
\cline{2-5} 
\cline{6-9} 
& B3LYP & B3LYP-D3 & PBE0 & BHHLYP & B3LYP & B3LYP-D3 & PBE0 & BHHLYP  \\ 
\hline 
MSE&	       0.003 	&	0.003 	&	-0.002 	&	-0.012 	&	0.003 	&	0.003 	&	-0.002 	&	-0.014 	\\ 
MAE&	       0.008 	&	0.008 	&	0.008 	&	0.013 	&	0.008 	&	0.008 	&	0.008 	&	0.015 	\\ 
rms&                0.013 	&	0.013 	&	0.012 	&	0.017 	&	0.013 	&	0.014 	&	0.013 	&	0.019 	\\ 
Max($-$)& 	-0.078 	&	-0.078 	&	-0.082 	&	-0.090 	&	-0.080 	&	-0.080 	&	-0.085 	&	-0.095 	\\ 
Max($+$)&	0.065 	&	0.065 	&	0.051 	&	0.025 	&	0.063 	&	0.063 	&	0.049 	&	0.035 	\\ 
\end{tabular} 
\end{ruledtabular} 
\end{center} 
\end{table*} 

\newpage 
\begin{table*} 
\caption{\label{table:training} 
Statistical errors (in kcal/mol) of the $\omega$B97 training set \cite{wB97X}, calculated using 
TAO-B3LYP, TAO-B3LYP-D3, TAO-PBE0, and TAO-BHHLYP (with the optimal $\theta$ values given in \Cref{table:opt_theta}). The $\theta = 0$ cases correspond to 
KS-B3LYP, KS-B3LYP-D3, KS-PBE0, and KS-BHHLYP, respectively.} 
\begin{ruledtabular} 
\begin{tabular*}{\textwidth}{llrrrrrrrr} 
& \multicolumn{4}{r}{KS-DFT} & \multicolumn{4}{r}{TAO-DFT}  \\ 
\cline{3-6} 
\cline{7-10} 
System & Error & B3LYP & B3LYP-D3 & PBE0 & BHHLYP & B3LYP & B3LYP-D3 & PBE0 & BHHLYP  \\ 
		\hline 
		G3/99 & MSE & -4.30 &	-1.99 &	3.94 	&	-29.55 &	0.90 	&	3.21 	&	11.48 	&	-11.32 \\ 
		(223) & MAE & 5.46 	&	3.64 	&	6.28 	&	29.68 &	5.25 	&	6.80 	&	13.34 	&	12.59 \\
		           & rms & 7.34 	&	5.23 	&	8.65 	&	34.13 &	6.97 	&	8.31 	&	17.16 	&	16.62 \\
		\hline
		IP     & MSE & 2.18 	&	2.17 	&	-0.13 &	-1.72 &	0.25 	&	0.24 &	-2.34 &	-5.66 \\
		(40)  & MAE & 3.68 	&	3.69 	&	3.33 	&	4.44 	&	4.25 	&	4.26 	&	4.37 	&	7.04 	\\
		          & rms & 4.81 	&	4.81 	&	3.98 	&	5.47 	&	5.30 	&	5.31 	&	5.27 	&	8.19 	\\		          
		\hline
		EA	& MSE & 1.71 	&	1.71 	&	-1.07 &	-4.79 &	-1.02 &	-1.02 &	-4.30 &	-9.98 \\
		(25)  & MAE & 2.38 	&	2.39 	&	3.10 	&	5.97 	&	3.48 	&	3.49 	&	4.63 	&	10.21 \\
		         & rms & 3.27 	&	3.29 	&	3.53 	&	6.84 	&	4.50 	&	4.52 	&	5.42 	&	11.38 \\
		\hline
		PA  & MSE & -0.77 	&	-0.66 &	0.18 &	-0.12 &	0.14 &	0.26 &	1.25 &	1.75 \\
		(8)  & MAE & 1.16 	&	1.07 	&	1.14 	&	1.55 	&	0.91 	&	1.01 	&	1.42 	&	2.02 	\\
		       & rms & 1.36 	&	1.33 	&	1.61 	&	1.78 	&	1.21 	&	1.27 	&	2.03 	&	2.63 	\\
		\hline
		NHTBH & MSE & -4.57 	&	-5.09 &	-3.13 &	0.52 &	-4.88 &	-5.39 &	-3.53 &	-0.51 \\
		(38)        & MAE & 4.69 	&	5.19 	&	3.63 	&	2.21 	&	5.08 	&	5.56 	&	4.18 	&	2.75 	\\
		               & rms & 5.71 	&	6.14 	&	4.63 	&	2.93 	&	6.02 	&	6.49 &	5.10 	&	3.29 \\
		\hline
		HTBH & MSE & -4.48 	&	-5.12 &	-4.60 &	0.58 &	-5.20 &	-5.84 &	-5.55 &	-1.43 \\
		(38)     & MAE & 4.56 	&	5.14 	&	4.60 	&	2.48 	&	5.20 	&	5.84 	&	5.55 	&	2.40 	\\
		            & rms & 5.10 	&	5.62 	&	4.88 	&	3.11 	&	5.79 	&	6.34 	&	5.80 	&	3.15 	\\
		\hline
		S22	& MSE & 3.95 	&	-0.02 &	2.50 	&	2.98 	&	2.74 &	-1.22 &	1.10 	&	0.18 \\
		(22)  & MAE & 3.95 	&	0.43 	&	2.52 	&	3.01 	&	2.76 	&	1.22 	&	1.49 	&	1.42 	\\
		         & rms & 5.17 	&	0.59 	&	3.62 	&	4.22 	&	3.98 	&	1.37 	&	2.40 	&	1.98 	\\
		\hline
		Total	 & MSE & -2.77 &	-1.80 &	1.55 &	-16.93 &	-0.34 &	0.63 	&	5.20 	&	-7.76 \\
		(394) & MAE & 4.75 	&	3.63 	&	5.05 	&	18.28 &	4.79 	&	5.69 	&	9.34 	&	9.11 	\\
		           & rms & 6.38 	&	5.03 	&	7.06 	&	25.85 &	6.27 	&	7.17 &	13.33 &	13.18 \\ 
\end{tabular*} 
\end{ruledtabular} 
\end{table*} 

\end{document}